\documentclass[aps,aps,twocolumn,superscriptaddress,dvips]{revtex4}
\usepackage{feynmp}
\usepackage{amsmath}
\usepackage{amssymb}
\usepackage{amsfonts}
\usepackage{epsfig}
\usepackage{graphicx}
\usepackage{subfigure}
\usepackage{hyperref}
\usepackage{bbm}
\usepackage{color}

\newcommand{\Rmnum}[1]{\expandafter\@slowromancap\romannumeral #1@}
\allowdisplaybreaks[4]
\begin{document}

\title{Fate of higher-order topological insulator under Coulomb interaction}

\author{Jing-Rong Wang}
\altaffiliation{Corresponding author: wangjr@hmfl.ac.cn}
\affiliation{High Magnetic Field Laboratory of Anhui Province,
Chinese Academy of Sciences, Hefei 230031, China}
\author{Chang-Jin Zhang}
\altaffiliation{Corresponding author: zhangcj@hmfl.ac.cn}
\affiliation{High Magnetic Field Laboratory of Anhui Province,
Chinese Academy of Sciences, Hefei 230031, China}
\affiliation{Institute of Physical Science and Information
Technology, Anhui University, Hefei 230601, China}

\begin{abstract}
In this article, we study the influence of long-range Coulomb interaction on three-dimensional second-order topological insulator (TI) by renormalization
group theory. We find that both the analysis method and conclusions in the recent Letter Phys. Rev. Lett. {\bf 127},
176601 (2021) are unreliable. There are two problems in this Letter. Firstly, the characteristic described by the RG flows $m\rightarrow\infty$
and $D\rightarrow0$ can not be used as the criterion for transition from second-order TI
to TI, since this characteristic could be essentially not induced by Coulomb interaction but only results from
the trivial power counting contribution of fermion action. Indeed, this characteristic is satisfied even for free second-order TI.
Second, the flow of $B$ is not paid attention, which is very important and should be seriously studied. In this article, we
analyze carefully the corrections for the flows of the model parameters induced by Coulomb interaction. We find that
the sign of $m$ changes but the sign of $B$ holds if the initial Coulomb strength is large enough, while the sign of $m$ holds
but the sign of $B$ changes if the initial Coulomb strength takes small values. These results indicate that second-order TI is
unstable to trivial band insulator not only under strong Coulomb interaction but also under weak Coulomb interaction. We also study the
effects of disorder scattering in second-order TI by renormalization group theory. According to the criterion in Phys. Rev. Lett. {\bf 127}, 176601 (2021),
weak disorder drives second-order TI to TI. However, we find that second-order TI is robust against weak disorder,
since weak disorder does not give qualitative modification for second-order TI. This result is consistent with recent studies based on other methods.
Interplay of Coulomb interaction and disorder in second-order TI is also investigated.
\end{abstract}

\maketitle

\section{Introduction}

The studies about topological materials including topological insulators (TIs), topological superconductors,
and various topological semimetals,  have became one of the most important fields in condensed matter physics
\cite{Hasan10, QiXL10, Vafek14, Wehling14, Sato17, Yan17, Hasan17, Armitage18, LvBQ21, Hasan21, Wieder21}.
Topological materials have critical potential applications, including quantum
computation, thermoelectric devices \emph{etc.}, due to their fantastic properties. Dirac, Weyl, and Majornana fermions have been observed in some
topological materials \cite{Hasan10, QiXL10, Vafek14, Wehling14, Sato17, Yan17, Hasan17, Armitage18, LvBQ21, Hasan21, Wieder21}.
These fermion excitations resemble the elementary particles in high-energy physics. Thus,
topological materials could provide a platform to simulate the concepts and phenomena in high-energy physics.
In some topological materials, there are also unusual fermion excitations, such as semi-Dirac fermions,
double-Weyl fermions, triple-Weyl fermions, multi-fold degenerate fermions \emph{etc.}, which have not counterparts
in high-energy physics \cite{LvBQ21, Hasan21, Wieder21}. These unusual fermion excitations could result in novel physical behaviors.

Recently, higher-order topological materials attracted a lot of interest \cite{XieBiYe21, Benalcazar17A, Benalcazar17B, Langbehn17, SongZhiDa17,
Schindler18, Ezawa18, ChenRui20, LiChangAn20, Roy21, FuBo21, HuYuSong21, WangC20, WangC21A, Ghorashi20, WangHaiXiao20, SzaboHODSM20, ZhangZhiQiang21,
GonzalezCuadra22}.
For $d$ dimensional TI, the system has $d-1$ gapless
Dirac edge states. For $d$ dimensional second-order TI, the system hosts $d-2$ gapless edge states. Concretely, three-dimensional (3D)
 second-order TI hosts 1D hinge states, and 2D second-order TI has 0D corner states.

Study about correlated interaction effects in topological materials is an important direction and attracted particular interest
\cite{Kotov12, Elias11, Siegel11, Yu13, Miao13, WangLiu12, Hofmann14, Goswami11, Hosur12, Tang18, Leaw19, Herbut06, Herbut09, Maciejko14, Roy16, Szabo21,
Moon13, Herbut14, YangNatPhys14, Abrikosov72, Isobe16, Cho16, WangLiuZhang17A, Lai15, Jian15, WangLiuZhang17B, ZhangShiXin17,
WangLiuZhang18, WangLiuZhang19, Han19, ZhangSX18, Roy18Birefringent, Kotov20, Roy17B, Roy18A, WangJing18, Roy18B, Boettcher20,
Savary14, Uryszek19, Sur19, Uryszek20, ZhaoPengLu21, LeeYuWenComment, LeeYuWen21, Roy16B, LiHeQiu22}.
For example, the theoretical studies showed that long-range Coulomb interaction induces singular fermion velocity renormalization for
2D Dirac fermions, which has been observed experimentally \cite{Kotov12, Elias11, Siegel11, Yu13, Miao13}.

Recently, Zhao \emph{et al.} studied the influence of long-range Coulomb interaction on 3D second-order TIs by renormalization group (RG) theory \cite{ZhaoPengLu21}.
They concluded that 3D second-order TIs are always unstable under Coulomb interaction. They showed that there are two types of transitions:
second-order TI to TI and second-order TI to trivial band insulator.

However, after careful studies, we find that both of the analysis method and conclusions in Ref.~\cite{ZhaoPengLu21} are unreliable.
There are two problems in Ref.~\cite{ZhaoPengLu21}. First, the characteristic described by the RG flows $m\rightarrow\infty$ and $D\rightarrow0$
can not be used as the criterion for transition from second-order TI
to TI, since this characteristic could be essentially not induced by Coulomb interaction but only results from
the trivial power counting contribution of fermion action. Actually, this characteristic is satisfied even for free second-order TI.
Second, the flow of $B$ is not paid attention, which is very important and should be seriously studied.

In the recent Comment \cite{LeeYuWenComment}, Lee and Yang have also pointed the problems in Ref.~\cite{ZhaoPengLu21} and indicated the
conclusions are misleading. The problems in Ref.~\cite{ZhaoPengLu21} pointed by Lee and Yang and the ones pointed by us are similar to each
other.

Whereas, there are also differences between the studies in the Comment \cite{LeeYuWenComment} and the studies by us. Based on further
calculations, they concluded that second-order TI is robust against weak Coulomb interaction. However, we find that the sign of $m$ holds
but the sign of the parameter $B$ changes for weak Coulomb interaction. Namely $mB$ changes under the weak Coulomb interaction. It represents
that second-order TI is unstable to trivial band insulator under weak Coulomb interaction.

We also study the influence of disorder on second-order TI. According to the criterion in Ref.~\cite{ZhaoPengLu21}, weak disorder drives
second-order TI to TI. However, we find that weak disorder does not give qualitative modification for second-order TI through RG analysis. Namely, second-order TI is robust
against weak disorder. Our result is consistent with recent studies about disorder effects in second-order TI based on other methods \cite{WangC20, WangC21A}.  The interplay
of long-range Coulomb interaction and disorder is also investigated.

The rest of paper is structured as follows. The model for second-order TI with long-range Coulomb interaction is defined in
Sec.~\ref{Sec:ModelCoulomb}. In Sec.~\ref{Sec:RGResultsCoulomb}, we present the RG analysis for influence of Coulomb interaction on second-order TI
based on numerical and analytical calculations.  We compare interaction effects in related systems and give discussions for some related questions in
Sec.~\ref{Sec:ResultsOtherAndDiscussions}. In Sec.~\ref{Sec:DisorderEffect}, we analyze the effects of disorder in second-order TI. The interplay of long-range
Coulomb interaction and disorder in second-order TI is studied in Sec.~\ref{Sec:Interplay}.   A brief summary is given in
Sec.~\ref{Sec:Summary}. The detailed calculations and derivations  are presented
in Appendices.

\section{Model \label{Sec:ModelCoulomb}}

The free action of fermions is
\begin{eqnarray}
S_{\Psi}=\int\frac{dk_{0}}{2\pi}\frac{d^{3}\mathbf{k}}{(2\pi)^{3}}\bar{\Psi}(k_{0},\mathbf{k})
\left(ik_{0}\gamma_{0}+\mathcal{H}_{f}
\right)\Psi(k_{0},\mathbf{k}),
\end{eqnarray}
where $\Psi$ is four-component spinor and $\bar{\Psi}=\Psi^{\dag}\gamma_{0}$.  The fermion
Hamiltonian density takes the form
\begin{eqnarray}
\mathcal{H}_{f}&=&i\left[v\left(k_{x}\gamma_{x}+k_{y}\gamma_{y}\right)+
v_{z}k_{z}\gamma_{z}+D\left(k_{x}^{2}-k_{y}^{2}\right)\gamma_{5}\right]\nonumber
\\
&&+m-B_{\bot}k_{\bot}^{2}
-B_{z}k_{z}^{2}. \label{Eq:HamiltonianFermion}
\end{eqnarray}
$v$, $v_{z}$, $D$, $B_{\bot}$, $B_{z}$ are model parameters.
If $mB_{\bot,z}>0$ and $D\neq0$, it corresponds to second-order TI.
If $mB_{\bot,z}>0$ and $D=0$, it corresponds to  TI. If $mB_{\bot,z}<0$ , it
corresponds to trivial band insulator. The matrices $\gamma_{0}$, $\gamma_{x}$,
$\gamma_{y}$, $\gamma_{z}$ and $\gamma_{5}$ satisfy the anticommuting
relation $\left\{\gamma_{\mu},\gamma_{\nu}\right\}=2\delta_{\mu\nu}$.

The long-range Coulomb
interaction between fermions can be written as
\begin{eqnarray}
H_{\mathrm{C}} = \frac{1}{4\pi}\int d^3\mathbf{x} d^3
\mathbf{x}'\rho(\mathbf{x}) \frac{e^2}{\epsilon\left|\mathbf{x} -
\mathbf{x}'\right|}\rho(\mathbf{x}'),
\end{eqnarray}
where $\rho(\mathbf{x}) = \Psi^{\dag}(\mathbf{x})
\Psi(\mathbf{x})$ is the fermion density operator, $e$
electric charge, and $\epsilon$  dielectric constant.
The Coulomb
interaction can be decoupled by introducing a bosonic field $\phi$
through Hubbard-Stratonovich transformation.
Accordingly, the Coulomb interaction between fermions can be described by the action of fermion-boson coupling as
\begin{eqnarray}
S_{\Psi\phi}&=&ig\int d\tau d^3\mathbf{x}\bar{\Psi}\gamma_{0}\Psi\phi,
\end{eqnarray}
where $g=e/\sqrt{\epsilon}$.
In energy-momentum space, it takes the form
\begin{eqnarray}
S_{\Psi\phi}&=&ig\int\frac{dk_{0,1}}{2\pi}\frac{d^3\mathbf{k}_{1}}{(2\pi)^{3}}\frac{dk_{0,2}}{2\pi}
\frac{d^3\mathbf{k}_{2}}{(2\pi)^{3}}\bar{\Psi}(k_{0,1},\mathbf{k}_{1})\gamma_{0}\nonumber
\\
&&\times\Psi(k_{0,2},\mathbf{k}_{2})\phi(k_{0,1}-k_{0,2},\mathbf{k}_{1}
-\mathbf{k}_{2}).
\end{eqnarray}
The free action of boson field $\phi$ can be written as
\begin{eqnarray}
S_{\phi}=\int\frac{dk_{0}}{2\pi}\frac{d^3\mathbf{k}}{(2\pi)^{3}}\phi(k_{0},\mathbf{k})
\left(k_{\bot}^{2}+\eta k_{z}^{2}
\right)\phi(k_{0},\mathbf{k}),
\end{eqnarray}
where $\eta$ is used to describe the anisotropy of $\phi$.

\section{RG Results \label{Sec:RGResultsCoulomb}}

In this article, we study the influence of long-range Coulomb interaction on second-order TI through RG method \cite{Shankar94}.
After detailed calculations and derivations shown in Appendices, we obtain the RG equations as following,
\begin{eqnarray}
\frac{dv}{d\ell}&=&\alpha\mathcal{R}_{v}v,
\\
\frac{dv_{z}}{d\ell}&=&\alpha\mathcal{R}_{v_{z}}v_{z},
\\
\frac{dm}{d\ell}
&=&m+\alpha\mathcal{R}_{m},
\\
\frac{dB_{\bot}}{d\ell}
&=&-B_{\bot}+\alpha\mathcal{R}_{B_{\bot}}
\\
\frac{dB_{z}}{d\ell}
&=&-B_{z}+\alpha\mathcal{R}_{B_{z}}
\\
\frac{dD}{d\ell}
&=&-D+\alpha \mathcal{R}_{D}D,
\\
\frac{d\alpha}{d\ell}
&=&-\alpha^{2}\mathcal{R}_{\alpha},
\\
\frac{d\zeta}{d\ell}
&=&\alpha\mathcal{R}_{\zeta}\zeta,
\end{eqnarray}
where $\ell$ is the RG running parameter. The strength of Coulomb interaction $\alpha$ and the parameter $\zeta$ are defined as
\begin{eqnarray}
\alpha&=&\frac{g^{2}}{4\pi^{2}v\sqrt{\eta}},
\\
\zeta&=&\frac{v_{z}}{v\sqrt{\eta}}.
\end{eqnarray}
The transformations
\begin{eqnarray}
\frac{m}{v\Lambda}\rightarrow m,
\\
\frac{B_{\bot}\Lambda}{v}\rightarrow B_{\bot},
\\
\frac{B_{z}\Lambda}{v\eta}\rightarrow B_{z},
\\
\frac{D\Lambda}{v}\rightarrow D,
\end{eqnarray}
have been utilized. $\mathcal{R}_{v}$, $\mathcal{R}_{v_{z}}$, $\mathcal{R}_{m}$, $\mathcal{R}_{B_{\bot}}$,
$\mathcal{R}_{B_{z}}$, $\mathcal{R}_{D}$, $\mathcal{R}_{\alpha}$, and $\mathcal{R}_{\zeta}$ are functions
of $m$, $B_{\bot}$, $B_{z}$, $D$, and $\zeta$, whose concrete expressions can be found in Appendix~\ref{Appendix:DerivationRGEs}.

For free fermions, the RG equations become
\begin{eqnarray}
\frac{dv^{f}}{d\ell}&=&0,
\\
\frac{dv_{z}^{f}}{d\ell}&=&0,
\\
\frac{dm^{f}}{d\ell}
&=&m,
\\
\frac{dB_{\bot}^{f}}{d\ell}
&=&-B_{\bot},
\\
\frac{dB_{z}^{f}}{d\ell}
&=&-B_{z},
\\
\frac{dD^{f}}{d\ell}
&=&-D,
\end{eqnarray}
where the superscript $f$ corresponds to free fermions.
The corresponding solutions are
\begin{eqnarray}
v^{f}(\ell)&=&v_{0}, \label{Eq:vFreeExp}
\\
v_{z}^{f}(\ell)&=&v_{z0}, \label{Eq:vzFreeExp}
\\
m^{f}(\ell)&=&m_{0}e^{\ell}, \label{Eq:mFreeExp}
\\
B_{\bot}^{f}(\ell)&=&B_{\bot0}e^{-\ell}, \label{Eq:BbotFreeExp}
\\
B_{z}^{f}(\ell)&=&B_{z0}e^{-\ell}, \label{Eq:BzFreeExp}
\\
D^{f}(\ell)&=&D_{0}e^{-\ell}, \label{Eq:DFreeExp}
\end{eqnarray}
where $v_{0}$, $v_{z0}$, $B_{\bot0}$, $B_{z0}$, and $D_{0}$ are initial
values.
It is easy to find that $v^{f}(\ell)$ and $v_{z}^{f}(\ell)$ are fixed,
and
\begin{eqnarray}
m^{f}(\ell)&\rightarrow&\infty, \label{Eq:mFreeExpLimit}
\\
B_{\bot}^{f}(\ell)&\rightarrow&0, \label{Eq:BbotFreeExpLimit}
\\
B_{z}^{f}(\ell)&\rightarrow&0,  \label{Eq:BzFreeExpLimit}
\\
D^{f}(\ell)&\rightarrow&0, \label{Eq:DFreeExpLimit}
\end{eqnarray}
in the lowest energy limit $\ell\rightarrow\infty$.
These results are directly related to the fact that $v$ and $v_{z}$
are coefficients of linear terms of momentum components, $m$ is coefficient of
zero power of momentum components, and $B_{\bot}$, $B_{z}$, $D$ are coefficients of
quadratic terms of momentum components.

\begin{figure}[htbp]
\center
\includegraphics[width=3.3in]{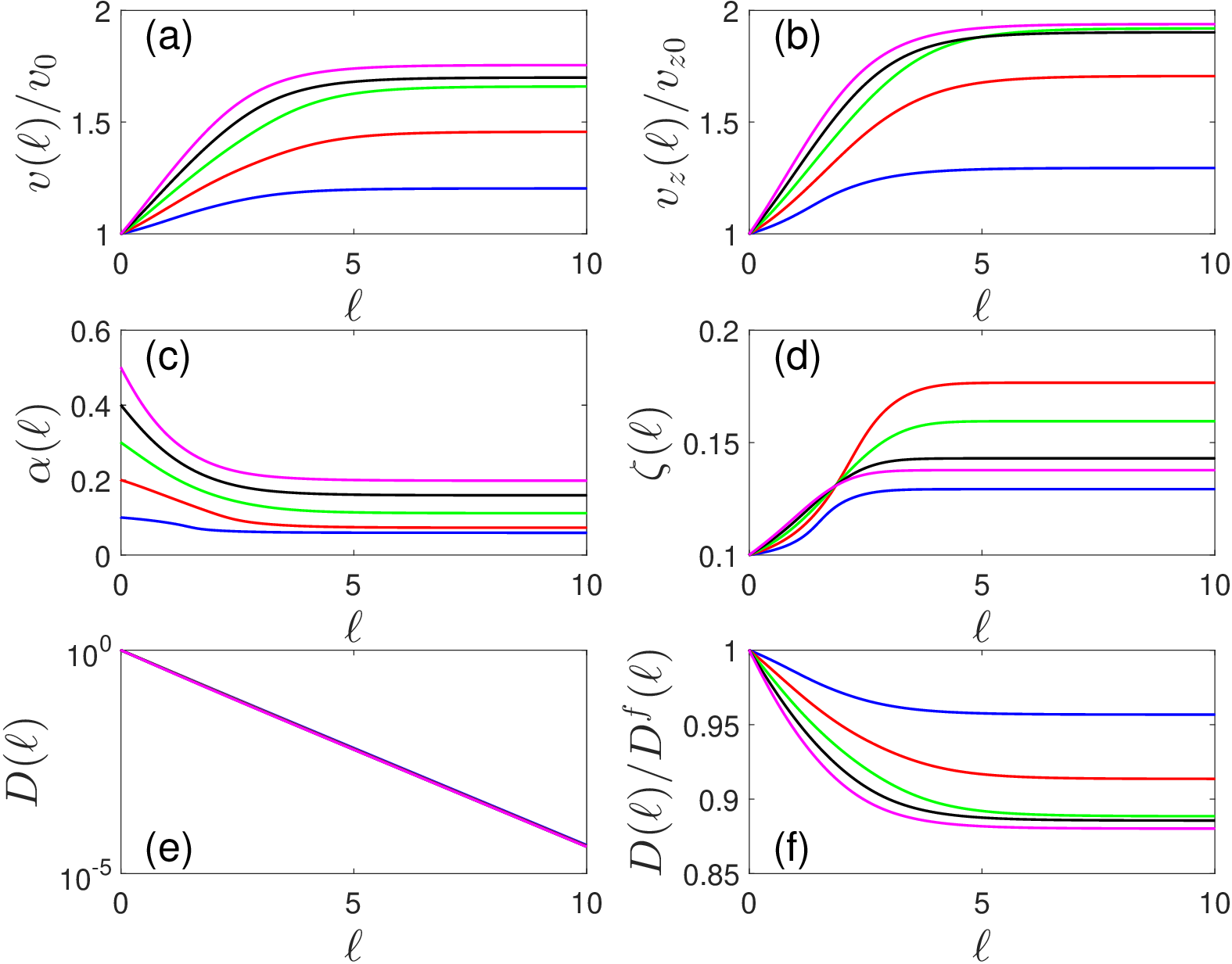}
\caption{(a)-(f): Flows of $v(\ell)$, $v_{z}(\ell)$, $\alpha(\ell)$, $\zeta(\ell)$, $D(\ell)$,
and $D(\ell)/D^{f}(\ell)$ with different initial values of Coulomb strength. Blue, red, green, black, and magenta lines
correspond to the initial values $\alpha_{0}=0.1, 0.2, 0.3, 0.4, 0.5$ respectively. $m_{0}=0.1$, $B_{\bot0}=1$, $B_{z0}=1$, $D_{0}=1$, $\zeta_{0}=0.1$ are taken. \label{Fig:VRGArrayA}}
\end{figure}

The physical meaning of these results shown in  Eqs.~(\ref{Eq:vFreeExp})-(\ref{Eq:DFreeExp}) and Eqs.~(\ref{Eq:mFreeExpLimit})-(\ref{Eq:DFreeExpLimit}) is:
Taking the linear terms of Hamiltonian as reference energy, the zero
power term becomes larger and larger, and the quadratic terms become smaller and smaller, with lowering of momentum. Namely,
\begin{eqnarray}
m/E_{linear}(\mathbf{k})&\rightarrow& \infty,
\\
Bk_{\bot}^2/E_{linear}(\mathbf{k})&\rightarrow& 0,
\\
Bk_{z}^2/E_{linear}(\mathbf{k})&\rightarrow& 0,
\\
D(k_{x}^{2}-k_{y}^{2})/E_{linear}(\mathbf{k})&\rightarrow& 0,
\end{eqnarray}
where
\begin{eqnarray}
E_{linear}(\mathbf{k})=\sqrt{v^{2}k_{\bot}^{2}+v_{z}^{2}k_{z}^{2}},
\end{eqnarray}
in the limit $k\rightarrow 0$.

In Ref.~\cite{ZhaoPengLu21}, the characteristic described by the RG flows
$m(\ell)\rightarrow\infty$ and $D(\ell)\rightarrow0$ in the limit $\ell\rightarrow\infty$
are used as the criterion for transition from second-order
TI to TI. It is clear that using this characteristic as the criterion for transition from second-order
TI to TI is invalid. This characteristic is even satisfied for free second-order TI.

\begin{figure}[htbp]
\center
\includegraphics[width=3.3in]{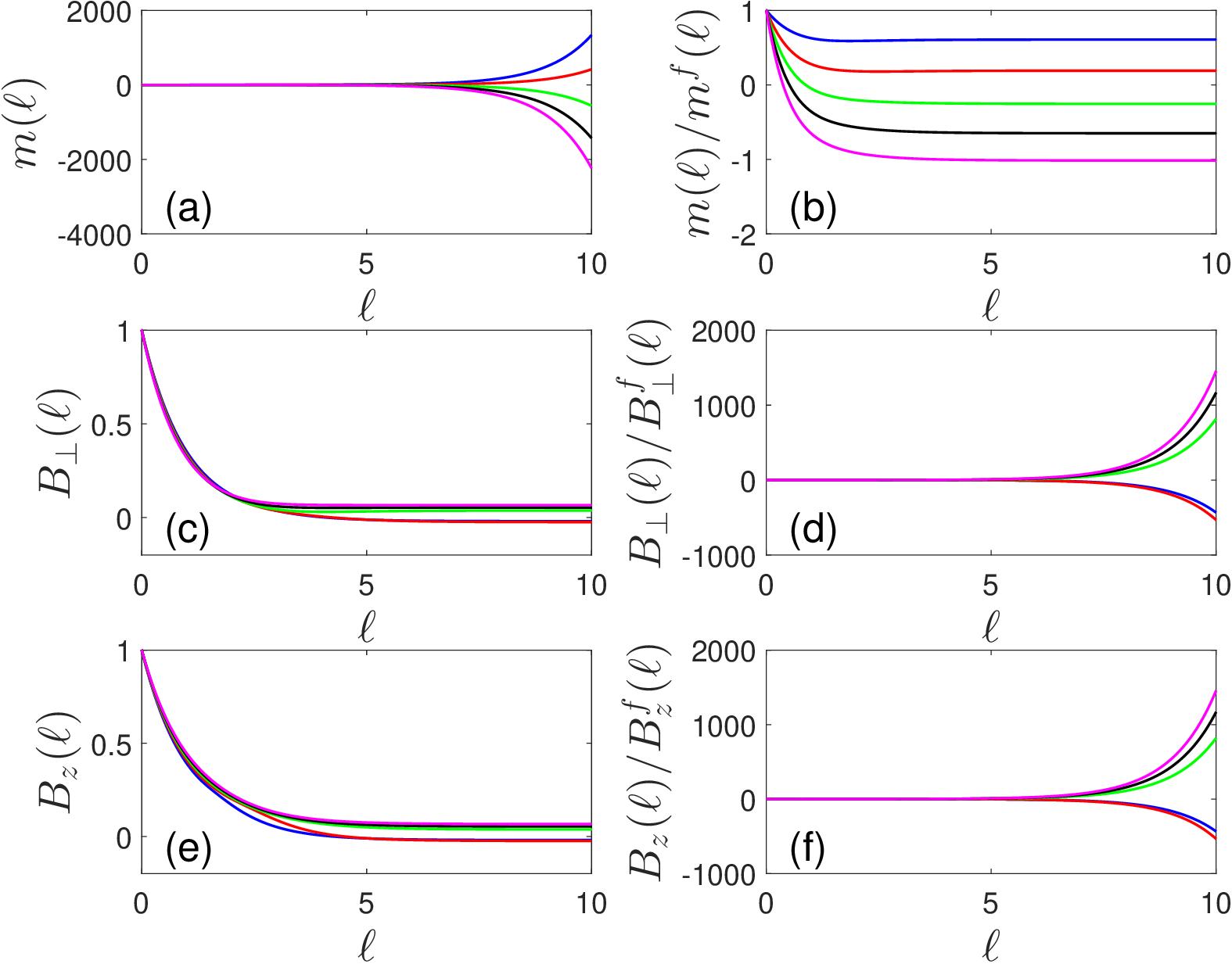}
\caption{(a)-(f): Flows of $m(\ell)$, $m(\ell)/m^{f}(\ell)$, $B_{\bot}(\ell)$, $B_{\bot}(\ell)/B_{\bot}^{f}(\ell)$, $B_{z}(\ell)$,
and $B_{z}(\ell)/B_{z}^{f}(\ell)$ with different initial values of Coulomb strength. Blue, red, green, black, and magenta lines
correspond to the initial values $\alpha_{0}=0.1, 0.2, 0.3, 0.4, 0.5$ respectively. $m_{0}=0.1$, $B_{\bot0}=1$, $B_{z0}=1$, $D_{0}=1$, $\zeta_{0}=0.1$ are taken. \label{Fig:VRGArrayB}}
\end{figure}

The flows of $v(\ell)$, $v_{z}(\ell)$, $m(\ell)$, $\alpha(\ell)$, $D(\ell)$, and $D(\ell)/D^{f}(\ell)$ with different initial values of Coulomb strength are presented in
Figs.~\ref{Fig:VRGArrayA}(a)-\ref{Fig:VRGArrayA}(f) respectively. We find that $v(\ell)/v_{0}$  and $v_{z}(\ell)/v_{z0}$ flow to positive constants.
Thus, Coulomb interaction results in quantitative corrections for $v$ and $v_{z}$. According to Figs.~\ref{Fig:VRGArrayA}(c) and \ref{Fig:VRGArrayA}(d),
\begin{eqnarray}
\alpha(\ell)&\rightarrow&\alpha^{*},
\\
\zeta(\ell)&\rightarrow&\zeta^{*},
\end{eqnarray}
where $\alpha^{*}$ and $\zeta^{*}$ are positive constants.
As shown in Fig.~\ref{Fig:VRGArrayA}(e), $D(\ell)$ approaches to zero quickly with lowering of energy scale.
From Fig.~\ref{Fig:VRGArrayA}(f),  we find that
$D(\ell)/D^{f}(\ell)$ flows to a positive constant value in the low energy regime. It indicates that $D(\ell)$ only
acquires quantitative correction in presence of Coulomb interaction, and takes qualitatively same behavior as $D^{f}(\ell)$ .
It means that qualitative behaviors of the term $D(k_{x}^{2}-k_{y}^{2})$ are not changed under Coulomb interaction. As the
term $D(k_{x}^{2}-k_{y}^{2})$ is not changed qualitatively by Coulomb interaction, the transition from
second-order TI to TI stated in Ref.~\cite{ZhaoPengLu21} does not exist.

We presented the flows of $m(\ell)$, $m(\ell)/m^{f}(\ell)$, $B_{\bot}(\ell)$,  $B_{\bot}(\ell)/B_{\bot}^{f}(\ell)$, $B_{z}(\ell)$,
$B_{z}(\ell)/B_{z}^{f}(\ell)$ with different initial values of Coulomb strength in Figs.~\ref{Fig:VRGArrayB}(a)-\ref{Fig:VRGArrayB}(f)
respectively. If the initial Coulomb strength is small, we find that
\begin{eqnarray}
m(\ell)/m^{f}(\ell)\rightarrow c_{m}^{*},\quad\mathrm{with}\quad c_{m}^{*}>0,
\end{eqnarray}
where $c_{m}^{*}$ is a positive constant. It means that $m(\ell)$ takes the qualitatively same behavior as $m^{f}(\ell)$.
According to Figs.~\ref{Fig:VRGArrayB}(c) and \ref{Fig:VRGArrayB}(e),
\begin{eqnarray}
B_{\bot}(\ell)&\rightarrow& c_{B_{\bot}}^{*},\quad \mathrm{with}\quad c_{B_{\bot}}^{*}<0,
\\
B_{z}(\ell)&\rightarrow& c_{B_{z}}^{*}, \quad \mathrm{with}\quad c_{B_{z}}^{*}<0,
\end{eqnarray}
with lowering of energy scale, where $c_{B_{\bot}}^{*}$ and $c_{B_{z}}^{*}$ are negative constants.
We can find that signs of $B_{\bot}(\ell)$ and $B_{z}(\ell)$ change. In the low energy regime, it is easy to obtain
\begin{eqnarray}
B_{\bot}(\ell)/B_{\bot}^{f}(\ell)&\sim& \frac{c_{B_{\bot}}^{*}}{B_{\bot0}}e^{\ell}\rightarrow-\infty, \label{Eq:BBotFlowWeak}
\\
B_{z}(\ell)/B_{z}^{f}(\ell)&\sim& \frac{c_{B_{z}}^{*}}{B_{z0}}e^{\ell}\rightarrow-\infty. \label{Eq:BzFlowWeak}
\end{eqnarray}
These results indicate that the behaviors of $B_{\bot}$ and $B_{z}$ are obviously modified by
Coulomb interaction. These results can be also noticed from Figs.~\ref{Fig:VRGArrayB}(d) and \ref{Fig:VRGArrayB}(f).
The behaviors shown in Eqs.~(\ref{Eq:BBotFlowWeak}) and (\ref{Eq:BzFlowWeak}) reveal that the positive quadratic terms
$B_{\bot}k_{\bot}^{2}$ and $B_{z}k_{z}^{2}$ become negative linear terms of momentum components induced by Coulomb interaction.

If the initial Coulomb strength is large enough, we notice that
\begin{eqnarray}
m(\ell)/m^{f}(\ell)\rightarrow c_{m}^{*},\quad\mathrm{with}\quad c_{m}^{*}<0,
\end{eqnarray}
where $c_{m}^{*}$ is a negative constant. The sign of $m(\ell)$ changes in this case.
As shown in Figs.~\ref{Fig:VRGArrayB}(c) and \ref{Fig:VRGArrayB}(e),
\begin{eqnarray}
B_{\bot}(\ell)&\rightarrow& c_{B_{\bot}}^{*},\quad \mathrm{with}\quad c_{B_{\bot}}^{*}>0,
\\
B_{z}(\ell)&\rightarrow& c_{B_{z}}^{*}, \quad \mathrm{with}\quad c_{B_{z}}^{*}>0,
\end{eqnarray}
where $c_{B_{\bot}}^{*}$ and $c_{B_{z}}^{*}$ are positive constants.
The signs of $B_{\bot}(\ell)$ and $B_{z}(\ell)$ hold. We can further get
\begin{eqnarray}
B_{\bot}(\ell)/B_{\bot}^{f}(\ell)&\sim& \frac{c_{B_{\bot}}^{*}}{B_{\bot0}}e^{\ell}\rightarrow\infty, \label{Eq:BBotFlowStrong}
\\
B_{z}(\ell)/B_{z}^{f}(\ell)&\sim& \frac{c_{B_{z}}^{*}}{B_{z0}}e^{\ell}\rightarrow\infty, \label{Eq:BzFlowStrong}
\end{eqnarray}
which can be also viewed from Figs.~\ref{Fig:VRGArrayB}(d) and \ref{Fig:VRGArrayB}(f).
The flows as shown in Eq.~(\ref{Eq:BBotFlowStrong}) and (\ref{Eq:BzFlowStrong}) mean that the positive quadratic terms
$B_{\bot}k_{\bot}^{2}$ and $B_{z}k_{z}^{2}$ are modified to positive linear terms of momentum components by
Coulomb interaction.

\begin{figure}[htbp]
\center
\includegraphics[width=3.3in]{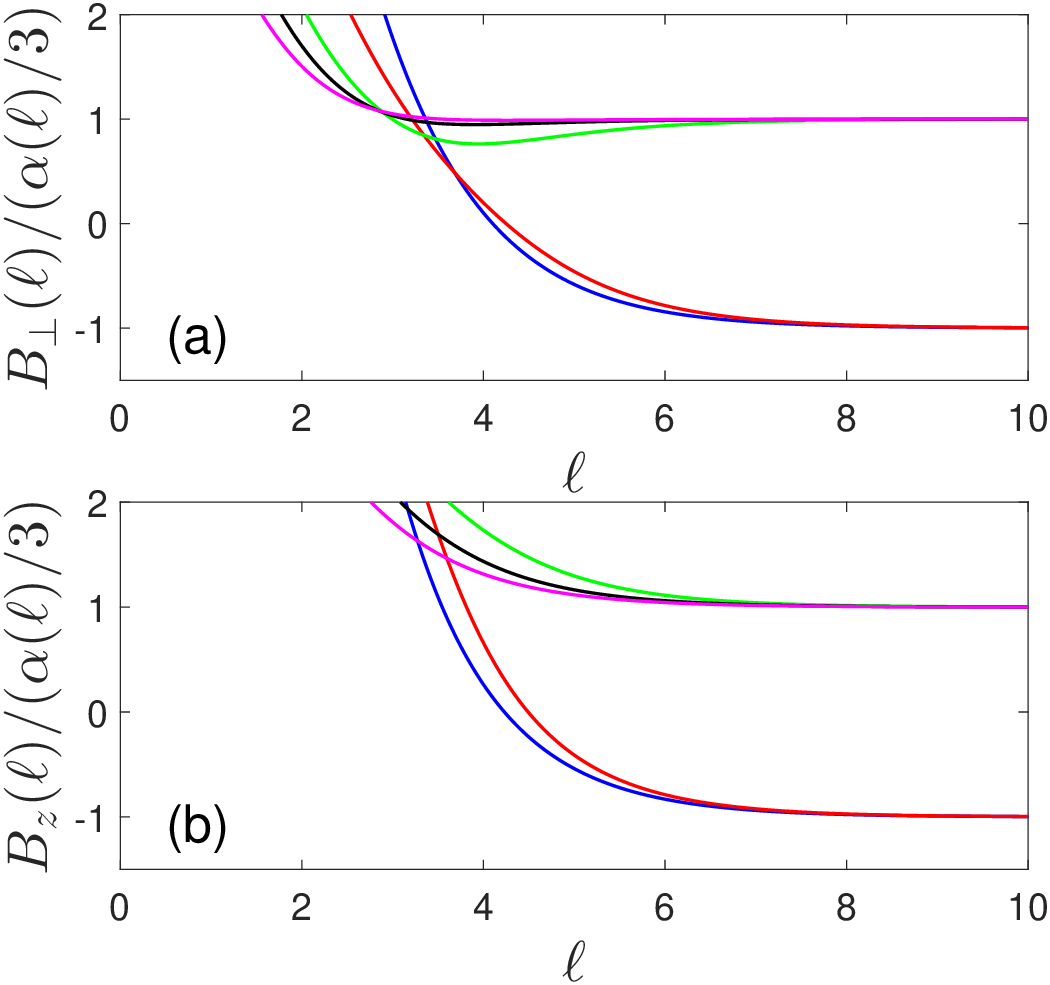}
\caption{(a) ad (b): Flows of $B_{\bot}(\ell)/(\alpha(\ell)/3)$ and $B_{z}(\ell)/(\alpha(\ell)/3)$  with different initial values of Coulomb strength. Blue, red, green, black, and magenta lines
correspond to the initial values $\alpha_{0}=0.1, 0.2, 0.3, 0.4, 0.5$ respectively. $m_{0}=0.1$, $B_{\bot0}=1$, $B_{z0}=1$, $D_{0}=1$, $\zeta_{0}=0.1$ are taken. \label{Fig:BbotBzAlphaRatioLimit}}
\end{figure}

In order to better understanding this question, we give analytical calculation for the asymptotic behaviors of $B_{\bot}(\ell)$ and $B_{z}(\ell)$.
We have find that $\alpha(\ell)\rightarrow\alpha^{*}$, $m(\ell)\rightarrow\infty$ for weak Coulomb interaction and $m(\ell)\rightarrow-\infty$
for strong enough Coulomb interaction. Accordingly, in the low energy regime,  the RG equations for $B_{\bot}$ and $B_{z}$ can be approximated by
\begin{eqnarray}
\frac{dB_{\bot}}{d\ell}&\sim&-B_{\bot}-\frac{1}{3}\alpha^{*}\frac{m}{|m|}\sim-B_{\bot}-\frac{1}{3}\alpha^{*}\mathrm{sgn}(m),
\\
\frac{dB_{z}}{d\ell}&\sim&-B_{z}-\frac{1}{3}\alpha^{*}\frac{m}{|m|}\sim-B_{z}-\frac{1}{3}\alpha^{*}\mathrm{sgn}(m).
\end{eqnarray}
Accordingly, we can find that
\begin{eqnarray}
B_{\bot}(\ell)\rightarrow-\frac{1}{3}\alpha^{*}\mathrm{sgn}(m), \label{Eq:BotAlphaRatioAsymptoticBehavior}
\\
B_{z}(\ell)\rightarrow-\frac{1}{3}\alpha^{*}\mathrm{sgn}(m),  \label{Eq:BzAlphaRatioAsymptoticBehavior}
\end{eqnarray}
in the limit $\ell\rightarrow\infty$.
The numerical results of $B_{\bot}(\ell)/(\alpha(\ell)/3)$ and $B_{z}(\ell)/(\alpha(\ell)/3)$  are shown in Fig.~\ref{Fig:BbotBzAlphaRatioLimit}.
We can find these numerical results verify the asymptotic behaviors as shown in Eqs.~(\ref{Eq:BotAlphaRatioAsymptoticBehavior}) and
(\ref{Eq:BzAlphaRatioAsymptoticBehavior}).

Therefore, we can find that the sign of $m$ holds  but the signs of $B_{\bot}$ and $B_{z}$ change for weak Coulomb interaction,
while the sign of $m$ changes but the signs of $B_{\bot}$ and $B_{z}$ hold if Coulomb interaction is strong enough. Namely, signs of
$mB_{\bot}$, $mB_{z}$ usually always change.  Thus, second-order TI usually becomes a trivial band insulator not only under strong
Coulomb interaction but also under weak Coulomb interaction.

Taking proper initial conditions, $m$, $B_{\bot}$, $B_{z}$ may flow to zero simultaneously.
In this case, the system may become a second-order Dirac semimetal (DSM).

In order to avoid possible misunderstanding, it should be strengthen that result $m(\ell)\rightarrow\infty$ does not mean the physical mass becomes to infinity,
but represents that the physical mass becomes larger and larger comparing to the linear terms of Hamiltonian with lowering
of momentum.  The physical mass for free second-order TI is a constant. The physical mass for second-order TI with weak Coulomb
interaction only receives quantitative correction and  is also a constant with same sign.  Thus, we can find that the RG
results shown in above will not break down at the scale where $m(\ell)$ is order of one.  Actually, the RG results are valid in the
limit $\ell\rightarrow\infty$, namely $k\rightarrow 0$.

\begin{figure}[htbp]
\center
\includegraphics[width=3.3in]{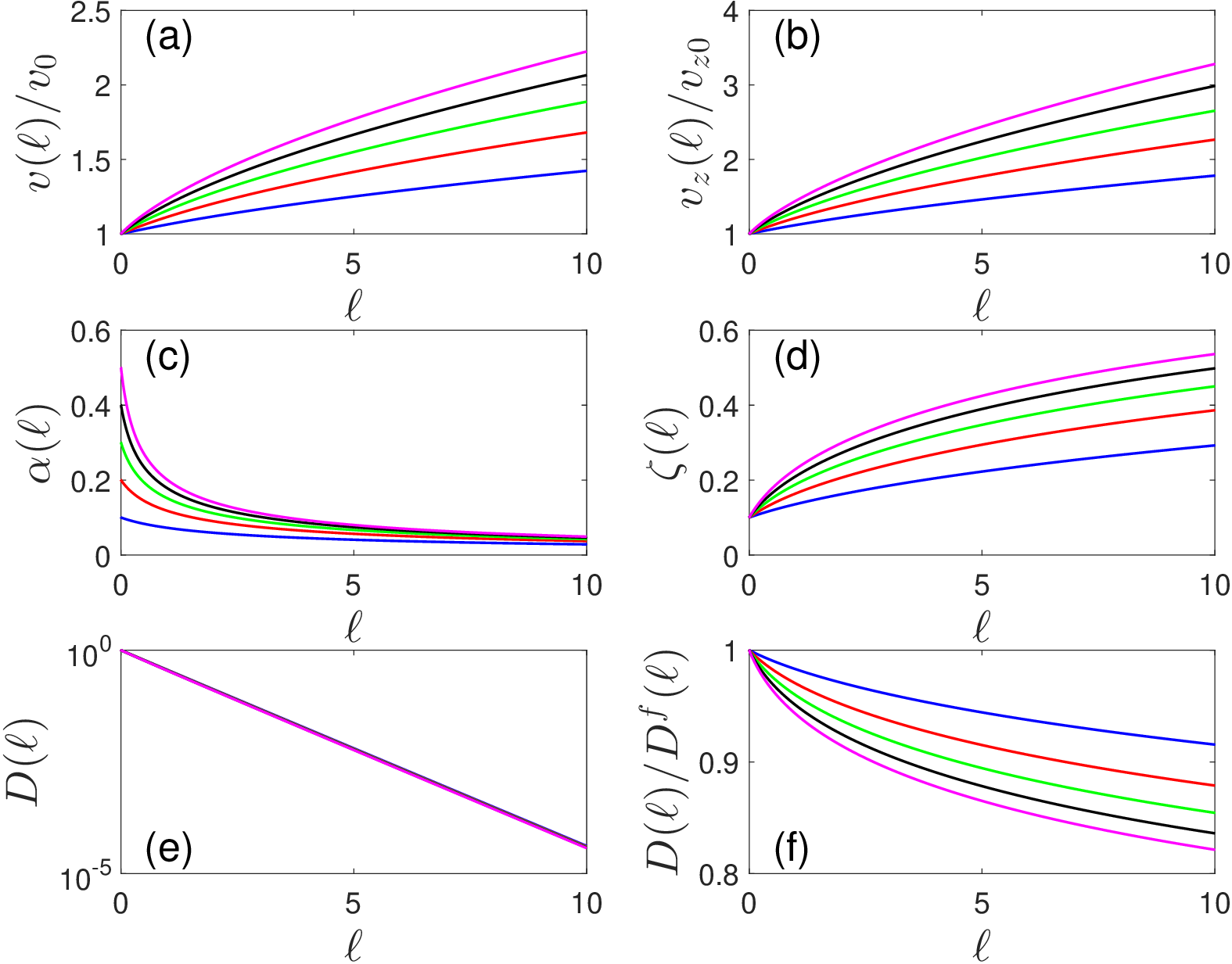}
\caption{(a)-(f): Flows of $v(\ell)$, $v_{z}(\ell)$, $\alpha(\ell)$, $\gamma(\ell)$, $D(\ell)$,
and $D(\ell)/D^{f}(\ell)$ for second-order DSM with different initial Coulomb strength. Blue, red, green, black, and magenta lines
correspond to the initial values $\alpha_{0}=0.1, 0.2, 0.3, 0.4, 0.5$ respectively. $D_{0}=1$, $\zeta_{0}=0.1$ are taken. \label{Fig:VRGArray2ndDSM}}
\end{figure}

\section{Results for other related systems and some discussions \label{Sec:ResultsOtherAndDiscussions}}

\subsection{Second-order Dirac semimetal}

Taking the initial conditions $m_{0}=0$, $B_{\bot0}=0$, and $B_{z0}=0$, we can obtain the RG equations  for
second-order DSM. The corresponding numerical results are shown in Fig.~\ref{Fig:VRGArray2ndDSM}. We notice that
$\alpha$ flows to zero slowly, and $v$ and $v_{z}$ increase slowly with lowering of energy scale. As shown in
Fig.~\ref{Fig:VRGArray2ndDSM}(f), $D(\ell)/D^{f}(\ell)$ decreases slowly with lowering of energy scale. These results
imply that the terms $vk_{\bot}$, $vk_{z}$, $D(k_{x}^{2}-k_{y}^{2})$ receive weak logarithmic-like corrections of momentum
components under Coulomb interaction. It indicates that the observable quantities such as density of states, specific heat,
compressibility acquire logarithmic-like corrections of energy or temperature, and second-order DSM is robust against Coulomb
interaction. These results are consistent with recent studies in Ref.~\cite{LeeYuWen21}.

\subsection{3D TI}

\begin{figure}[htbp]
\center
\includegraphics[width=3.3in]{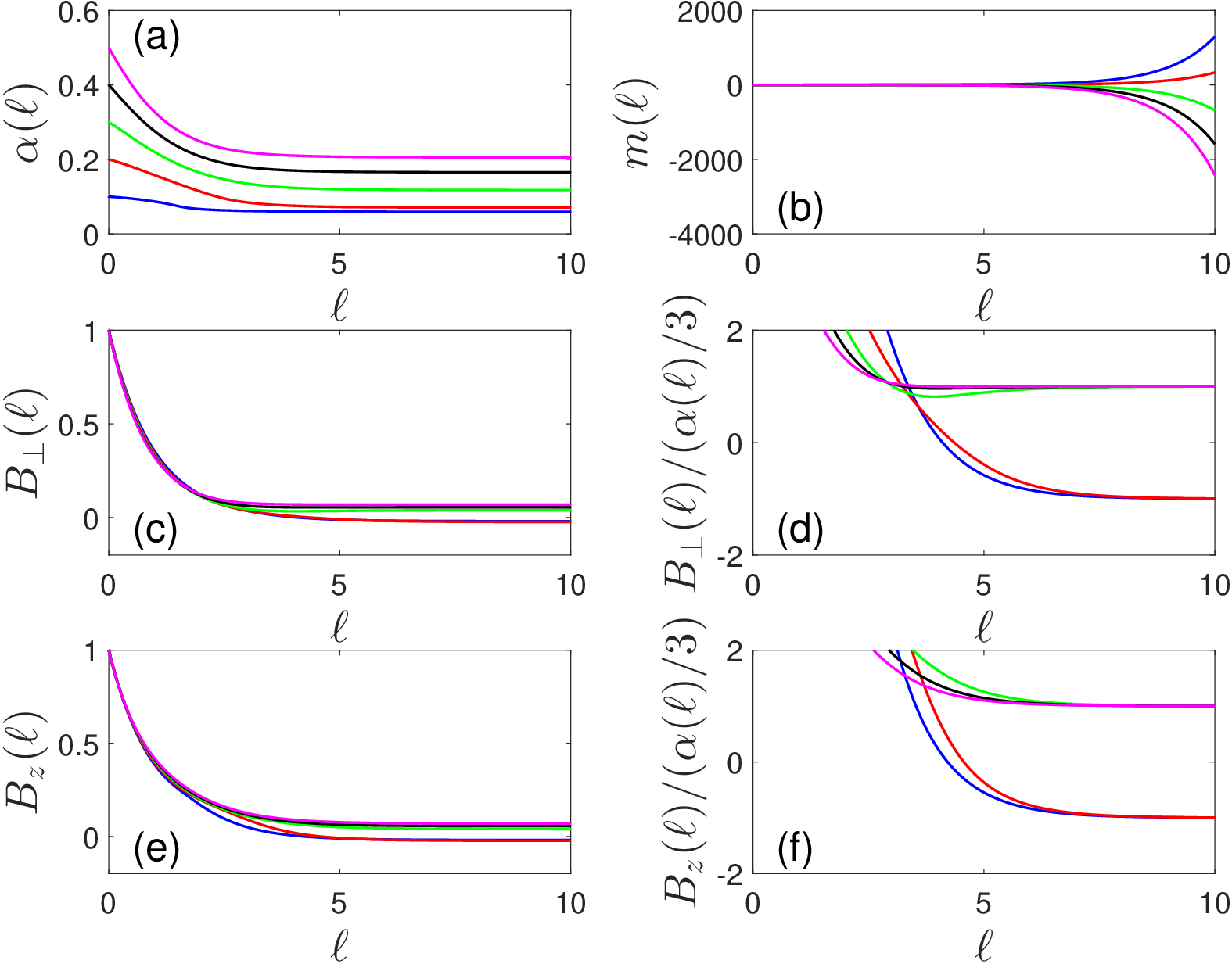}
\caption{(a)-(f): Flows of $\alpha(\ell)$, $m(\ell)$, $B_{\bot}(\ell)$, $B_{\bot}(\ell)/(\alpha(\ell)/3)$, $B_{z}(\ell)$,
and $B_{z}(\ell)/(\alpha(\ell)/3)$ for TI with different initial values of Coulomb strength. Blue, red, green, black, and magenta lines
correspond to the initial values $\alpha_{0}=0.1, 0.2, 0.3, 0.4, 0.5$ respectively. $m_{0}=0.1$, $B_{\bot0}=1$, $B_{z0}=1$, $\zeta_{0}=0.1$ are taken. \label{Fig:VRG1stTI}}
\end{figure}

Taking the initial value $D_{0}=0$, we get the RG equations for TI.
From numerical and analytical calculations, we also find that $\alpha$ approaches to constant value $\alpha^{*}$, and
\begin{eqnarray}
B_{\bot}(\ell)\rightarrow-\frac{1}{3}\alpha^{*}\mathrm{sgn}(m), \label{Eq:BotAlphaRatioAsymptoticBehavior1stTI}
\\
B_{z}(\ell)\rightarrow-\frac{1}{3}\alpha^{*}\mathrm{sgn}(m). \label{Eq:BzAlphaRatioAsymptoticBehavior1stTI}
\end{eqnarray}
The flows of $\alpha(\ell)$, $m(\ell)$, $B_{\bot}(\ell)$, $B_{\bot}(\ell)/(\alpha(\ell)/3)$, $B_{z}(\ell)$,
and $B_{z}(\ell)/(\alpha(\ell)/3)$ for TI are presented in Fig.~\ref{Fig:VRG1stTI}.
These results indicate that TI is also usually unstable to trivial band insulator
not only under strong Coulomb interaction but also under weak Coulomb interaction.

In the study by Goswami \emph{et al.} \cite{Goswami11}, the
RG equations for the TI with long-range Coulomb interaction are given. However, the RG flow
of the parameter $B$ is not analyzed carefully and changing of sign of $B$ is not noticed in Ref.~\cite{Goswami11}.
Through numerical and analytical calculations for the RG equations shown in Ref.~\cite{Goswami11}, we  obtain that
$\alpha$ flows to a constant value $\alpha^{*}$, and
\begin{eqnarray}
B(\ell)\rightarrow \frac{\pi}{3}\alpha^{*}\mathrm{sgn}(m),
\end{eqnarray}
in the limit $\ell\rightarrow\infty$ in their notations. It should be noticed that the definition of $B$ in  Ref.~\cite{Goswami11} is different from ours with
a minus. Namely, $mB<0$ corresponds to TI and $mB>0$  stands for normal insulator in Ref.~\cite{Goswami11}. Whereas, $mB_{\bot,z}>0$
correspond to TI and $mB_{\bot,z}<0$ stand for normal insulator in our article. Additionally, the definition of $\alpha$ in our
article is different form the one in Ref.~\cite{Goswami11} with a factor $1/\pi$. Thus, the results we obtain through
the RG equations in Ref.~\cite{Goswami11} are completely consistent with the ones shown in Eqs.~(\ref{Eq:BotAlphaRatioAsymptoticBehavior1stTI}) and
(\ref{Eq:BzAlphaRatioAsymptoticBehavior1stTI}).

\subsection{Short-range four-fermion interactions}

Roy \emph{et al.} have showed the phase diagrams of TI under short-range four-fermions
interactions \cite{Roy16B}. Influence of short-range four-fermion interactions on second-order TI is also interesting. For
short-range interactions with the form
\begin{eqnarray}
\mathcal{L}_{fourfermion}=\sum_{i}\lambda_{i}\left(\bar{\Psi}\Gamma_{i}\Psi\right)^{2},
\end{eqnarray}
through RG calculation, we find that $B_{\bot,z}(\ell)$ and $D(\ell)$ take the behaviors
\begin{eqnarray}
B_{\bot}(\ell)&=&B_{\bot0}e^{-\ell},
\\
B_{z}(\ell)&=&B_{z0}e^{-\ell},
\\
D_{z}(\ell)&=&D_{0}e^{-\ell},
\end{eqnarray}
which are same as the free system, and $m$ will not change sign  under weak short-range four-fermion interactions.
Thus, second-order TI is robust against weak short-range four-fermion interactions. If the strength of short-range four-fermion interaction
is large enough, some quantum phase transitions could be triggered. The detailed studies about second-order TI under short-range four-fermion
interactions are left in further studies.

The important point we should strength here is that weak long-range Coulomb interaction is different from weak short-range four-fermion
interactions. Weak long-range Coulomb interaction drives second-order TI to trivial band insulator, whereas second-order TI is robust against weak short-range four-fermion
interactions.

\subsection{Comments for the arguments in a recent reference}

In a recent reference, Liu \emph{et al.} give some arguments for our results about sign change of $B_{\bot, z}$, and they believe that second-order TI is robust
against weak Coulomb interaction \cite{LiHeQiu22}. We should indicate that their arguments are unfounded. First, as shown in Sec.~\ref{Sec:RGResultsCoulomb},
through numerical and analytical calculations, we have clearly showed that $B_{\bot,z}$ indeed change sign under weak long-range Coulomb interaction.
Second, the Coulomb strength $\alpha(\ell)$ decreases from the initial value $\alpha_{0}$ gradually and approaches to a small finite value $\alpha^{*}$ in the
lowest energy limit $\ell\rightarrow\infty$.  Thus, the RG is controlled and the corresponding results are valid in the lowest energy regime. Their argument
that RG becomes invalid in large $\ell$  (low energy regime) is incorrect. Third, their arguments about the high order terms of momenta are invalid. Liu  \emph{et al.}
considered that $B_{\bot}k_{\bot}^{2}$, $B_{z}k_{z}^{2}$ terms come form the expansion of $\sum_{i}t_{i}\cos(k_{i}a)$ as
\begin{eqnarray}
\sum_{i}t_{i}\cos(k_{i}a)=\mathrm{const}-B_{\bot}k_{\bot}^{2}-B_{z}k_{z}^{2}+O(k^4,k^6),
\end{eqnarray}
and higher order terms beyond $k^{2}$ terms exist in general. They stated that even if $B_{\bot,z}$ change sign, then the high order terms $k^{4}$, $k^{6}$
can not be ignored. However, we should indicate that their arguments are unfounded. It is easy to find that $\sum_{i}t_{i}\cos(k_{i}a)$ can be expand as
\begin{eqnarray}
&&\sum_{i}t_{i}\cos(k_{i}a)
\\
&=&\mathrm{const}-B_{\bot}k_{\bot}^{2}-B_{z}k_{z}^{2}\nonumber
\\
&&+B_{\bot(4)}\left(k_{x}^{4}+k_{y}^{4}\right)+B_{z(4)}k_{z}^{4}\nonumber
\\
&&-B_{\bot(6)}\left(k_{x}^{6}+k_{6}^{6}\right)-B_{z(6)}k_{z}^{6}\nonumber
\\
&&...+(-1)^{n}\left[B_{\bot(2n)}\left(k_{x}^{2n}+k_{y}^{2n}\right)+B_{z(2n)}k_{z}^{2n}\right]\nonumber
\\
&&...,
\end{eqnarray}
where the coefficients  $B_{\bot}$, $B_{z}$, $B_{\bot(4)}$, $B_{z(4)}$, $B_{\bot(6)}$, $B_{z(6)}$ \emph{etc.} are all positive. We can find the terms
$k_{i}^{2n}$ with $n$ being odd number are negative, whereas the terms $k_{i}^{2n}$  with $n$ being even number are positive. These terms could be modified by
long-range Coulomb interaction. From the numerical and analytical calculations shown in Sec.~\ref{Sec:RGResultsCoulomb}, we have showed that $B_{\bot}$ and $B_{z}$ change sign under the weak Coulomb interaction.
The high order terms $k_{i}^{4}$, $k_{i}^{6}$  could be also modified by long-range Coulomb interaction. Through tedious calculations, we find that the RG equations for
$B_{\bot(4)}$, $B_{z(4)}$, $B_{\bot(6)}$, and $B_{z(6)}$ in the low energy regime can be approximately written as
\begin{eqnarray}
\frac{dB_{\bot(4)}}{d\ell}&\sim&-3B_{\bot(4)}+\frac{1}{5}\alpha^{*}\mathrm{sgn}(m), \label{Eq:RGEBBot4}
\\
\frac{dB_{z(4)}}{d\ell}&\sim&-3B_{z(4)}+\frac{1}{5}\alpha^{*} \mathrm{sgn}(m),\label{Eq:RGEBz4}
\\
\frac{dB_{\bot(6)}}{d\ell}&\sim&-5B_{\bot(6)}-\frac{1}{7}\alpha^{*}\mathrm{sgn}(m), \label{Eq:RGEBBot6}
\\
\frac{dB_{z(6)}}{d\ell}&\sim&-5B_{z(6)}-\frac{1}{7}\alpha^{*}\mathrm{sgn}(m), \label{Eq:RGEBz6}
\end{eqnarray}
where the transformations
\begin{eqnarray}
\frac{B_{\bot(4)}\Lambda^{3}}{v}\rightarrow B_{\bot(4)},
\\
\frac{B_{z(4)}\Lambda^{3}}{v\eta^{2}}\rightarrow B_{z(4)},
\\
\frac{B_{\bot(6)}\Lambda^{5}}{v}\rightarrow B_{\bot(6)},
\\
\frac{B_{z(6)}\Lambda^{5}}{v\eta^{3}}\rightarrow B_{z(6)},
\end{eqnarray}
have been utilized. Solving Eqs.~(\ref{Eq:RGEBBot4})-(\ref{Eq:RGEBz6}), we obtain
\begin{eqnarray}
B_{\bot(4)}(\ell)\rightarrow\frac{1}{15}\alpha^{*}\mathrm{sgn}(m),
\\
B_{z(4)}(\ell)\rightarrow\frac{1}{15}\alpha^{*} \mathrm{sgn}(m),
\\
B_{\bot(6)}(\ell)\rightarrow-\frac{1}{35}\alpha^{*}\mathrm{sgn}(m),
\\
B_{z(6)}(\ell)\rightarrow-\frac{1}{35}\alpha^{*}\mathrm{sgn}(m),
\end{eqnarray}
in the lowest energy limit $\ell\rightarrow\infty$. We can find that for weak Coulomb interaction $B_{\bot(4)}$ and $B_{z(4)}$ are still positive,
whereas $B_{\bot(6)}$ and $B_{z(6)}$ become to negative. Further, we notice that the RG equations for the coefficients of high order terms can be
generally expressed by
\begin{eqnarray}
\frac{dB_{\bot(2n)}}{d\ell}&\sim&-(2n-1)B_{\bot(2n)}+\frac{(-1)^{n}}{2n+1}\alpha^{*}\mathrm{sgn}(m), \label{Eq:RGEBBot2n}
\\
\frac{dB_{z(2n)}}{d\ell}&\sim&-(2n-1)B_{z(2n)}+\frac{(-1)^{n}}{2n+1}\alpha^{*} \mathrm{sgn}(m),\label{Eq:RGEBz2n}
\end{eqnarray}
where the transformations
\begin{eqnarray}
\frac{B_{\bot(2n)}\Lambda^{2n-1}}{v}\rightarrow B_{\bot(2n)},
\\
\frac{B_{z(2n)}\Lambda^{2n-1}}{v\eta^{n}}\rightarrow B_{z(2n)},
\end{eqnarray}
have been employed.  From Eqs.~(\ref{Eq:RGEBBot2n}) and (\ref{Eq:RGEBz2n}), we get
\begin{eqnarray}
B_{\bot(2n)}(\ell)\rightarrow(-1)^{n}\frac{1}{(2n+1)(2n-1)}\alpha^{*}\mathrm{sgn}(m),
\\
B_{z(2n)}(\ell)\rightarrow(-1)^{n}\frac{1}{(2n+1)(2n-1)}\alpha^{*} \mathrm{sgn}(m),
\end{eqnarray}
in the limit $\ell\rightarrow\infty$.
Thus, under weak Coulomb interaction, $B_{\bot,z(2n)}$ change sign if $n$ is odd number, but hold sign if $n$ is even number. From the above results we know
that the original negative terms $k_{i}^{2n}$ with $n$ being odd number become positive, and the original positive terms $k_{i}^{2n}$ with $n$ being even number
are still positive, under weak Coulomb interaction. Therefore, we find that the system becomes trivial insulator under the influence of weak long-range Coulomb interaction even if
the high order terms $k_{i}^{4}$, $k_{i}^{6}$ \emph{etc.} are considered.

\section{Influence of Disorder scattering \label{Sec:DisorderEffect}}

In Ref.~\cite{ZhaoPengLu21}, Zhao \emph{et al.} also studied the influence of disorder on 3D second-order TI based on RG method. Their studies about disorder effects in
second-order TI have several  problems. First, they used the invalid criterion for transition from second-order TI to TI to study the influence of
disorder on second-order TI. According to their criterion, considering only disorder, weak disorder drives second-order TI to TI.
However, recent studies about disorder effects in 3D second-order TI through other methods showed that 3D second-order TI is robust against weak disorder \cite{WangC20, WangC21A}.

Second, generation new type of disorder by one type of disorder is not considered in Ref.~\cite{ZhaoPengLu21}.
If the fermion Hamiltonian of a system satisfies
\begin{eqnarray}
\mathcal{H}_{f}(\mathbf{k})+\mathcal{H}_{f}(-\mathbf{k})=0,  \label{Eq:HamiltonianKNegativeKZero}
\end{eqnarray}
one type of disorder can exist solely. For examples, in DSM  or Weyl semimetal (WSM), the
Hamiltonian satisfies Eq.~(\ref{Eq:HamiltonianKNegativeKZero}), thus one type of disorder can exist solely
\cite{Roy16, Ludwig94, Evers08, Foster12, Roy14, Syzranov16, Sbierski16}.
However, if the Hamiltonian of a system satisfies
\begin{eqnarray}
\mathcal{H}_{f}(\mathbf{k})+\mathcal{H}_{f}(-\mathbf{k})\neq0. \label{Eq:HamiltonianKNegativeKNonZero}
\end{eqnarray}
Some type of disorder can not exist solely and generate other types of disorder.
For examples, the generation effect of new disorder results from
Eq.~(\ref{Eq:HamiltonianKNegativeKNonZero}) was considered in the
studies about disorder effects in 2D semi-DSM \cite{Carpentier13}, 3D anisotropic WSM \cite{Roy18AWSM, LuoXL18, WangLiWangZhang21},
3D double-WSM \cite{Bera16, WangLiuZhang17B}, and Luttinger semimetal \cite{Nandkishore17, Mandal18}, in which
the Hamiltonian
satisfies Eq.~(\ref{Eq:HamiltonianKNegativeKNonZero}).  We can find that the
Hamiltonian of second-order TI as shown in Eq.~(\ref{Eq:HamiltonianFermion}) satisfies Eq.~(\ref{Eq:HamiltonianKNegativeKNonZero}).
Therefore, the generation of new type of disorder by one type of disorder should be carefully studied.

Third, generation of new type of disorder by two types of disorder was not considered in Ref.~\cite{ZhaoPengLu21}.
If two types of disorder exist initially, third type of disorder usually is generated. This effect could be quite
important. For example, for 2D DSM, if any two types of disorder among random chemical potential, random vector potential, and random
mass exist initially, the third one is generated and the the system becomes compressible diffusive metal \cite{Evers08, Foster12}.
In Ref.~\cite{ZhaoPengLu21}, they considered random chemical potential and random mass. Other types of disorder could be generated by
random chemical potential and random mass.  Whereas, this effect was not considered in Ref.~\cite{ZhaoPengLu21}.

In order to avoid the second and third problems above-mentioned, we find that twelve kinds of disorder should be  considered for second-order TI. The corresponding
action for fermion-disorder coupling reads
\begin{eqnarray}
S_{\mathrm{dis}} &=&\int d\tau d^3\mathbf{x}
\bar{\Psi}\bigg[V_{C}(\mathbf{x})\gamma_{0}+V_{M}(\mathbf{x})\mathbbm{1}_{4\times4}\nonumber
\\
&&+V_{AC}(\mathbf{x})\gamma_{0}\gamma_{5}
+\sum_{j=x,y}V_{SO\bot}(\mathbf{x})\gamma_{0}\gamma_{j}\nonumber
\\
&&+V_{SOz}(\mathbf{x})\gamma_{0}\gamma_{z}
+V_{PM}(\mathbf{x})i\gamma_{5}\nonumber
\\
&&+V_{MN\bot}(\mathbf{x})\left(i\gamma_{y}\gamma_{z}+i\gamma_{z}\gamma_{x}\right)+V_{MNz}(\mathbf{x})i\gamma_{x}\gamma_{y}\nonumber
\\
&&+V_{AMN\bot}(\mathbf{x})\sum_{j=x,y}i\gamma_{5}\gamma_{j}+V_{AMNz}(\mathbf{x})i\gamma_{5}\gamma_{z}\nonumber
\\
&&+V_{CR\bot}(\mathbf{x})\sum_{j=x,y}i\gamma_{j}+V_{CRz}(\mathbf{x})i\gamma_{z}\bigg]\Psi.
\end{eqnarray}
$V_{C}(\mathbf{x})$ corresponds to random chemical potential. $V_{M}(\mathbf{x})$ stands for random mass.
$V_{AC}(\mathbf{x})$ represents random axial chemical potential. $V_{SO\bot}$ and $V_{SOz}$ correspond to random spin-orbital scattering  within $xy$ plane
and along $z$ axis. $V_{PM}(\mathbf{x})$ stands for random pseudo mass. $V_{MN\bot}$ and $V_{MNz}$ represent random magnetization within $xy$ plane and along $z$ axis.
$V_{AMN\bot}$ and $V_{AMNz}$ are corresponding to random axial magnetization within $xy$ plane and along $z$ axis.$V_{CR\bot}$ and $V_{CRz}$ stand for
random current within $xy$ plane and along $z$ axis.

The quenched random field $V_{j}$ is taken as a Gaussian white noise
distribution that satisfies $\langle V_{j}(\mathbf{x})\rangle = 0$
and $\langle V_{j}(\mathbf{x})V_{j}(\mathbf{x}')\rangle =
\Delta_{j}\delta^{3}(\mathbf{x}-\mathbf{x}')$. The disorder
can be treated by the replica method. The effective action in replica formalism
can be written as
\begin{widetext}
\begin{eqnarray}
S_{\mathrm{dis}}&=&-\frac{\Delta_{C}}{2}\int d\tau
d\tau'd^3\mathbf{x} \left(\bar{\Psi}_{a}\gamma_{0}
\Psi_{a}\right)_{\tau}\left(\bar{\Psi}_{b}\gamma_{0}\Psi_{b}\right)_{\tau'}-\frac{\Delta_{M}}{2}\int d\tau
d\tau'd^3\mathbf{x} \left(\bar{\Psi}_{a}
\Psi_{a}\right)_{\tau}\left(\bar{\Psi}_{b}\Psi_{b}\right)_{\tau'}\nonumber
\\
&&-\frac{\Delta_{AC}}{2}\int d\tau
d\tau'd^3\mathbf{x} \left(\bar{\Psi}_{a}\gamma_{0}\gamma_{5}
\Psi_{a}\right)_{\tau}\left(\bar{\Psi}_{b}\gamma_{0}\gamma_{5}\Psi_{b}\right)_{\tau'}
-\frac{\Delta_{SO\bot}}{2}\sum_{j=x,y}\int d\tau
d\tau'd^3\mathbf{x} \left(\bar{\Psi}_{a}\gamma_{0}\gamma_{j}
\Psi_{a}\right)_{\tau}\left(\bar{\Psi}_{b}\gamma_{0}\gamma_{j}\Psi_{b}\right)_{\tau'}\nonumber
\\
&&-\frac{\Delta_{SOz}}{2}\int d\tau
d\tau'd^3\mathbf{x} \left(\bar{\Psi}_{a}\gamma_{0}\gamma_{z}
\Psi_{a}\right)_{\tau}\left(\bar{\Psi}_{b}\gamma_{0}\gamma_{z}\Psi_{b}\right)_{\tau'}\nonumber
-\frac{\Delta_{PM}}{2}\int d\tau
d\tau'd^3\mathbf{x} \left(\bar{\Psi}_{a}i\gamma_{5}
\Psi_{a}\right)_{\tau}\left(\bar{\Psi}_{b}i\gamma_{5}\Psi_{b}\right)_{\tau'}\nonumber
\\
&&-\frac{\Delta_{MN\bot}}{2}\sum_{j=x,y}\int d\tau
d\tau'd^3\mathbf{x}\left[\left(\bar{\Psi}_{a}i\gamma_{y}\gamma_{z}
\Psi_{a}\right)_{\tau}\left(\bar{\Psi}_{b}i\gamma_{y}\gamma_{z}\Psi_{b}\right)_{\tau'}
+\left(\bar{\Psi}_{a}i\gamma_{z}\gamma_{x}
\Psi_{a}\right)_{\tau}\left(\bar{\Psi}_{b}i\gamma_{z}\gamma_{x}\Psi_{b}\right)_{\tau'}\right]\nonumber
\\
&&-\frac{\Delta_{MNz}}{2}\int d\tau
d\tau'd^3\mathbf{x}\left(\bar{\Psi}_{a}i\gamma_{x}\gamma_{y}
\Psi_{a}\right)_{\tau}\left(\bar{\Psi}_{b}i\gamma_{x}\gamma_{y}\Psi_{b}\right)_{\tau'}
-\frac{\Delta_{AMN\bot}}{2}\sum_{j=x,y}\int d\tau
d\tau'd^3\mathbf{x}\left(\bar{\Psi}_{a}i\gamma_{5}\gamma_{j}
\Psi_{a}\right)_{\tau}\nonumber
\\
&&\times\left(\bar{\Psi}_{b}i\gamma_{5}\gamma_{j}\Psi_{b}\right)_{\tau'}-\frac{\Delta_{AMNz}}{2}\int d\tau
d\tau'd^3\mathbf{x}\left(\bar{\Psi}_{a}i\gamma_{5}\gamma_{z}
\Psi_{a}\right)_{\tau}\left(\bar{\Psi}_{b}i\gamma_{5}\gamma_{z}\Psi_{b}\right)_{\tau'}\nonumber
\\
&&-\frac{\Delta_{CR\bot}}{2}\sum_{j=x,y}\int d\tau
d\tau'd^3\mathbf{x}\left(\bar{\Psi}_{a}i\gamma_{j}
\Psi_{a}\right)_{\tau}\left(\bar{\Psi}_{b}i\gamma_{j}\Psi_{b}\right)_{\tau'}-\frac{\Delta_{CRz}}{2}\int d\tau
d\tau'd^3\mathbf{x}\left(\bar{\Psi}_{a}i\gamma_{z}
\Psi_{a}\right)_{\tau}\left(\bar{\Psi}_{b}i\gamma_{z}\Psi_{b}\right)_{\tau'},
\end{eqnarray}
with $a,b = 1,2,..,n$ being the replica
indices. At the end of the calculation, the limit $n\rightarrow 0$
will be taken.

Through the detailed calculations shown in the Appendices, we get the RG equations
as following
\begin{eqnarray}
\frac{dv}{d\ell}&=&-C_{0}^{dis}v,
\\
\frac{dv_{z}}{d\ell}&=&-C_{0}^{dis}v_{z},
\\
\frac{dm}{d\ell}
&=&m+C_{dis}^{m}m,
\\
\frac{dB_{\bot}}{d\ell}
&=&-B_{\bot},
\\
\frac{dB_{z}}{d\ell}
&=&-B_{z},
\\
\frac{dD}{d\ell}
&=&-D,
\\
\frac{d\Delta_{C}}{d\ell}
&=&-\Delta_{C}+2\Delta_{C}\left(\Delta_{C}+\Delta_{M}+\Delta_{AC}+2\Delta_{SO\bot}
+\Delta_{SOz}+\Delta_{PM}+2\Delta_{MN\bot}+\Delta_{MNz}+2\Delta_{AMN\bot}\right.\nonumber
\\
&&\left.
+\Delta_{AMNz}+2\Delta_{CR\bot}+\Delta_{CRz}\right)\left(\mathcal{G}_{1}^{\bot}
+\mathcal{G}_{1}^{z}+\mathcal{G}_{1}^{D}+\mathcal{G}_{1}^{m}\right)\nonumber
\\
&&+4\left[\left(\Delta_{M}\Delta_{SO\bot}+\Delta_{PM}\Delta_{MN\bot}+\Delta_{AMN\bot}\Delta_{CRz}+\Delta_{AMNz}\Delta_{CR\bot}\right)\mathcal{G}_{1}^{\bot}
+\left(\Delta_{M}\Delta_{SOz}+\Delta_{PM}\Delta_{MNz}\right.\right.\nonumber
\\
&&\left.+2\Delta_{AMN\bot}\Delta_{CR\bot}\right)\mathcal{G}_{1}^{z}+\left(\Delta_{C}\Delta_{PM}+2\Delta_{SO\bot}\Delta_{AMN\bot}+\Delta_{SOz}\Delta_{AMNz}\right)\mathcal{G}_{1}^{D}\nonumber
\\
&&\left.+\left(\Delta_{C}\Delta_{M}+2\Delta_{MN\bot}\Delta_{AMN\bot}+\Delta_{MNz}\Delta_{AMNz}\right)\mathcal{G}_{1}^{m}\right],
\\
\frac{d\Delta_{M}}{d\ell}
&=&-\Delta_{M}+2\Delta_{M}\left(\Delta_{C}
+\Delta_{M}-\Delta_{AC}-2\Delta_{SO\bot}-\Delta_{SOz}-\Delta_{PM}+2\Delta_{MN\bot}+\Delta_{MNz}+2\Delta_{AMN\bot}\right.\nonumber
\\
&&\left.+\Delta_{AMNz}-2\Delta_{CR\bot}-2C_{CRz}\right)
\left(-\mathcal{G}_{1}^{\bot}
-\mathcal{G}_{1}^{z}-\mathcal{G}_{1}^{D}+\mathcal{G}_{1}^{m}\right)\nonumber
\\
&&+4\left[\left(\Delta_{C}\Delta_{SO\bot}+\Delta_{PM}\Delta_{AMN\bot}+\Delta_{MN\bot}\Delta_{CRz}+\Delta_{MNz}\Delta_{CR\bot}\right)\mathcal{G}_{1}^{\bot}
+\left(\Delta_{C}\Delta_{SOz}+\Delta_{PM}\Delta_{AMNz}\right.\right.\nonumber
\\
&&\left.+2\Delta_{MN\bot}\Delta_{CR\bot}\right)\mathcal{G}_{1}^{z}
+\left(\Delta_{M}\Delta_{PM}+2\Delta_{SO\bot}\Delta_{MN\bot}+\Delta_{SOz}\Delta_{MNz}\right)\mathcal{G}_{1}^{D}\nonumber
\\
&&+\left(\Delta_{C}^{2}+\Delta_{M}^{2}+\Delta_{AC}^{2}+2\Delta_{SO\bot}^{2}+\Delta_{SOz}^{2}+\Delta_{PM}^{2}
+2\Delta_{MN\bot}^{2}+\Delta_{MNz}^{2}+2\Delta_{AMN\bot}^{2}+\Delta_{AMNz}^{2}\right.\nonumber
\\
&&\left.\left.+2\Delta_{CR\bot}^{2}+\Delta_{CRz}^{2}\right)\mathcal{G}_{1}^{m}\right],
\\
\frac{d\Delta_{AC}}{d\ell}&=&-\Delta_{AC}+2\Delta_{AC}\left(-\Delta_{C}+\Delta_{M}-\Delta_{AC}+2\Delta_{SO\bot}+\Delta_{SOz}+\Delta_{PM}+2\Delta_{MN\bot}+\Delta_{MNz}
\right.\nonumber
\\
&&\left.-2\Delta_{AMN\bot}-\Delta_{AMNz}-2\Delta_{CR\bot}-\Delta_{CRz}\right)
\left(-\mathcal{G}_{1}^{\bot}
-\mathcal{G}_{1}^{z}+\mathcal{G}_{1}^{D}+\mathcal{G}_{1}^{m}\right)\nonumber
\\
&&+4\left[\left(\Delta_{SO\bot}\Delta_{SOz}+\Delta_{MN\bot}\Delta_{MNz}+\Delta_{AMN\bot}\Delta_{AMNz}+\Delta_{CR\bot}\Delta_{CRz}\right)\mathcal{G}_{1}^{\bot}\right.\nonumber
\\
&&+2\left(\Delta_{SO\bot}^{2}+\Delta_{MN\bot}^{2}+\Delta_{AMN\bot}^{2}+\Delta_{CR\bot}^{2}\right)\mathcal{G}_{1}^{z}
+\left(\Delta_{AC}\Delta_{PM}+2\Delta_{SO\bot}\Delta_{CR\bot}+\Delta_{SOz}\Delta_{CRz}\right)\mathcal{G}_{1}^{D}\nonumber
\\
&&\left.+\left(\Delta_{M}\Delta_{AC}+\Delta_{MNz}\Delta_{CRz}+2\Delta_{MN\bot}\Delta_{CR\bot}\right)\mathcal{G}_{1}^{m}\right],
\\
\frac{d\Delta_{SO\bot}}{d\ell}&=&-\Delta_{SO\bot}+2\Delta_{SO\bot}\left(-\Delta_{C}
+\Delta_{M}+\Delta_{AC}
+\Delta_{SOz}-\Delta_{PM}-\Delta_{MNz}+\Delta_{AMNz}-\Delta_{CRz}\right)
\nonumber
\\
&&\times\left(
-\mathcal{G}_{1}^{z}
-\mathcal{G}_{1}^{D}+\mathcal{G}_{1}^{m}\right)\nonumber
\\
&&+2\left[\left(\Delta_{C}\Delta_{M}+\Delta_{AC}\Delta_{SOz}+\Delta_{PM}\Delta_{CRz}+4\Delta_{MN\bot}\Delta_{AMN\bot}+\Delta_{MNz}\Delta_{AMNz}\right)\mathcal{G}_{1}^{\bot}
\right.\nonumber
\\
&&+2\left(\Delta_{AC}\Delta_{SO\bot}+\Delta_{PM}\Delta_{CR\bot}+\Delta_{MN\bot}\Delta_{AMNz}+\Delta_{MNz}\Delta_{AMN\bot}\right)\mathcal{G}_{1}^{z}\nonumber
\\
&&+2\left(\Delta_{C}\Delta_{AMN\bot}+\Delta_{M}\Delta_{MN\bot}+\Delta_{AC}\Delta_{CR\bot}+\Delta_{SO\bot}\Delta_{PM}\right)\mathcal{G}_{1}^{D}\nonumber
\\
&&\left.+2\left(\Delta_{M}\Delta_{SO\bot}+\Delta_{PM}\Delta_{MN\bot}+\Delta_{AMN\bot}\Delta_{CRz}+\Delta_{AMNz}\Delta_{CR\bot}\right)\mathcal{G}_{1}^{m}\right],
\\
\frac{d\Delta_{SOz}}{d\ell}&=&-\Delta_{SOz}+2\Delta_{SOz}\left(-\Delta_{C}
+\Delta_{M}+\Delta_{AC}+2\Delta_{SO\bot}-\Delta_{SOz}-\Delta_{PM}
-2\Delta_{MN\bot}+\Delta_{MNz}\right.\nonumber
\\
&&\left.+2\Delta_{AMN\bot}
-\Delta_{AMNz}-2\Delta_{CR\bot}+\Delta_{CRz}
\right)\left(-\mathcal{G}_{1}^{\bot}
+\mathcal{G}_{1}^{z}-\mathcal{G}_{1}^{D}
+\mathcal{G}_{1}^{m}\right)\nonumber
\\
&&+4\left[\left(\Delta_{AC}\Delta_{SO\bot}+\Delta_{PM}\Delta_{CR\bot}+\Delta_{MN\bot}\Delta_{AMNz}+\Delta_{MNz}\Delta_{AMN\bot}\right)\mathcal{G}_{1}^{\bot}\right.\nonumber
\\
&&+\left(\Delta_{C}\Delta_{M}+2\Delta_{MN\bot}\Delta_{AMN\bot}+\Delta_{MNz}\Delta_{AMNz}\right)\mathcal{G}_{1}^{z}\nonumber
\\
&&+\left(\Delta_{C}\Delta_{AMNz}+\Delta_{M}\Delta_{MNz}+\Delta_{AC}\Delta_{CRz}+\Delta_{SOz}\Delta_{PM}\right)\mathcal{G}_{1}^{D}\nonumber
\\
&&\left.+\left(\Delta_{M}\Delta_{SOz}+\Delta_{PM}\Delta_{MNz}+2\Delta_{AMN\bot}\Delta_{CR\bot}\right)\mathcal{G}_{1}^{m}\right],
\\
\frac{d\Delta_{PM}}{d\ell}&=&-\Delta_{PM}+2\Delta_{PM}\left(-\Delta_{C}
+\Delta_{M}+\Delta_{AC}-2\Delta_{SO\bot}-\Delta_{SOz}-\Delta_{PM}+2\Delta_{MN\bot}+\Delta_{MNz}
\right.\nonumber
\\
&&\left.-2\Delta_{AMN\bot}-\Delta_{AMNz}+2\Delta_{CR\bot}+\Delta_{CRz}\right)\left(\mathcal{G}_{1}^{\bot}
+\mathcal{G}_{1}^{z}-\mathcal{G}_{1}^{D}+\mathcal{G}_{1}^{m}\right)\nonumber
\\
&&+4\left[\left(\Delta_{C}\Delta_{MN\bot}+\Delta_{M}\Delta_{AMN\bot}+\Delta_{SO\bot}\Delta_{CRz}+\Delta_{SOz}\Delta_{CR\bot}\right)\mathcal{G}_{1}^{\bot}\right.\nonumber
\\
&&+\left(\Delta_{C}\Delta_{MNz}+\Delta_{M}\Delta_{AMNz}+2\Delta_{SO\bot}\Delta_{CR\bot}\right)\mathcal{G}_{1}^{z}+\left(\Delta_{C}^{2}+\Delta_{M}^{2}+\Delta_{AC}^{2}+2\Delta_{SO\bot}^{2}+\Delta_{SOz}^{2}\right.\nonumber
\\
&&\left.+\Delta_{PM}^{2}+2\Delta_{MN\bot}^{2}+\Delta_{MNZ}^{2}+2\Delta_{AMN\bot}^{2}+\Delta_{AMNz}^{2}+2\Delta_{CR\bot}^{2}+\Delta_{CRz}^{2}\right)\mathcal{G}_{1}^{D}\nonumber
\\
&&\left.+\left(\Delta_{M}\Delta_{PM}+2\Delta_{SO\bot}\Delta_{MN\bot}+\Delta_{SOz}\Delta_{MNz}\right)\mathcal{G}_{1}^{m}\right],
\\
\frac{d\Delta_{MN\bot}}{d\ell}&=&-\Delta_{MN\bot}
+2\Delta_{MN\bot}
\left(\Delta_{C}+\Delta_{M}-\Delta_{AC}+\Delta_{SOz}-\Delta_{PM}-\Delta_{MNz}-\Delta_{AMNz}
+\Delta_{CRz}\right)\nonumber
\\
&&\times\left(\mathcal{G}_{1}^{z}-\mathcal{G}_{1}^{D}+\mathcal{G}_{1}^{m}\right)\nonumber
\\
&&+2\left[\left(\Delta_{C}\Delta_{PM}+\Delta_{M}\Delta_{CRz}+\Delta_{AC}\Delta_{MNz}+4\Delta_{SO\bot}\Delta_{AMN\bot}+\Delta_{SOz}\Delta_{AMNz}\right)\mathcal{G}_{1}^{\bot}\right.\nonumber
\\
&&+2\left(\Delta_{M}\Delta_{CR\bot}+\Delta_{AC}\Delta_{MN\bot}+\Delta_{SO\bot}\Delta_{AMNz}+\Delta_{SOz}\Delta_{AMN\bot}\right)\mathcal{G}_{1}^{z}\nonumber
\\
&&+2\left(\Delta_{M}\Delta_{SO\bot}+\Delta_{PM}\Delta_{MN\bot}+\Delta_{AMN\bot}\Delta_{CRz}+\Delta_{AMNz}\Delta_{CR\bot}\right)\mathcal{G}_{1}^{D}\nonumber
\\
&&\left.+2\left(\Delta_{C}\Delta_{AMN\bot}+\Delta_{M}\Delta_{MN\bot}+\Delta_{AC}\Delta_{CR\bot}+\Delta_{SO\bot}\Delta_{PM}\right)\mathcal{G}_{1}^{m}\right],
\\
\frac{d\Delta_{MNz}}{d\ell}&=&-\Delta_{MNz}+2\Delta_{MNz}
\left(\Delta_{C}+\Delta_{M}-\Delta_{AC}+2\Delta_{SO\bot}-\Delta_{SOz}-\Delta_{PM}-2\Delta_{MN\bot}+\Delta_{MNz}
\right.\nonumber
\\
&&\left.-2\Delta_{AMN\bot}+\Delta_{AMNz}+2\Delta_{CR\bot}-\Delta_{CRz}\right)\left(\mathcal{G}_{1}^{\bot}-\mathcal{G}_{1}^{z}-\mathcal{G}_{1}^{D}+\mathcal{G}_{1}^{m}\right)\nonumber
\\
&&+4\left[\left(\Delta_{M}\Delta_{CR\bot}+\Delta_{AC}\Delta_{MN\bot}+\Delta_{SO\bot}\Delta_{AMNz}+\Delta_{SOz}\Delta_{AMN\bot}\right)\mathcal{G}_{1}^{\bot}\right.\nonumber
\\
&&+\left(\Delta_{C}\Delta_{PM}+2\Delta_{SO\bot}\Delta_{AMN\bot}+\Delta_{SOz}\Delta_{AMNz}\right)\mathcal{G}_{1}^{z}\nonumber
\\
&&+\left(\Delta_{M}\Delta_{SOz}+\Delta_{PM}\Delta_{MNz}+2\Delta_{AMN\bot}\Delta_{CR\bot}\right)\mathcal{G}_{1}^{D}\nonumber
\\
&&\left.+\left(\Delta_{C}\Delta_{AMNz}+\Delta_{M}\Delta_{MNz}+\Delta_{AC}\Delta_{CRz}+\Delta_{SOz}\Delta_{PM}\right)\mathcal{G}_{1}^{m}\right],
\\
\frac{d\Delta_{AMN\bot}}{d\ell}&=&-\Delta_{AMN\bot}+2\Delta_{AMN\bot}\left(\Delta_{C}+\Delta_{M}+\Delta_{AC}-\Delta_{SOz}+\Delta_{PM}-\Delta_{MNz}-\Delta_{AMNz}
-\Delta_{CRz}\right)\nonumber
\\
&&\times\left(
-\mathcal{G}_{1}^{z}+\mathcal{G}_{1}^{D}+\mathcal{G}_{1}^{m}\right)\nonumber
\\
&&+2\left[\left(\Delta_{C}\Delta_{CRz}+\Delta_{M}\Delta_{PM}+\Delta_{AC}\Delta_{AMNz}+4\Delta_{SO\bot}\Delta_{MN\bot}+\Delta_{SOz}\Delta_{MNz}\right)\mathcal{G}_{1}^{\bot}\right.\nonumber
\\
&&+2\left(\Delta_{C}\Delta_{CR\bot}+\Delta_{AC}\Delta_{AMN\bot}+\Delta_{SO\bot}\Delta_{MNz}+\Delta_{SOz}\Delta_{MN\bot}\right)\mathcal{G}_{1}^{z}\nonumber
\\
&&+2\left(\Delta_{C}\Delta_{SO\bot}+\Delta_{PM}\Delta_{AMN\bot}+\Delta_{MN\bot}\Delta_{CRz}+\Delta_{MNz}\Delta_{CR\bot}\right)\mathcal{G}_{1}^{D}\nonumber
\\
&&\left.+2\left(\Delta_{C}\Delta_{MN\bot}+\Delta_{M}\Delta_{AMN\bot}+\Delta_{SO\bot}\Delta_{CRz}+\Delta_{SOz}\Delta_{CR\bot}\right)\mathcal{G}_{1}^{m}\right],
\\
\frac{d\Delta_{AMNz}}{d\ell}&=&-\Delta_{AMNz}+2\Delta_{AMNz}\left(\Delta_{C}+\Delta_{M}+\Delta_{AC}-2\Delta_{SO\bot}+\Delta_{SOz}+\Delta_{PM}
-2\Delta_{MN\bot}+\Delta_{MNz}\right.\nonumber
\\
&&\left.-2\Delta_{AMN\bot}+\Delta_{AMNz}-2\Delta_{CR\bot}+\Delta_{CRz}\right)
\left(-\mathcal{G}_{1}^{\bot}
+\mathcal{G}_{1}^{z}+\mathcal{G}_{1}^{D}+\mathcal{G}_{1}^{m}\right)\nonumber
\\
&&+4\left[\left(\Delta_{C}\Delta_{CR\bot}+\Delta_{AC}\Delta_{AMN\bot}+\Delta_{SO\bot}\Delta_{MNz}+\Delta_{SOz}\Delta_{MN\bot}\right)\mathcal{G}_{1}^{\bot}\right.\nonumber
\\
&&+\left(\Delta_{M}\Delta_{PM}+2\Delta_{SO\bot}\Delta_{MN\bot}+\Delta_{SOz}\Delta_{MNz}\right)\mathcal{G}_{1}^{z}\nonumber
\\
&&+\left(\Delta_{C}\Delta_{SOz}+\Delta_{PM}\Delta_{AMNz}+2\Delta_{MN\bot}\Delta_{CR\bot}\right)\mathcal{G}_{1}^{D}\nonumber
\\
&&\left.+\left(\Delta_{C}\Delta_{MNz}+\Delta_{M}\Delta_{AMNz}+2\Delta_{SO\bot}\Delta_{CR\bot}\right)\mathcal{G}_{1}^{m}\right],
\\
\frac{d\Delta_{CR\bot}}{d\ell}&=&-\Delta_{CR\bot}+2\Delta_{CR\bot}\left(-\Delta_{C}
+\Delta_{M}-\Delta_{AC}-\Delta_{SOz}+\Delta_{PM}-\Delta_{MNz}
+\Delta_{AMNz}
+\Delta_{CRz}\right)\nonumber
\\
&&\times\left(\mathcal{G}_{1}^{z}+\mathcal{G}_{1}^{D}+\mathcal{G}_{1}^{m}\right)
\nonumber
\\
&&+2\left[\left(\Delta_{C}\Delta_{AMNz}+\Delta_{M}\Delta_{MNz}+\Delta_{AC}\Delta_{CRz}+\Delta_{SOz}\Delta_{PM}\right)\mathcal{G}_{1}^{\bot}\right.\nonumber
\\
&&+2\left(\Delta_{C}\Delta_{AMN\bot}+\Delta_{M}\Delta_{MN\bot}+\Delta_{AC}\Delta_{CR\bot}+\Delta_{SO\bot}\Delta_{PM}\right)\mathcal{G}_{1}^{z}\nonumber
\\
&&+2\left(\Delta_{AC}\Delta_{SO\bot}+\Delta_{PM}\Delta_{CR\bot}+\Delta_{MN\bot}\Delta_{AMNz}+\Delta_{MNz}\Delta_{AMN\bot}\right)\mathcal{G}_{1}^{D}\nonumber
\\
&&\left.+2\left(\Delta_{M}\Delta_{CR\bot}+\Delta_{AC}\Delta_{MN\bot}+\Delta_{SO\bot}\Delta_{AMNz}+\Delta_{SOz}\Delta_{AMN\bot}\right)\mathcal{G}_{1}^{m}\right],
\\
\frac{d\Delta_{CRz}}{d\ell}&=&-\Delta_{CRz}+2\Delta_{CRz}
\left(-\Delta_{C}+\Delta_{M}-\Delta_{AC}-2\Delta_{SO\bot}+\Delta_{SOz}+\Delta_{PM}
-2\Delta_{MN\bot}+\Delta_{MNz}\right.\nonumber
\\
&&\left.+2\Delta_{AMN\bot}-\Delta_{AMNz}+2\Delta_{CR\bot}
-\Delta_{CRz}\right)\left(\mathcal{G}_{1}^{\bot}
-\mathcal{G}_{1}^{z}+\mathcal{G}_{1}^{D}+\mathcal{G}_{1}^{m}\right)\nonumber
\\
&&+4\left[\left(\Delta_{C}\Delta_{AMN\bot}+\Delta_{M}\Delta_{MN\bot}+\Delta_{AC}\Delta_{CR\bot}+\Delta_{SO\bot}\Delta_{PM}\right)\mathcal{G}_{1}^{\bot}\right.\nonumber
\\
&&+\left(\Delta_{AC}\Delta_{SOz}+\Delta_{PM}\Delta_{CRz}+2\Delta_{MN\bot}\Delta_{AMN\bot}\right)\mathcal{G}_{1}^{D}\nonumber
\\
&&\left.+\left(\Delta_{M}\Delta_{CRz}+\Delta_{AC}\Delta_{MNz}+2\Delta_{SO\bot}\Delta_{AMN\bot}\right)\mathcal{G}_{1}^{m}\right],
\end{eqnarray}
\end{widetext}
The transformations
\begin{eqnarray}
\Delta_{j}\frac{\Lambda}{2\pi^{2} v^{2}\sqrt{\eta}}\rightarrow\Delta_{j}
\end{eqnarray}
have been used. The expressions of $C_{0}^{dis}$, $C_{m}^{dis}$, $\mathcal{G}_{1}^{\bot}$, $\mathcal{G}_{1}^{z}$, $\mathcal{G}_{1}^{D}$, and $\mathcal{G}_{1}^{m}$
can be found in Appendix~\ref{Appendix:DerivationRGEs}.

\begin{figure*}[htbp]
\center
\includegraphics[width=6.8in]{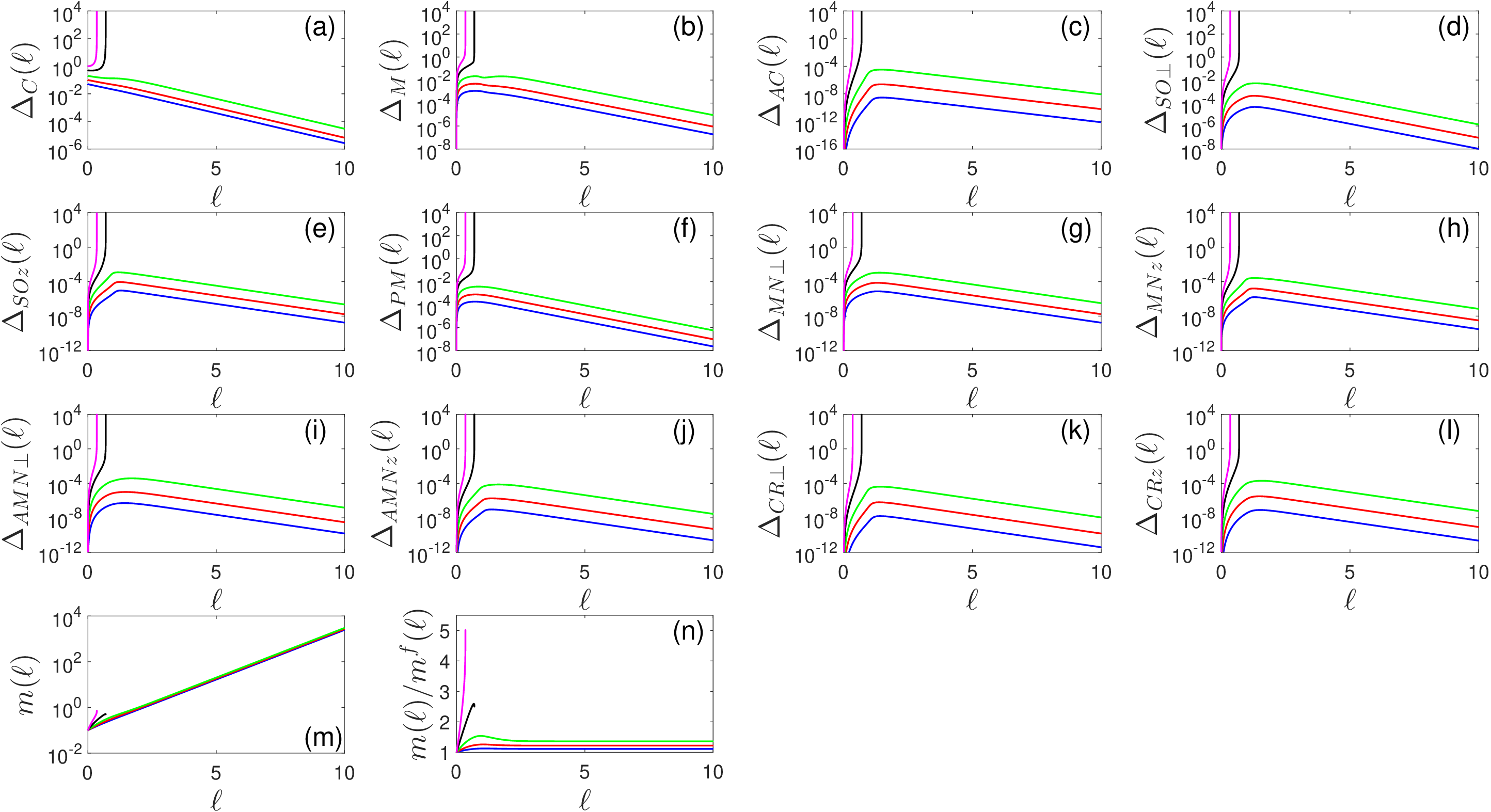}
\caption{(a)-(m): Flows of parameters $\Delta_{C}(\ell)$, $\Delta_{M}(\ell)$, $\Delta_{AC}(\ell)$, $\Delta_{SO\bot}(\ell)$, $\Delta_{SOz}(\ell)$, $\Delta_{PM}(\ell)$, $\Delta_{MN\bot}(\ell)$, $\Delta_{MNz}(\ell)$,
$\Delta_{AMN\bot}(\ell)$, $\Delta_{AMNz}(\ell)$, $\Delta_{CR\bot}(\ell)$, $\Delta_{CRz}(\ell)$, $m(\ell)$ and $m(\ell)/m^{f}(\ell)$. Blue, red, green, black, and magenta lines
correspond to the initial values $\Delta_{C0}=0.05, 0.1, 0.2, 0.5, 1.0$ respectively. Initial values of other disorder parameters are taken to be zero.
$m_{0}=0.1$, $B_{\bot0}=1$, $B_{z0}=1$, $D_{0}=1$ are taken. \label{Fig:DisorderAndmIniDeltaC}}
\end{figure*}

It is easy to find that
\begin{eqnarray}
B_{\bot}(\ell)&=&B_{\bot,0}e^{-\ell},
\\
B_{z}(\ell)&=&B_{z,0}e^{-\ell},
\\
D(\ell)&=&D_{0}e^{-\ell},
\end{eqnarray}
which take the same behaviors as the clean second-order TI.

Considering random chemical potential initially with different initial values $\Delta_{C0}$, the flows of different disorder coupling parameters are shown in
Figs.~\ref{Fig:DisorderAndmIniDeltaC}(a)-\ref{Fig:DisorderAndmIniDeltaC}(l). The flows of $m(\ell)$ and $m(\ell)/m^{f}(\ell)$ are depicted in Figs.~\ref{Fig:DisorderAndmIniDeltaC}(m) and \ref{Fig:DisorderAndmIniDeltaC}(n).  If $\Delta_{C0}$ is
small, $\Delta_{C}(\ell)$ approaches to zero in the lowest energy limit $\ell\rightarrow\infty$, other disorder coupling parameters $\Delta_{M}$, $\Delta_{AC}$, $\Delta_{SO\bot}$, $\Delta_{SOz}$,
$\Delta_{PM}$, $\Delta_{MN\bot}$, $\Delta_{MNz}$, $\Delta_{AMN\bot}$, $\Delta_{AMNz}$, $\Delta_{CR\bot}$, and $\Delta_{CRz}$ increase from zero at beginning and also
approach to zero in the lowest energy limit.  In this case, as shown in Fig.~\ref{Fig:DisorderAndmIniDeltaC}(m), $m(\ell)\rightarrow\infty$ in the lowest energy limit.

In Ref.~\cite{ZhaoPengLu21}, the behaviors $D(\ell)\rightarrow0$ and $m(\ell)\rightarrow\infty$ in the lowest energy limit are used as the criterion
for the transition from second-order TI to TI. According to their criterion, weak disorder drives second-order TI to TI. However, recent studies based on other methods
showed that second-order TI is robust against weak disorder \cite{WangC20, WangC21A}.  This contradiction is just due to that the criterion used in Ref.~\cite{ZhaoPengLu21}
is invalid.

As depicted in Fig.~\ref{Fig:DisorderAndmIniDeltaC}(m),
$m(\ell)/m^{f}(\ell)$ approaches to a positive constant value in the lowest energy limit if $\Delta_{C0}$ is small.
This indicates that $m(\ell)$ takes the qualitatively same behavior as the clean second-order TI. Additionally, $B_{\bot}(\ell)$, $B_{z}(\ell)$, $D(\ell)$ take the same
behaviors as clean second-order TI. These behaviors clearly reflect that weak disorder does not induce qualitative modification for second-order TI. Namely, second-order
TI is robust against weak disorder.

If $\Delta_{C0}$ is large enough, we can find that $\Delta_{C}$ approaches to infinity at a finite running parameter $\ell_{c}$.
$\Delta_{M}$, $\Delta_{AC}$, $\Delta_{SO\bot}$, $\Delta_{SOz}$,
$\Delta_{PM}$, $\Delta_{MN\bot}$, $\Delta_{MNz}$, $\Delta_{AMN\bot}$, $\Delta_{AMNz}$, $\Delta_{CR\bot}$, and $\Delta_{CRz}$ grow from zero and also flow to infinity
at $\ell_{c}$. As shown in Figs.~\ref{Fig:DisorderAndmIniDeltaC}(m) and \ref{Fig:DisorderAndmIniDeltaC}(n), we notice that $m(\ell)$ approaches to finite value when $\ell\rightarrow\ell_{c}$.
These behaviors represent that second-order TI is driven to a diffusive metal phase.

\begin{figure}[htbp]
\center
\includegraphics[width=3.3in]{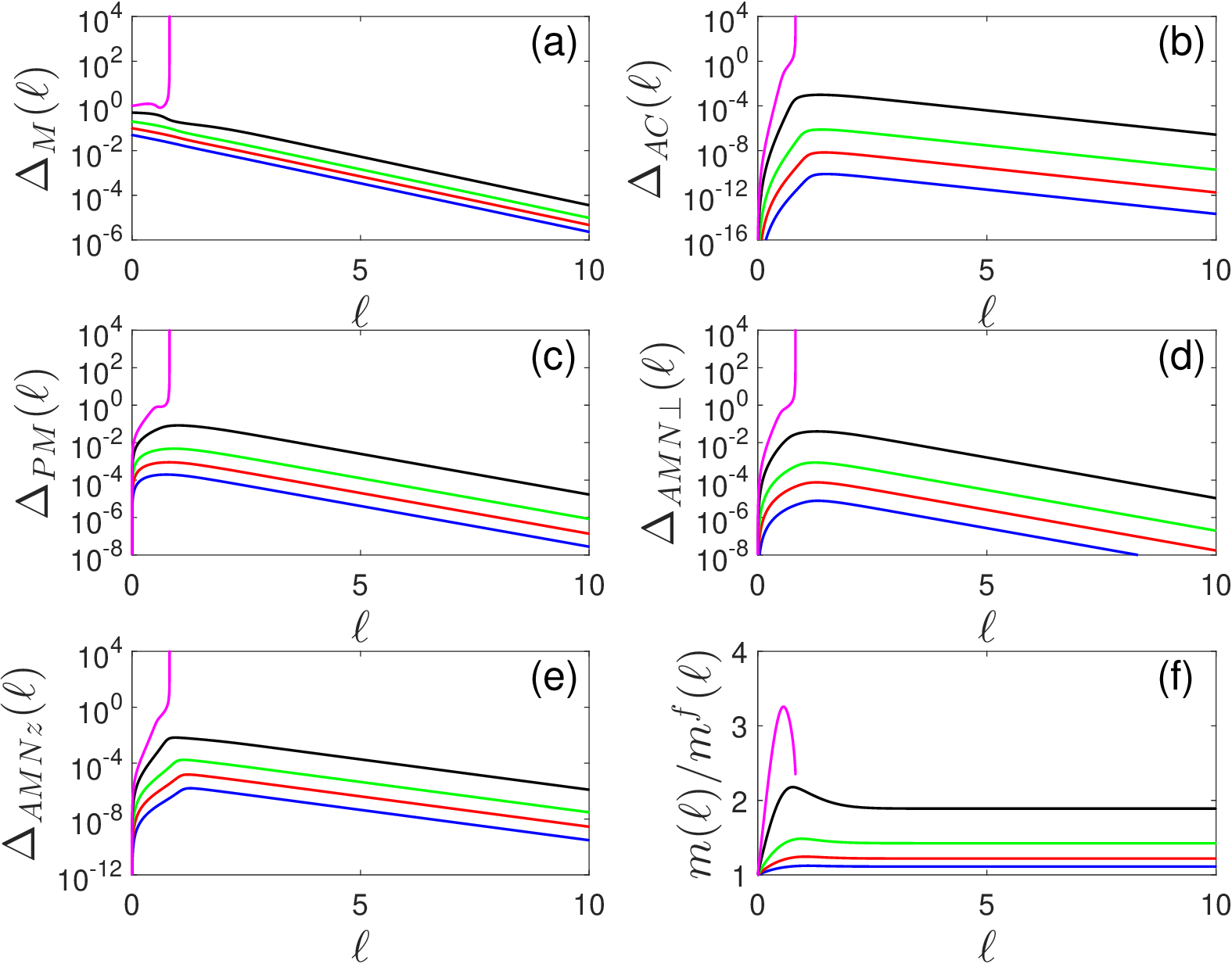}
\caption{(a)-(f): Flows of parameters $\Delta_{M}(\ell)$, $\Delta_{AC}(\ell)$, $\Delta_{PM}(\ell)$, $\Delta_{AMN\bot}(\ell)$, $\Delta_{AMNz}(\ell)$, and $m(\ell)/m^{f}(\ell)$. Blue, red, green, black, and magenta lines
correspond to the initial values $\Delta_{M0}=0.05, 0.1, 0.2, 0.5, 1.0$ respectively. Initial values of other disorder parameters are taken to be zero.
$m_{0}=0.1$, $B_{\bot0}=1$, $B_{z0}=1$, $D_{0}=1$ are taken. \label{Fig:DisorderAndmIniDeltaM}}
\end{figure}

Considering random mass initially with different initial values $\Delta_{M0}$, the flows of disorder coupling parameters $\Delta_{M}$, $\Delta_{AC}$, $\Delta_{PM}$, $\Delta_{AMN\bot}$,
and $\Delta_{AMNz}$ are displayed in
Figs.~\ref{Fig:DisorderAndmIniDeltaM}(a)-\ref{Fig:DisorderAndmIniDeltaM}(e). $\Delta_{C}$, $\Delta_{SO\bot}$, $\Delta_{SOz}$, $\Delta_{MN\bot}$, $\Delta_{MNz}$, $\Delta_{CR\bot}$, $\Delta_{CRz}$
are not generated, and equal to the initial value zero. Thus, the flows of these seven disorder coupling parameters are not shown in Fig.~\ref{Fig:DisorderAndmIniDeltaM}.
The flow of $m(\ell)/m^{f}(\ell)$ is shown in \ref{Fig:DisorderAndmIniDeltaM}(f). If $\Delta_{M0}$ is small enough, $\Delta_{M}$ flows to zero in the lowest energy limit $\ell\rightarrow\infty$.
$\Delta_{AC}$, $\Delta_{PM}$, $\Delta_{AMN\bot}$, $\Delta_{AMNz}$ increase from zero at the beginning, but become to decrease when $\ell$ is large enough and
 finally approach to zero in the lowest energy limit.  $m(\ell)/m^{f}(\ell)$ approaches to a positive constant which is greater than 1. It represents that disorder only induces quantitative increment
 but does not result in qualitative modification for $m$. Thus, second-order TI is stable in this case.

 If $\Delta_{M0}$ is large enough, $\Delta_{M}$ flows to infinity when $\ell$ approaches to a critical value $\ell_{c}$. $\Delta_{M}$, $\Delta_{AC}$, $\Delta_{PM}$, $\Delta_{AMN\bot}$, and $\Delta_{AMNz}$
are generated from zero and finally also approach to infinity when $\ell\rightarrow\ell_{c}$. $m(\ell)/m^{f}(\ell)$ flows to a finite value in the limit $\ell\rightarrow\ell_{c}$. These behaviors should
represent that the system is driven to a diffusive metal state. Whereas, it was shown that  the system can not be driven to diffusive metal phase even $\Delta_{M0}$ is large in Ref.~\cite{ZhaoPengLu21}.
This is due to that the generation of new types of disorder by $\Delta_{M}$ is not considered in Ref.~\cite{ZhaoPengLu21}

\begin{figure}[htbp]
\center
\includegraphics[width=3.3in]{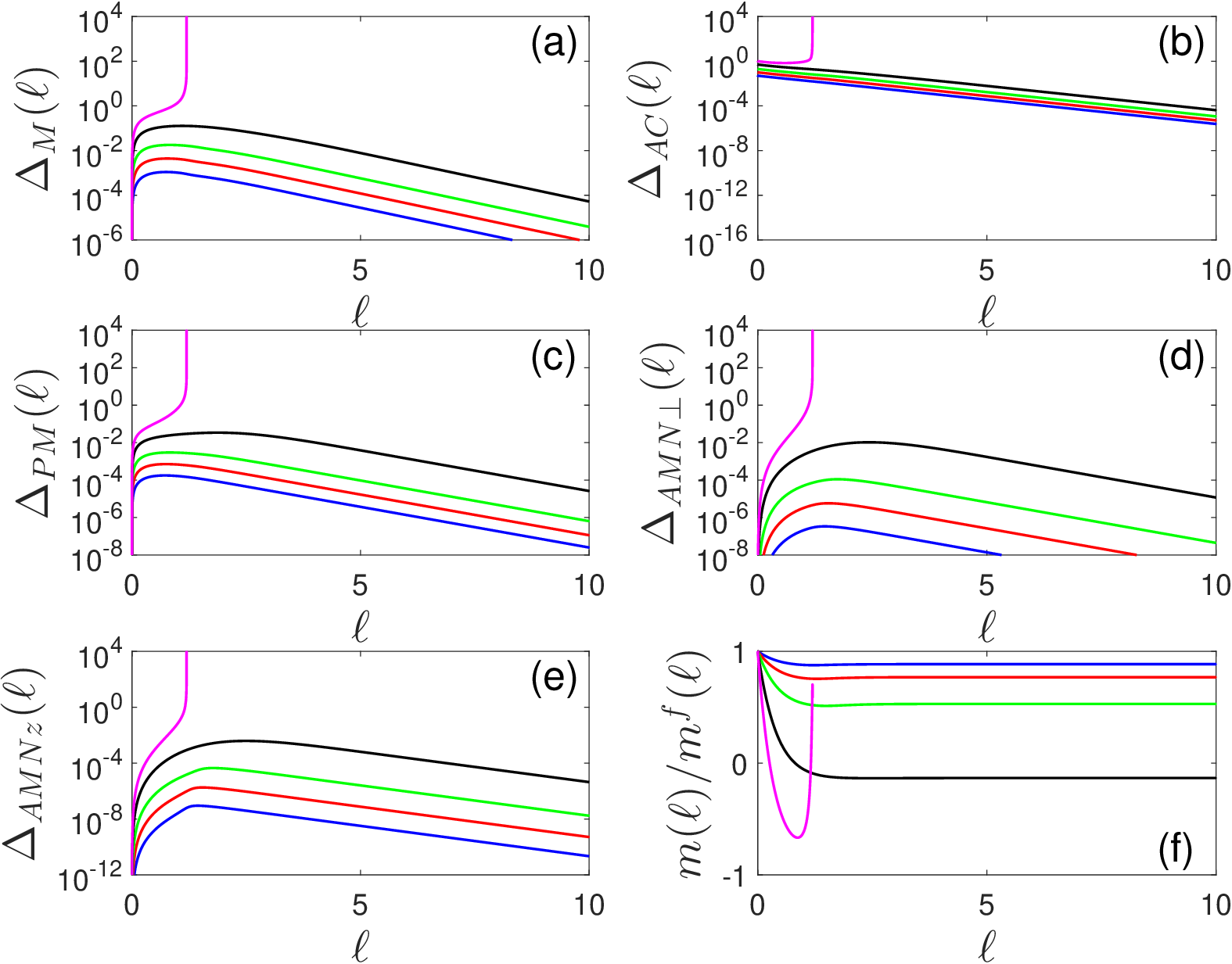}
\caption{(a)-(f): Flows of parameters $\Delta_{M}(\ell)$, $\Delta_{AC}(\ell)$, $\Delta_{PM}(\ell)$, $\Delta_{AMN\bot}(\ell)$, $\Delta_{AMNz}(\ell)$, and $m(\ell)/m^{f}(\ell)$. Blue, red, green, black, and magenta lines
correspond to the initial values $\Delta_{AC0}=0.05, 0.1, 0.2, 0.5, 1.0$ respectively. Initial values of other disorder parameters are taken to be zero.
$m_{0}=0.1$, $B_{\bot0}=1$, $B_{z0}=1$, $D_{0}=1$ are taken. \label{Fig:DisorderAndmIniDeltaAC}}
\end{figure}

\begin{figure}[htbp]
\center
\includegraphics[width=3.3in]{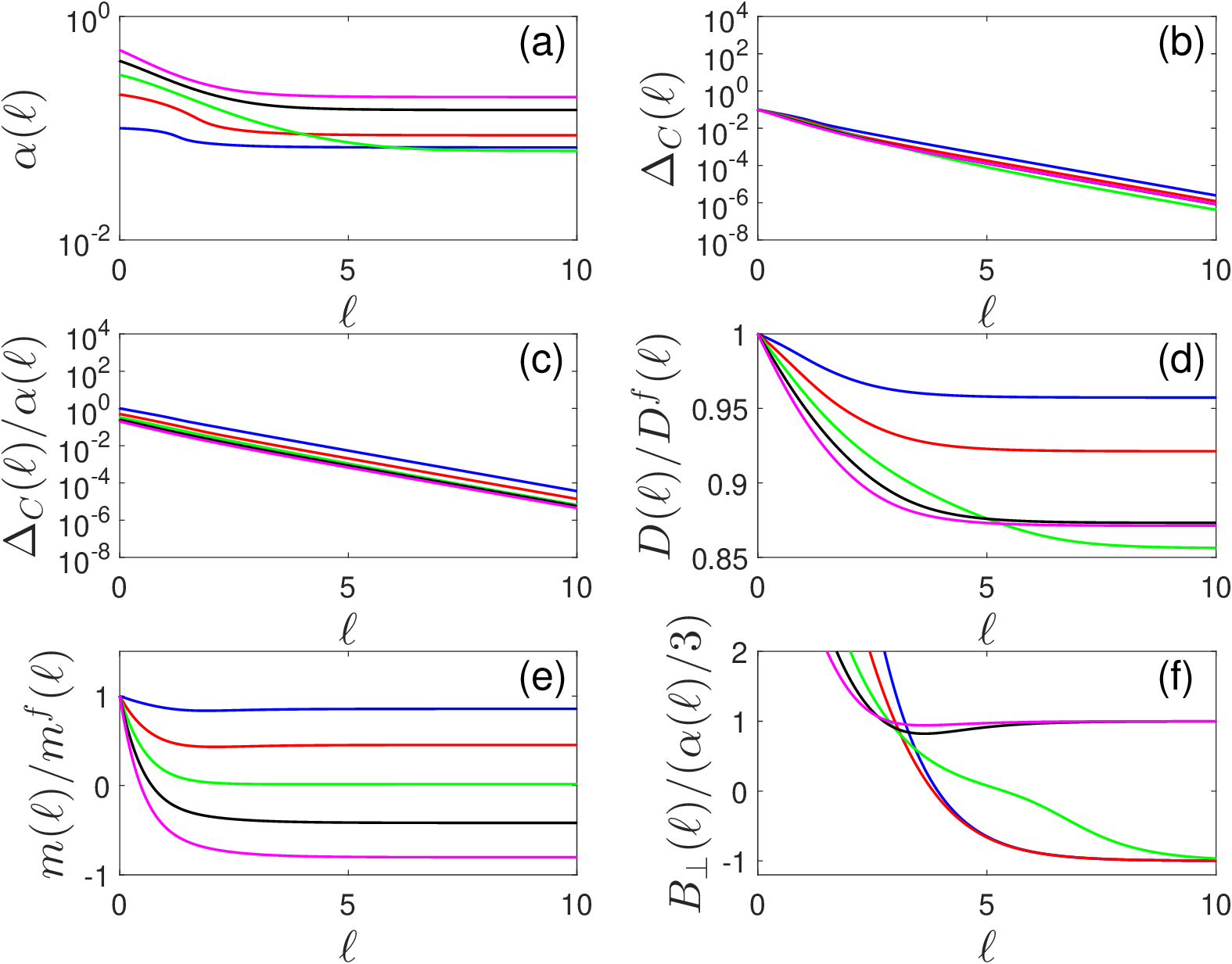}
\caption{(a)-(f): Flows of parameters $\alpha(\ell)$, $\Delta_{C}(\ell)$, $\Delta_{C}(\ell)/\alpha(\ell)$, $D(\ell)$, $m(\ell)/m^{f}(\ell)$, and $B_{\bot}(\ell)/(\alpha(\ell)/3)$ initially
considering both of long-range Coulomb interaction and random chemical potential. Blue, red, green, black, and magenta lines
correspond to the initial values $\alpha_{0}=0.1, 0.2, 0.3, 0.4, 0.5$ respectively. $\Delta_{C0}=0.1$, and initial values of other disorder parameters are taken to be zero.
$m_{0}=0.1$, $B_{\bot0}=1$, $B_{z0}=1$, $D_{0}=1$ are taken. \label{Fig:Interplay1}}
\end{figure}

In proper initial conditions, we notice that the second-order TI is driven to trivial band insulator with intermediate disorder strength. For example, considering random axial chemical initially  with
different values of $\Delta_{AC0}$, the flows of $\Delta_{M}$, $\Delta_{AC}$, $\Delta_{PM}$, $\Delta_{AMN\bot}$,
and $\Delta_{AMNz}$ are shown in
Figs.~\ref{Fig:DisorderAndmIniDeltaAC}(a)-\ref{Fig:DisorderAndmIniDeltaAC}(e). If $\Delta_{AC0}$ is small, the disorder coupling parameters  $\Delta_{M}$, $\Delta_{AC}$, $\Delta_{PM}$, $\Delta_{AMN\bot}$,
and $\Delta_{AMNz}$ all flow to zero in the lowest energy limit $\ell\rightarrow\infty$. According to Fig.~\ref{Fig:DisorderAndmIniDeltaAC}(f), $m(\ell)/m^{f}(\ell)$ flows to a positive constant value
which is smaller than one. It represents that $m$ only decreases
quantitatively but not receive qualitative modification. Thus, the system is still in  second-order TI phase with small $\Delta_{AC0}$. If $\Delta_{AC0}$ is large enough,
$\Delta_{M}$, $\Delta_{AC}$, $\Delta_{PM}$, $\Delta_{AMN\bot}$, and $\Delta_{AMNz}$ all approach to infinity at a finite running paramter $\ell_{c}$. $m(\ell)/m^{f}(\ell)$ approaches to finite value at same
$\ell_{c}$. In this case, the system is driven to diffusive metal phase. If $\Delta_{AC0}$ takes intermediate values, $\Delta_{M}$, $\Delta_{AC}$, $\Delta_{PM}$, $\Delta_{AMN\bot}$,
and $\Delta_{AMNz}$ all flow to zero in the lowest energy limit $\ell\rightarrow\infty$. Whereas, $m(\ell)/m^{f}(\ell)$ flows to a negative constant value. It represents that the system becomes a trivial band
insulator.

\section{Interplay of Coulomb interaction and disorder\label{Sec:Interplay}}

\begin{figure}[htbp]
\center
\includegraphics[width=3.3in]{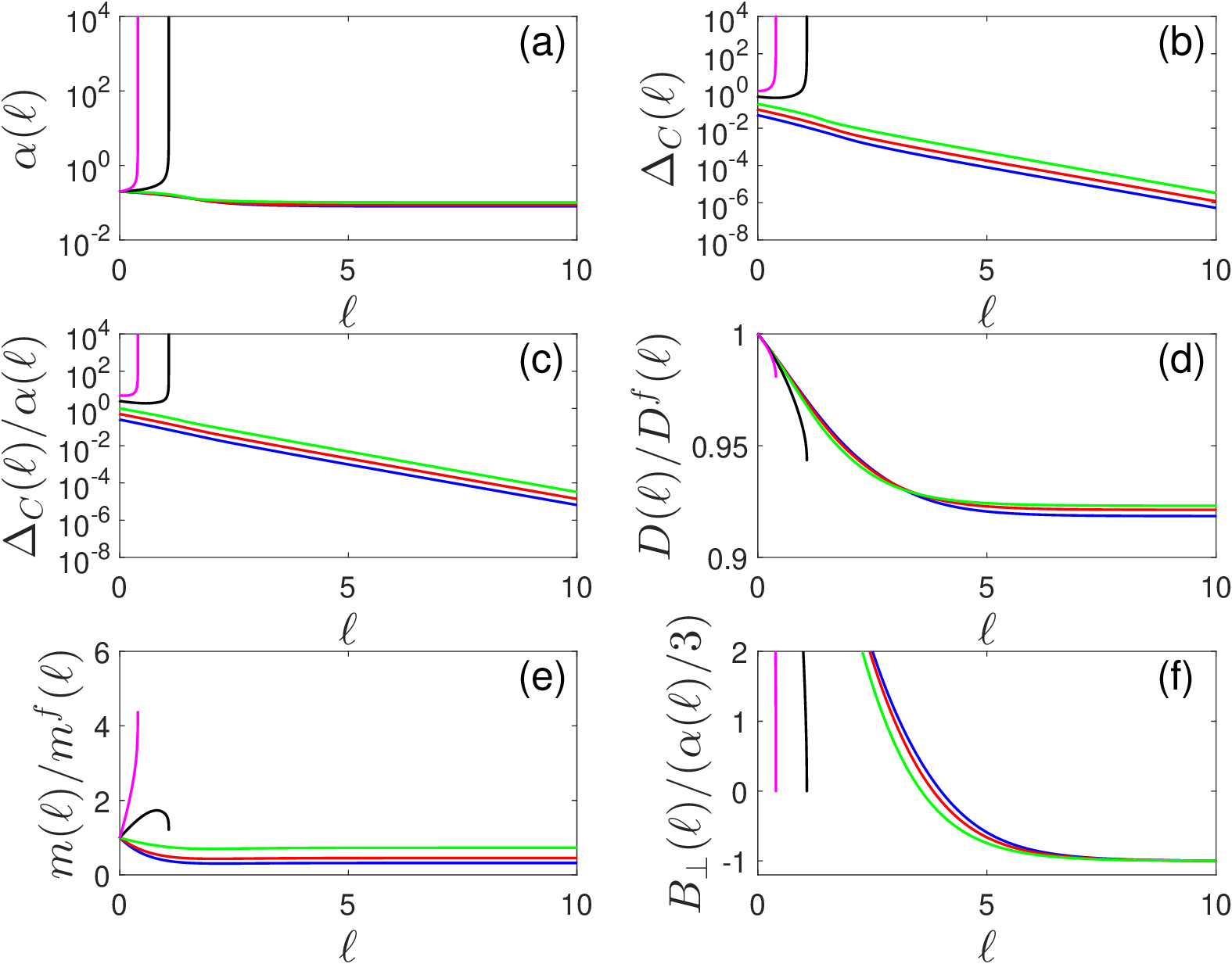}
\caption{(a)-(f): Flows of parameters $\alpha(\ell)$, $\Delta_{C}(\ell)$, $\Delta_{C}(\ell)/\alpha(\ell)$, $D(\ell)$, $m(\ell)/m^{f}(\ell)$, and $B_{\bot}(\ell)/(\alpha(\ell)/3)$ initially
considering both of long-range Coulomb interaction and random chemical potential. Blue, red, green, black, and magenta lines
correspond to the initial values $\Delta_{C0}=0.05, 0.1, 0.2, 0.5, 1.0$ respectively. Initial values of other disorder parameters are taken to be zero.
$\alpha_{0}=0.2$, $m_{0}=0.1$, $B_{\bot0}=1$, $B_{z0}=1$, $D_{0}=1$ are taken. \label{Fig:Interplay2}}
\end{figure}

In this section, we study the interplay of long-range Coulomb interaction and disorder. The RG equations for the model
parameters considering both of Coulomb interaction and disorder can be found in Appendix~\ref{Appendix:DerivationRGEs}. Considering initially both of
long-range Coulomb interaction and random chemical potential, we show the flows of $\alpha(\ell)$, $\Delta_{C}(\ell)$, $\Delta_{C}(\ell)/\alpha(\ell)$,
$D(\ell)$, $m(\ell)/m^{f}(\ell)$, and $B_{\bot}(\ell)/(\alpha(\ell)/3)$ in Figs.~\ref{Fig:Interplay1} and \ref{Fig:Interplay2} with different initial
conditions. The flow of $B_{z}(\ell)/(\alpha(\ell)/3)$ is not shown since it take the qualitatively same behavior as $B_{\bot}(\ell)/(\alpha(\ell)/3)$.
If $\Delta_{C0}$ and $\alpha_{0}$ both are small, $\alpha(\ell)$ flows to a constant value $\alpha^{*}$, $\Delta_{C}(\ell)$ approaches to zero,
$D(\ell)/D^{f}(\ell)$ approaches to a positive constant value, $m(\ell)/m^{f}(\ell)$ flows to a positive constant value, and
$B_{\bot}(\ell)/(\alpha(\ell)/3)\rightarrow -1$ in the lowest energy limit. These behaviors represent that second-order TI is driven to trivial band insulator.
For a given $\Delta_{C0}$, if $\alpha_{0}$ is large enough, we can find that $\alpha(\ell)$ flows to a constant value $\alpha^{*}$, $\Delta_{C}(\ell)$ approaches to zero,
$D(\ell)/D^{f}(\ell)$ approaches to a positive constant value, $m(\ell)/m^{f}(\ell)$ flows to a negative constant value, and $B_{\bot}(\ell)/(\alpha(\ell)/3)\rightarrow 1$ in the limit $\ell\rightarrow\infty$.
These results indicate that second-order TI becomes to trivial band insulator.  For a given $\alpha_{0}$, if $\Delta_{C0}$ is large enough, we notice that $\alpha(\ell)\rightarrow\infty$,
$\Delta_{C}(\ell)\rightarrow\infty$, $\Delta_{C}(\ell)/\alpha(\ell)\rightarrow\infty$, $D(\ell)/D^{f}(\ell)$, $m(\ell)/m^{f}(\ell)$,
and $B_{\bot}(\ell)$ flow to a finite values
at a finite running parameter $\ell_{c}$. These behaviors represent that second-order TI is driven to diffusive metal phase.

\section{Summary \label{Sec:Summary}}

In this article, we study the influence of Coulomb interaction on second-order TI by RG theory. We show that both the analysis method and conclusions
in recent studies in Ref.~\cite{ZhaoPengLu21} are unreliable. First, the criterion for transition from second-order TI
to TI adopted in Ref.~\cite{ZhaoPengLu21} is invalid. Second, the flow of $B$ is important, but is not paid attention in Ref.~\cite{ZhaoPengLu21}.
Through analyzing the corrections for flows of model parameters induced by Coulomb interaction, we find that second-order TI is unstable to trivial
band insulator not only under strong Coulomb interaction but also under weak Coulomb interaction. We also study the effects of disorder in second-order TI.
According to the  criterion adopted in Ref.~\cite{ZhaoPengLu21}, weak disorder will drive second-order TI to TI. Whereas, we find that weak disorder
does not result in qualitative modification for the Hamiltonian of second-order TI. It indicates that second-order TI is robust against weak disorder.
This result is consistent with recent studies about disorder effects in second-order TI through other methods \cite{WangC20, WangC21A}.  We also obtain the behaviors of
second-order TI considering both of long-range Coulomb interaction and disorder.

For very strong Coulomb interaction, the particle-hole pairs may be formed, and second-order TI may become an excitonic insulator. In this article,
we do not consider this possibility. This possibility could be studied by self-consistent Dyson-Schwinger equations \cite{WangLiu12, WangLiuZhang17A,
Carrington16, Carrington18}.

\section*{ACKNOWLEDGEMENTS}

We acknowledge the support from the National Natural Science
Foundation of China under Grants 11974356, and U1832209. A portion of this work
was supported by the High Magnetic Field Laboratory of Anhui Province under
Grant AHHM-FX-2020-01.

\appendix

\section{The propagators of fermion and boson}

The propagator of fermion is given by
\begin{eqnarray}
G_{0}\left(k_{0},\mathbf{k}\right)=\frac{1}{ik_{0}\gamma_{0}+\mathcal{H}_{f}},
\label{Eq:FermionPropagator}
\end{eqnarray}
where
\begin{eqnarray}
\mathcal{H}_{f}&=&i\left[v\left(k_{x}\gamma_{x}+k_{y}\gamma_{y}\right)
+v_{z}\gamma_{z}+D\left(k_{x}^{2}-k_{y}^{2}\right)\gamma_{5}\right]\nonumber
\\
&&+m-B_{\bot}k_{\bot}^{2}-B_{z}k_{z}^{2},
\end{eqnarray}
with $k_{\bot}^{2}=k_{x}^{2}+k_{y}^{2}$.
The propagator of boson field $\phi$ can be written as
\begin{eqnarray}
D_{0}\left(k_{0},\mathbf{k}\right)=\frac{1}{k_{\bot}^{2}+\eta k_{z}^{2}}. \label{Eq:BosonPropagator}
\end{eqnarray}

\section{The self-energy of fermion}

\begin{figure}[htbp]
\center
\includegraphics[width=3in]{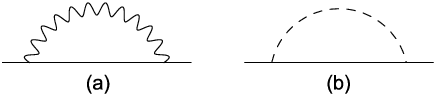}
\caption{Self-energy of fermions due to (a) Coulomb interaction and
(b) disorder. The solid line represents the
fermion propagator, and the wavy line stands for the boson
propagator that is equivalent to the Coulomb interaction function.
The dashed line denotes disorder scattering.
\label{Fig:FermionSelfEnergy}}
\end{figure}

\subsection{The self-energy of fermion induced by Coulomb interaction}

As shown in Fig.~\ref{Fig:FermionSelfEnergy}, the self-energy of fermion induced by Coulomb interaction is given by
\begin{eqnarray}
\Sigma\left(k_{0},\mathbf{k}\right)&=&-g^2\int_{-\infty}^{+\infty}\frac{dq_{0}}{2\pi}\int'\frac{d^3\mathbf{q}}{(2\pi)^{3}}
\gamma_{0}G_{0}(q_{0},\mathbf{q})\gamma_{0}\nonumber
\\
&&\times D_{0}\left(k_{0}-q_{0},\mathbf{k}-\mathbf{q}\right). \label{Eq:SelfEnergyExpression}
\end{eqnarray}
$\int'$ represents that a proper momentum shell will be chosen.
Substituting Eqs.~(\ref{Eq:FermionPropagator}) and (\ref{Eq:BosonPropagator}) into Eq.~(\ref{Eq:SelfEnergyExpression}),
to the leading order, we obtain
\begin{eqnarray}
\Sigma\left(k_{0},\mathbf{k}\right)
&=&-\frac{g^2}{2}\int'\frac{d^3\mathbf{q}}{(2\pi)^{3}}
\frac{\left(m-B_{\bot}q_{\bot}^{2}-B_{z}q_{z}^{2}\right)}
{E_{\mathbf{q}}\left(q_{\bot}^{2}+\eta q_{z}^{2}\right)}\nonumber
\\
&&-ivk_{x}\gamma_{x}g^2\int'\frac{d^3\mathbf{q}}{(2\pi)^{3}}
\frac{q_{x}^{2}}
{E_{\mathbf{q}}\left(q_{\bot}^{2}+\eta q_{z}^{2}\right)^{2}}\nonumber
\\
&&-ivk_{y}\gamma_{y}g^2\int'\frac{d^3\mathbf{q}}{(2\pi)^{3}}
\frac{q_{y}^{2}}
{E_{\mathbf{q}}\left(q_{\bot}^{2}+\eta q_{z}^{2}\right)^{2}}\nonumber
\\
&&-iv_{z}k_{z}\gamma_{z}\eta g^2\int'\frac{d^3\mathbf{q}}{(2\pi)^{3}}
\frac{q_{z}^{2}}
{E_{\mathbf{q}}\left(q_{\bot}^{2}+\eta q_{z}^{2}\right)^{2}}\nonumber
\\
&&+\left(k_{x}^{2}+k_{y}^{2}\right)\frac{g^2}{2}\int'\frac{d^3\mathbf{q}}{(2\pi)^{3}}
\frac{m-B_{\bot}q_{\bot}^{2}-B_{z}q_{z}^{2}}
{E_{\mathbf{q}}\left(q_{\bot}^{2}+\eta q_{z}^{2}\right)^{2}}\nonumber
\\
&&+ k_{z}^{2}\eta \frac{g^2}{2}\int'\frac{d^3\mathbf{q}}{(2\pi)^{3}}
\frac{m-B_{\bot}q_{\bot}^{2}-B_{z}q_{z}^{2}}
{E_{\mathbf{q}}\left(q_{\bot}^{2}+\eta q_{z}^{2}\right)^{2}}\nonumber
\\
&&-iDk_{x}^{2}\gamma_{5}2g^2\int'\frac{d^3\mathbf{q}}{(2\pi)^{3}}
\frac{\left(q_{x}^{2}-q_{y}^{2}\right)q_{x}^{2}}
{E_{\mathbf{q}}\left(q_{\bot}^{2}+\eta q_{z}^{2}\right)^{3}}\nonumber
\\
&&-iDk_{y}^{2}\gamma_{5}2g^2\int'\frac{d^3\mathbf{q}}{(2\pi)^{3}}
\frac{\left(q_{x}^{2}-q_{y}^{2}\right)q_{y}^{2}}
{E_{\mathbf{q}}\left(q_{\bot}^{2}+\eta q_{z}^{2}\right)^{3}}\nonumber
\\
&&-k_{x}^{2}2g^2\int'\frac{d^3\mathbf{q}}{(2\pi)^{3}}
\frac{\left[m-B_{\bot}q_{\bot}^{2}-B_{z}q_{z}^{2}\right]q_{x}^{2}}
{E_{\mathbf{q}}\left(q_{\bot}^{2}+\eta q_{z}^{2}\right)^{3}}\nonumber
\\
&&-k_{y}^{2}2g^2\int'\frac{d^3\mathbf{q}}{(2\pi)^{3}}
\frac{\left[m-B_{\bot}q_{\bot}^{2}-B_{z}q_{z}^{2}\right]q_{y}^{2}}
{E_{\mathbf{q}}\left(q_{\bot}^{2}+\eta q_{z}^{2}\right)^{3}}\nonumber
\\
&&-k_{z}^{2}2\eta^{2}g^2\int'\frac{d^3\mathbf{q}}{(2\pi)^{3}}
\frac{\left[m-B_{\bot}q_{\bot}^{2}-B_{z}q_{z}^{2}\right]q_{z}^{2}}
{E_{\mathbf{q}}\left(q_{\bot}^{2}+\eta q_{z}^{2}\right)^{3}},\nonumber
\\
\end{eqnarray}
where
\begin{eqnarray}
E_{\mathbf{q}}&=&\left[v^{2}q_{\bot}^{2}+v_{z}^{2}q_{z}^{2}+D^{2}\left(q_{x}^{2}-q_{y}^{2}\right)^{2}\right.\nonumber
\\
&&\left.+\left(m-B_{\bot}q_{\bot}^{2}-B_{z}q_{z}^{2}\right)^{2}\right]^{\frac{1}{2}}.
\end{eqnarray}

Using the transformation
\begin{eqnarray}
\mathbf{q}'=\left(q_{x},q_{y},\sqrt{\eta}q_{z}\right), \label{Eq:CoordinatTransformation}
\end{eqnarray}
and employing the RG scheme
\begin{eqnarray}
\int\frac{d^3\mathbf{q}'}{(2\pi)^{3}}=\frac{1}{8\pi^{3}}\int_{0}^{\pi}\sin(\varphi)d\varphi\int_{0}^{2\pi}d\theta\int_{b\Lambda}^{\Lambda}dq'q'^{2},
\label{Eq:RGScheme}
\end{eqnarray}
where $b=e^{-\ell}$ with $\ell$ being RG running parameter and $\Lambda$ the momentum cutoff,
we finally obtain
\begin{eqnarray}
\Sigma\left(k_{0},\mathbf{k}\right)
&=&-mC_{m}\ell-iv\left(k_{x}\gamma_{x}+k_{y}\gamma_{y}\right)C_{v}\ell\nonumber
\\
&&-iv_{z}k_{z}\gamma_{z} C_{v_{z}}\ell
-B_{\bot}\left(k_{x}^{2}+k_{y}^{2}\right)C_{B_{\bot}}\ell
\nonumber
\\
&&-B_{z}k_{z}^{2}C_{B_{z}} \ell-iD\left(k_{x}^{2}-k_{y}^{2}\right)\gamma_{5}C_{D}\ell,
\end{eqnarray}
where
\begin{eqnarray}
C_{m}&=&\alpha \frac{v\Lambda}{m}\left[\frac{m}{v\Lambda}\left(\mathcal{F}_{0}^{\bot}+\mathcal{F}_{0}^{z}\right)
-\frac{B_{\bot}\Lambda}{v}\mathcal{F}_{0}^{\bot}\right.\nonumber
\\
&&\left.-\frac{B_{z}\Lambda}{v\eta} \mathcal{F}_{0}^{z}\right],
\\
C_{v}&=&\alpha\mathcal{F}_{0}^{\bot},
\\
C_{v_{z}}&=&2\alpha\mathcal{F}_{0}^{z},
\\
C_{B_{\bot}}&=&\alpha \frac{v}{B_{\bot}\Lambda}
\left[\frac{m}{v\Lambda}\left(\mathcal{F}_{1}^{\bot}+\mathcal{F}_{1}^{z}\right)-\frac{B_{\bot}\Lambda}{v}\mathcal{F}_{1}^{\bot}\right.\nonumber
\\
&&\left.-\frac{B_{z}\Lambda}{v\eta} \mathcal{F}_{1}^{z}\right],
\\
C_{B_{z}}&=&\alpha\frac{ \eta v}{B_{z}\Lambda}\left\{\frac{m}{v\Lambda}\left[\mathcal{F}_{0}^{\bot}+\mathcal{F}_{0}^{z}
-2\left(\mathcal{F}_{1}^{\bot}+\mathcal{F}_{1}^{z}\right)\right]-\frac{B_{\bot}\Lambda}{v}\right.\nonumber
\\
&&\left.\times\left(\mathcal{F}_{0}^{\bot}-2\mathcal{F}_{1}^{\bot}\right)-\frac{B_{z}\Lambda}{v\eta}\left(\mathcal{F}_{0}^{z}-2\mathcal{F}_{1}^{z}\right)\right\},
\\
C_{D}&=&\alpha \mathcal{F}_{1}^{D},
\end{eqnarray}
with
\begin{eqnarray}
\alpha=\frac{g^{2}}{4\pi^{2}v\sqrt{\eta}}.
\end{eqnarray}
$\mathcal{F}_{0}^{\bot}$, $\mathcal{F}_{0}^{z}$, $\mathcal{F}_{1}^{\bot}$, $\mathcal{F}_{1}^{z}$, $\mathcal{F}_{1}^{D}$ are given by
\begin{eqnarray}
\mathcal{F}_{0}^{\bot}&=&\frac{1}{4\pi}\int_{0}^{\pi}\sin(\varphi)d\varphi\int_{0}^{2\pi}d\theta\frac{\sin^{2}(\varphi)}{\Xi},
\\
\mathcal{F}_{0}^{z}&=&\frac{1}{4\pi}\int_{0}^{\pi}\sin(\varphi)d\varphi\int_{0}^{2\pi}d\theta\frac{\cos^{2}(\varphi)}{\Xi},
\\
\mathcal{F}_{1}^{\bot}&=&\frac{1}{4\pi}\int_{0}^{\pi}\sin(\varphi)d\varphi\int_{0}^{2\pi}d\theta\frac{-\sin^{2}(\varphi)\cos(2\varphi)}{\Xi},
\\
\mathcal{F}_{1}^{z}&=&\frac{1}{4\pi}\int_{0}^{\pi}\sin(\varphi)d\varphi\int_{0}^{2\pi}d\theta\frac{-\cos^{2}(\varphi)\cos(2\varphi)}{\Xi},
\\
\mathcal{F}_{1}^{D}&=&\frac{1}{2\pi}\int_{0}^{\pi}\sin(\varphi)d\varphi\int_{0}^{2\pi}d\theta\frac{\sin^{4}(\varphi)\cos^{2}(2\theta)}{\Xi},
\end{eqnarray}
where
\begin{eqnarray}
\Xi&=&\left[\sin^{2}(\varphi)+\left(\frac{v_{z}}{v\sqrt{\eta}}\right)^{2}\cos^{2}(\varphi)\right.\nonumber
\\
&&+\frac{D^{2}}{v^{2}}\Lambda^{2}\sin^{4}(\varphi)\cos^{2}(2\theta)\nonumber
\\
&&\left.+\left(\frac{m}{v\Lambda}-\frac{B_{\bot}\Lambda}{v}\sin^{2}(\varphi)-\frac{B_{z}\Lambda}{v\eta}\cos^{2}(\varphi)\right)^{2}\right]^{\frac{1}{2}}. \label{Eq:XiExpression}
\end{eqnarray}
We should notice that there is a relation
\begin{eqnarray}
\mathcal{F}_{1}^{\bot}+\mathcal{F}_{1}^{z}=\mathcal{F}_{0}^{\bot}-\mathcal{F}_{0}^{z}.
\end{eqnarray}

\subsection{The self-energy of fermion induced by disorder scattering}

As depicted in Fig.~\ref{Fig:FermionSelfEnergy}, the self-energy of fermions induced by disorder scattering is expressed by
\begin{eqnarray}
\Sigma_{dis}(k_{0})
&=&\sum_{j}\Delta_{j}\int'\frac{d^3\mathbf{k}}{(2\pi)^{3}}\Gamma_{j}G_{0}\left(k_{0},\mathbf{k}\right)\Gamma_{j}.  \label{Eq:FermionSelfEnergyDisExpression}
\end{eqnarray}
Substituting Eq.~(\ref{Eq:FermionPropagator}) into Eq.~(\ref{Eq:FermionSelfEnergyDisExpression}), we obtain
\begin{eqnarray}
\Sigma_{dis}(k_{0})
&=&-ik_{0}\gamma_{0}C_{0}^{dis}\ell-mC_{m}^{dis}\ell,
\end{eqnarray}
where
\begin{eqnarray}
C_{0}^{dis}&=&\left(\Delta_{C}+\Delta_{M}+\Delta_{AC}+2\Delta_{SO\bot}+\Delta_{SOz}+\Delta_{PM}\right.\nonumber
\\
&&+2\Delta_{MN\bot}+\Delta_{MNz}+2\Delta_{AMN\bot}+\Delta_{AMNz}\nonumber
\\
&&\left.+2\Delta_{CR\bot}
+\Delta_{CRz}\right)\frac{\Lambda}{2\pi^{2}v^{2}\sqrt{\eta}}
\left(\mathcal{G}_{0}^{\bot}+\mathcal{G}_{0}^{z}\right), \label{Eq:C0Dis}
\\
C_{m}^{dis}&=&-\left(\Delta_{C}+\Delta_{M}-\Delta_{AC}-2\Delta_{SO\bot}-\Delta_{SOz}
\right.\nonumber
\\
&&-\Delta_{PM}+2\Delta_{MN\bot}+\Delta_{MNz}+2\Delta_{AMN\bot}\nonumber
\\
&&\left.+\Delta_{AMNz}-2\Delta_{CR\bot}-\Delta_{CRz}\right)\frac{\Lambda}{2\pi^{2}v^{2}\sqrt{\eta}}\frac{v\Lambda}{m}\nonumber
\\
&&\times\left[\frac{m}{v\Lambda}\left(\mathcal{G}_{0}^{\bot}+\mathcal{G}_{0}^{z}\right)
-\frac{B_{\bot}\Lambda}{v}\mathcal{G}_{0}^{\bot}
-\frac{B_{z}\Lambda}{v\eta}\mathcal{G}_{0}^{z}\right], \label{Eq:C0mDis}
\end{eqnarray}
with
\begin{eqnarray}
\mathcal{G}_{0}^{\bot}&=&\frac{1}{4\pi}\int_{0}^{\pi}d\varphi\sin(\varphi)\int_{0}^{2\pi}d\theta \frac{\sin^{2}(\varphi)}
{\Xi^{2}},
\\
\mathcal{G}_{0}^{z}&=&\frac{1}{4\pi}\int_{0}^{\pi}d\varphi\sin(\varphi)\int_{0}^{2\pi}d\theta\frac{\cos^{2}(\varphi)}
{\Xi^{2}}.
\end{eqnarray}
$\Xi$ is given by Eq.~(\ref{Eq:XiExpression}).

\begin{figure}[htbp]
\center
\includegraphics[width=1.8in]{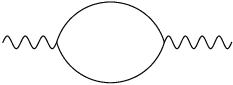}
\caption{Self-energy of bosonic field.
\label{Fig:BosonSelfEnergy}}
\end{figure}

\begin{widetext}

\section{The self-energy of boson}

As shown in Fig.~\ref{Fig:BosonSelfEnergy}, the self-energy of boson is defined as
\begin{eqnarray}
\Pi(k_{0},\mathbf{k})=g^{2}\int\frac{dq_{0}}{2\pi}\int'\frac{d^3\mathbf{q}}{(2\pi)^{3}}
\mathrm{Tr}\left[\gamma_{0}G_{0}(q_{0},\mathbf{q})\gamma_{0}G_{0}\left(k_{0}+q_{0},\mathbf{k},\mathbf{q}\right)\right].
\label{Eq:BosonPropagatorDefinition}
\end{eqnarray}
Substituting Eq.~(\ref{Eq:FermionPropagator}) into Eq.~(\ref{Eq:BosonPropagatorDefinition}), we arrive
\begin{eqnarray}
\Pi(k_{0},\mathbf{k})
&=&4g^{2}\int\frac{dq_{0}}{2\pi}\int'\frac{d^3\mathbf{q}}{(2\pi)^{3}}
\frac{1}
{\left(q_{0}^{2}+E_{\mathbf{q}}^{2}\right)\left[\left(k_{0}+q_{0}\right)^{2}+E_{\mathbf{k}+\mathbf{q}}^{2}\right]}
\Big\{-q_{0}\left(k_{0}+q_{0}\right)+v^{2}q_{x}\left(k_{x}+q_{x}\right)+v^{2}q_{y}\left(k_{y}+q_{y}\right)\nonumber
\\
&&+v_{z}^{2}q_{z}\left(k_{z}+q_{z}\right)+D^{2}\left(q_{x}^{2}-q_{y}^{2}\right)
\left[\left(k_{x}+q_{x}\right)^{2}-\left(k_{y}+q_{y}\right)^{2}\right]
+\left(m-B_{i}q_{i}^{2}\right)\left[m-B_{i}\left(k_{i}+q_{i}\right)^{2}\right]\Big\}.
\end{eqnarray}
Taking $k_{0}=0$, and performing further simplification, we get
\begin{eqnarray}
\Pi(\mathbf{k})
&=&-g^{2}\frac{1}{2}\int'\frac{d^3\mathbf{q}}{(2\pi)^{3}}\frac{1}{E_{\mathbf{q}}^{3}}F_{1}(\mathbf{k},\mathbf{q})+g^{2}\frac{1}{2}\int'\frac{d^3\mathbf{q}}{(2\pi)^{3}}
\frac{1}{E_{\mathbf{q}}^{5}}F_{1}^{2}(\mathbf{k},\mathbf{q})\nonumber
+g^{2}\int'\frac{d^3\mathbf{q}}{(2\pi)^{3}}\frac{1}{E_{\mathbf{q}}^{3}}F_{2}(\mathbf{k},\mathbf{q})\nonumber
\\
&&-g^{2}\frac{3}{4}\int'\frac{d^3\mathbf{q}}{(2\pi)^{3}}
\frac{1}{E_{\mathbf{q}}^{5}}F_{1}(\mathbf{k},\mathbf{q})F_{2}(\mathbf{k},\mathbf{q}),
\end{eqnarray}
where
\begin{eqnarray}
F_{1}(\mathbf{k},\mathbf{q})&=&\left(2v^{2}q_{x}+4D^{2}q_{x}^{3}-4D^{2}q_{x}q_{y}^{2}-4mB_{\bot}q_{x}
+4B_{\bot}^{2}q_{x}q_{\bot}^{2}+4B_{\bot}B_{z}q_{x}q_{z}^{2}\right)k_{x}\nonumber
\\
&&+\left(2v^{2}q_{y}+4D^{2}q_{y}^{3}-4D^{2}q_{y}q_{x}^{2}-4mB_{\bot}q_{y}+4B_{\bot}^{2}q_{y}q_{\bot}^{2}
+4B_{\bot}B_{z}q_{y}q_{z}^{2}\right)k_{y}\nonumber
\\
&&+\left(2v_{z}^{2}q_{z}-4mB_{z}q_{z}+4B_{z}^{2}q_{z}^{3}+4B_{z}B_{\bot}q_{z}q_{\bot}^{2}\right)k_{z}\nonumber
\\
&&+\left(v^{2}+6D^{2}q_{x}^{2}-2D^{2}q_{y}^{2}+4B_{\bot}^{2}q_{x}^{2}-2mB_{\bot}+2B_{\bot}^{2}q_{\bot}^{2}
+2B_{\bot}B_{z}q_{z}^{2}\right)k_{x}^{2}\nonumber
\\
&&+\left(v^{2}+6D^{2}q_{y}^{2}-2D^{2}q_{x}^{2}+4B_{\bot}^{2}q_{y}^{2}-2mB_{\bot}+2B_{\bot}^{2}q_{\bot}^{2}
+2B_{\bot}B_{z}q_{z}^{2}\right)k_{y}^{2}\nonumber
\\
&&+\left(v_{z}^{2}+4B_{z}^{2}q_{z}^{2}-2mB_{z}+2B_{\bot}B_{z}q_{\bot}^{2}+2B_{z}^{2}q_{z}^{2}\right)k_{z}^{2}\nonumber
\\
&&+\left(-8D^{2}q_{x}q_{y}+8B_{\bot}^{2}q_{x}q_{y}\right)k_{x}k_{y}+8B_{\bot}B_{z}q_{x}q_{z}k_{x}k_{z}+8B_{\bot}B_{z}q_{y}q_{z}k_{y}k_{z},
\end{eqnarray}
and
\begin{eqnarray}
F_{2}(\mathbf{k},\mathbf{q})&=&\left(v^{2}q_{x}+2D^{2}q_{x}^{3}-2D^{2}q_{x}q_{y}^{2}-2mB_{\bot}q_{x}
+2B_{\bot}^{2}q_{\bot}^{2}q_{x}+2B_{\bot}B_{z}q_{z}^{2}q_{x}\right)k_{x}\nonumber
\\
&&+\left(v^{2}q_{y}+2D^{2}q_{y}^{3}-2D^{2}q_{y}q_{x}^{2}-2mB_{\bot}q_{y}+2B_{\bot}^{2}q_{\bot}^{2}q_{y}+2B_{\bot}B_{z}q_{z}^{2}q_{y}\right)k_{y}\nonumber
\\
&&+\left(v_{z}^{2}q_{z}-2mB_{z}q_{z}+2mB_{\bot}B_{z}q_{\bot}^{2}q_{z}+2B_{z}^{2}q_{z}^{3}\right)k_{z}\nonumber
\\
&&+\left(D^{2}q_{x}^{2}-D^{2}q_{y}^{2}-mB_{\bot}+B_{\bot}^{2}q_{\bot}^{2}+B_{\bot}B_{z}q_{z}^{2}\right)k_{x}^{2}\nonumber
\\
&&+\left(-D^{2}q_{x}^{2}+D^{2}q_{y}^{2}-mB_{\bot}+B_{\bot}^{2}q_{\bot}^{2}+B_{\bot}B_{z}q_{z}^{2}\right)k_{y}^{2}\nonumber
\\
&&+\left(-mB_{z}+B_{\bot}B_{z}q_{\bot}^{2}+B_{z}^{2}q_{z}^{2}\right)k_{z}^{2}.
\end{eqnarray}
Retaining the quadratic order of $k_{i}$, $\Pi(\mathbf{k})$ can be further written as
\begin{eqnarray}
&&\Pi(\mathbf{k})
\\
&=&k_{x}^{2}\frac{1}{2}g^{2}\int'\frac{d^3\mathbf{q}}{(2\pi)^{3}}
\left\{-\frac{1}{E_{\mathbf{q}}^{3}}
\left(v^{2}+4D^{2}q_{x}^{2}+4B_{\bot}^{2}q_{x}^{2}\right)+\frac{1}{E_{\mathbf{q}}^{5}}
\left[v^{2}+2D^{2}\left(q_{x}^{2}-q_{y}^{2}\right)-2mB_{\bot}
+2B_{\bot}^{2}q_{\bot}^{2}+2B_{\bot}B_{z}q_{z}^{2}\right]^{2}q_{x}^{2}\right\}\nonumber
\\
&&+k_{y}^{2}\frac{1}{2}g^{2}\int'\frac{d^3\mathbf{q}}{(2\pi)^{3}}
\left\{-\frac{1}{E_{\mathbf{q}}^{3}}\left(v^{2}+4D^{2}q_{y}^{2}+4B_{\bot}^{2}q_{y}^{2}\right)
+\frac{1}{E_{\mathbf{q}}^{5}}\left[v^{2}+2D^{2}\left(q_{y}^{2}-q_{x}^{2}\right)-2mB_{\bot}+2B_{\bot}^{2}q_{\bot}^{2}
+2B_{\bot}B_{z}q_{z}^{2}\right]^{2}q_{y}^{2}\right\}\nonumber
\\
&&+k_{z}^{2}\frac{1}{2}g^{2}\int'\frac{d^3\mathbf{q}}{(2\pi)^{3}}
\left[-\frac{1}{E_{\mathbf{q}}^{3}}\left(v_{z}^{2}+4B_{z}^{2}q_{z}^{2}\right)
+\frac{1}{E_{\mathbf{q}}^{5}}\left(v_{z}^{2}-2mB_{z}+2B_{z}^{2}q_{z}^{2}+2B_{z}B_{\bot}
q_{\bot}^{2}\right)^{2}q_{z}^{2}\right].
\end{eqnarray}
Adopting the transformation as shown in Eq.~(\ref{Eq:CoordinatTransformation}), and utilizing the RG scheme as shown in Eq.~(\ref{Eq:RGScheme}),
we obtain
\begin{eqnarray}
\Pi(\mathbf{k})&=&-C_{x}k_{x}^{2}\ell-C_{y}k_{y}^{2}\ell-\eta C_{z} k_{z}^{2}\ell,
\end{eqnarray}
where
\begin{eqnarray}
C_{x}&=&\alpha \mathcal{F}_{2}^{x},
\\
C_{y}&=&\alpha \mathcal{F}_{2}^{y},
\\
C_{z}&=&\alpha \mathcal{F}_{2}^{z}.
\end{eqnarray}
$\mathcal{F}_{2}^{x}$, $\mathcal{F}_{2}^{y}$, $\mathcal{F}_{2}^{z}$ are expressed by
\begin{eqnarray}
\mathcal{F}_{2}^{x}&=&\frac{1}{4\pi}\int_{0}^{\pi}\sin(\varphi)d\varphi
\int_{0}^{2\pi}d\theta
\left\{\frac{1+4\left(\left(\frac{D\Lambda}{v}\right)^{2}+\left(\frac{B_{\bot}\Lambda}{v}\right)^{2}\right)
\sin^{2}(\varphi)\cos^{2}(\theta)}{\Xi^{3}}
\right.\nonumber
\\
&&\left.-\frac{\left[1+2\left(\frac{D\Lambda}{v}\right)^{2}\sin^{2}(\varphi)\cos(2\theta)-2\frac{m}{v\Lambda}\frac{B_{\bot}\Lambda}{v}
+2\left(\frac{B_{\bot}\Lambda}{v}\right)^{2}\sin^{2}(\varphi)+2\frac{B_{\bot}\Lambda}{v}\frac{B_{z}\Lambda}{v\eta}\cos^{2}(\varphi)\right]^{2}\sin^{2}(\varphi)
\cos^{2}(\theta)}
{\Xi^{5}}
\right\},
\\
\mathcal{F}_{2}^{y}&=&\frac{1}{4\pi}\int_{0}^{\pi}\sin(\varphi)d\varphi
\int_{0}^{2\pi}d\theta
\left\{\frac{1+4\left(\left(\frac{D\Lambda}{v}\right)^{2}+\left(\frac{B_{\bot}\Lambda}{v}\right)^{2}\right)\sin^{2}(\varphi)\sin^{2}(\theta)}{\Xi^{3}}\right.\nonumber
\\
&&\left.-\frac{\left[1-2\left(\frac{D\Lambda}{v}\right)^{2}\sin^{2}(\varphi)\cos(2\theta)-2\frac{m}{v\Lambda}
\frac{B_{\bot}\Lambda}{v}+2\left(\frac{B_{\bot}\Lambda}{v}\right)^{2}\sin^{2}(\varphi)
+2\frac{B_{\bot}\Lambda}{v}\frac{B_{z}\Lambda}{v\eta}\cos^{2}(\varphi)\right]^{2}\sin^{2}(\varphi)\sin^{2}(\theta)}{\Xi^{5}}\right\},
\\
\mathcal{F}_{2}^{z}&=&\frac{1}{4\pi}\int_{0}^{\pi}\sin(\varphi)d\varphi
\int_{0}^{2\pi}d\theta\left\{\frac{\left(\frac{v_{z}}{v\sqrt{\eta}}\right)^{2}+4\left(\frac{B_{z}\Lambda}{v\eta}\right)^{2}\cos^{2}(\varphi)}{\Xi^{3}}\right.\nonumber
\\
&&\left.-\frac{\left(\left(\frac{v_{z}}{v\sqrt{\eta}}\right)^{2}-2\frac{m}{v\Lambda}\frac{B_{z}\Lambda}{v\eta}
+2\left(\frac{B_{z}\Lambda}{v\eta}\right)^{2}\cos^{2}(\varphi)+2\frac{B_{z}\Lambda}{v\eta}\frac{B_{\bot}\Lambda}{v}
\sin^{2}(\varphi)\right)^{2}\cos^{2}(\varphi)}{\Xi^{5}}\right\},
\end{eqnarray}
\end{widetext}
where $\Xi$ is given by Eq.~(\ref{Eq:XiExpression}).
It easy to verify that $\mathcal{F}_{2}^{x}=\mathcal{F}_{2}^{y}$.  Then we can define
\begin{eqnarray}
\mathcal{F}_{2}^{\bot}\equiv\mathcal{F}_{2}^{x}=\mathcal{F}_{2}^{y},
\end{eqnarray}
and
\begin{eqnarray}
C_{\bot}=\alpha\mathcal{F}_{2}^{\bot}.
\end{eqnarray}
Accordingly, the polarization can be written as
\begin{eqnarray}
\Pi(\mathbf{k})&=&-C_{\bot}k_{\bot}^{2}\ell-\eta C_{z} k_{z}^{2}\ell.
\end{eqnarray}

\section{The corrections to fermion-boson coupling vertex}

\begin{figure}[htbp]
\center
\includegraphics[width=2.7in]{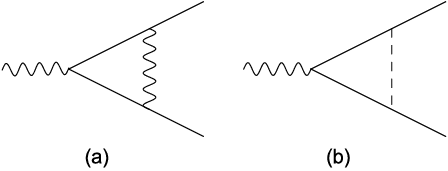}
\caption{Corrections to fermion-boson coupling due to (a) Coulomb
interaction and (b) disorder. \label{Fig:CoulombVertexCorrection}}
\end{figure}

As shown in Fig.~\ref{Fig:CoulombVertexCorrection}(a), the correction to fermion-boson coupling vertex induced by Coulomb interaction takes the form
\begin{eqnarray}
\Gamma^{(1)}&=&-ig^{3}\int_{-\infty}^{+\infty}\frac{dq_{0}}{2\pi}\int'\frac{d^3\mathbf{q}}{(2\pi)^{3}}\gamma_{0}
G_{0}(q_{0},\mathbf{q})\nonumber
\\
&&\times\gamma_{0}G_{0}\left(q_{0},\mathbf{q}\right)\gamma_{0}
D_{0}(q_{0},\mathbf{q}). \label{Eq:VertexCorrectionCoulombA}
\end{eqnarray}
Substituting the expressions of fermion propagator and boson propagator into Eq.~(\ref{Eq:VertexCorrectionCoulombA}), we find
\begin{eqnarray}
\Gamma^{(1)}
&=&0,
\end{eqnarray}
which indicates
\begin{eqnarray}
\delta g^{(1)}=0.
\end{eqnarray}

The correction to fermion-boson coupling induced by disorder scattering as shown in Fig.~\ref{Fig:CoulombVertexCorrection}(b) can be written as
\begin{eqnarray}
\Gamma^{(2)}
&=&ig\sum_{j}\Delta_{j}\int'\frac{d^3\mathbf{k}}{(2\pi)^{3}}\Gamma_{j}G_{0}(0,\mathbf{k})\gamma_{0}\nonumber
\\
&&\times G_{0}(0,\mathbf{k})\Gamma_{j} \label{Eq:VertexCorrectionCoulombB}
\end{eqnarray}
Substituting Eq.~(\ref{Eq:FermionPropagator}) into Eq.~(\ref{Eq:VertexCorrectionCoulombB}), one can get
\begin{eqnarray}
\Gamma^{(2)}
&=&ig\gamma_{0}C_{0}^{dis}\ell,
\end{eqnarray}
where $C_{0}^{dis}$ is given by Eq.~(\ref{Eq:C0Dis}). Thus,
\begin{eqnarray}
\delta g^{(2)}&=&gC_{0}^{dis}\ell.
\end{eqnarray}

Then we obtain
\begin{eqnarray}
\delta g&=&\delta g^{(1)}+\delta g^{(2)}=gC_{0}^{dis}\ell.
\end{eqnarray}

\section{Corrections to fermion-disorder coupling vertex}

\begin{figure}[htbp]
\center
\includegraphics[width=3.3in]{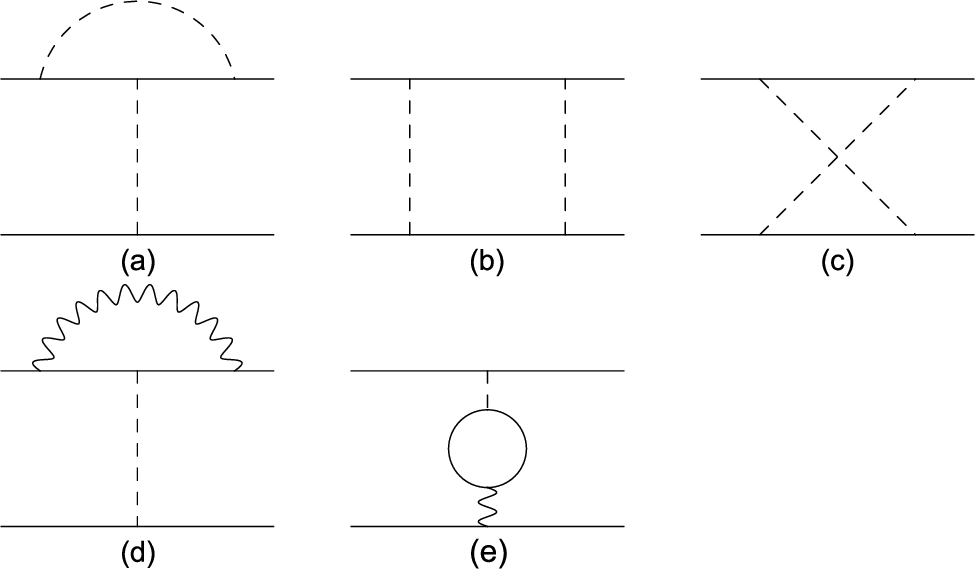}
\caption{One-loop Feynman diagrams for the corrections to the
fermion-disorder coupling. \label{Fig:VertexCorrection}}
\end{figure}

The correction induced by disorder scattering from Fig.~\ref{Fig:VertexCorrection}(a) can be written as
\begin{eqnarray}
W^{(1)}&=&\sum_{i}W_{i}^{(1)}, \label{Eq:VertexCorrectionDisW1}
\end{eqnarray}
where
\begin{eqnarray}
W_{i}^{(1)}&=&\Delta_{i}\sum_{j}\Delta_{j}\left(\bar{\Psi}_{a}\Gamma_{i}\Psi\right)\int'\frac{d^3\mathbf{k}}{(2\pi)^{3}}
\left[\bar{\Psi}_{b}\Gamma_{j}G_{0}(0,\mathbf{k})\Gamma_{i}\right.\nonumber
\\
&&\left.\times G_{0}(0,\mathbf{k})\Gamma_{j}\Psi_{b}\right]. \label{Eq:VertexCorrectionDisW1i}
\end{eqnarray}

The correction generated by disorder scattering from Figs.~\ref{Fig:VertexCorrection}(b) and \ref{Fig:VertexCorrection}(c) takes the  form
\begin{eqnarray}
W^{(2)+(3)}&=&\sum_{i}\sum_{i\le j}W_{ij}^{(2)+(3)}, \label{Eq:VertexCorrectionW23}
\end{eqnarray}
where
\begin{eqnarray}
W_{ij}^{(2)+(3)}&=&\Delta_{i}\Delta_{j}\int'\frac{d^3\mathbf{k}}{(2\pi)^{3}}\left[\bar{\Psi}_{a}\Gamma_{i}G_{0}(0,\mathbf{k})\Gamma_{j}\Psi_{a}\right]\nonumber
\\
&&\times\left\{\bar{\Psi}_{b}\left[\Gamma_{j}G_{0}(0,\mathbf{k})\Gamma_{i}\right.\right.\nonumber
\\
&&\left.\left.+\Gamma_{i}G_{0}(0,-\mathbf{k})\Gamma_{j}\right]\Psi_{b}\right\}. \label{Eq:VertexCorrectionW23ij}
\end{eqnarray}

Fig.~\ref{Fig:VertexCorrection}(d) induces the correction
\begin{eqnarray}
W^{(4)}&=&\sum_{i}W_{i}^{(4)}, \label{Eq:VertexCorrectionDisW4}
\end{eqnarray}
where
\begin{eqnarray}
W_{i}^{(4)}&=&-\Delta_{i}g^{2}\left(\bar{\Psi}_{a}\Gamma_{i}\Psi_{a}\right)\int\frac{dq_{0}}{2\pi}
\int'\frac{d^3\mathbf{q}}{(2\pi)^{3}}\left[\bar{\Psi}_{b}\gamma_{0}\right.\nonumber
\\
&&\left.\times G_{0}(q_{0},\mathbf{q})\Gamma_{i}G_{0}(q_{0},\mathbf{q})\gamma_{0}\Psi_{b}\right]
D_{0}(q_{0},\mathbf{q}). \label{Eq:VertexCorrectionDisW4i}
\end{eqnarray}

The contribution from Fig.~\ref{Fig:VertexCorrection}(e) is given by
\begin{eqnarray}
W_{i}^{(5)}&=&\Delta_{i}g^{2}\left(\bar{\Psi}_{a}\Gamma_{i}\Psi_{a}\right)\int\frac{dq_{0}}{2\pi}\int'\frac{d^3\mathbf{q}}{(2\pi)^{3}}
\mathrm{Tr}\left[\Gamma_{i}G_{0}(q_{0},\mathbf{q})\right.\nonumber
\\
&&\left.\times\gamma_{0}G_{0}(q_{0},\mathbf{k}+\mathbf{q})\right]D_{0}(\mathbf{k})
\left(\bar{\Psi}_{b}\gamma_{0}\Psi_{b}\right).  \label{Eq:VertexCorrectionDisW5i}
\end{eqnarray}

\subsection{Correction induced by disorder scattering from Feynman diagram \ref{Fig:VertexCorrection}(a)}

Substituting the Eq.~(\ref{Eq:FermionPropagator}) into Eq.~(\ref{Eq:VertexCorrectionDisW1i}), we arrive
\begin{eqnarray}
W_{i}^{(1)}
&=&\frac{\Lambda}{2\pi^{2}v^{2}\sqrt{\eta}}\ell\Delta_{i}\left(\bar{\Psi}_{a}\Gamma_{i}\Psi_{a}\right)\sum_{j}\Delta_{j}\nonumber
\\
&&\times\bigg\{\bar{\Psi}_{b}\Gamma_{j}\Big[-\frac{1}{2}\left(\gamma_{x}\Gamma_{i}\gamma_{x}+\gamma_{y}\Gamma_{i}\gamma_{y}\right)\mathcal{G}_{1}^{\bot}
\nonumber
\\
&&-\gamma_{z}\Gamma_{i}\gamma_{z}\mathcal{G}_{1}^{z}-\gamma_{5}\Gamma_{i}\gamma_{5}\mathcal{G}_{1}^{D}
+\Gamma_{i}\mathcal{G}_{1}^{m}\Big]\Gamma_{j}\Psi_{b}\bigg\},
\end{eqnarray}
with
\begin{eqnarray}
\mathcal{G}_{1}^{\bot}&=&\frac{1}{4\pi}\int_{0}^{\pi}d\varphi\sin(\varphi)\int_{0}^{2\pi}d\theta \frac{\sin^{2}(\varphi)}
{\Xi^{4}},
\\
\mathcal{G}_{1}^{z}&=&\frac{1}{4\pi}\int_{0}^{\pi}d\varphi\sin(\varphi)\int_{0}^{2\pi}d\theta \frac{\left(\frac{v_{z}}{v\sqrt{\eta}}\right)^{2}\cos^{2}(\varphi)}
{\Xi^{4}},
\\
\mathcal{G}_{1}^{D}&=&\frac{1}{4\pi}\int_{0}^{\pi}d\varphi\sin(\varphi)\int_{0}^{2\pi}d\theta\nonumber
\\
&&\times\frac{\frac{D^{2}\Lambda^{2}}{v^{2}}\sin^{4}(\varphi)\left(\cos^{2}(\theta)-\sin^{2}(\theta)\right)^{2}}
{\Xi^{4}},
\\
\mathcal{G}_{1}^{m}&=&\frac{1}{4\pi}\int_{0}^{\pi}d\varphi\sin(\varphi)\int_{0}^{2\pi}d\theta\nonumber
\\
&&\times\frac{\left(\frac{m}{v\Lambda}-\frac{B_{\bot}\Lambda}{v}\sin^{2}(\varphi)
-\frac{B_{z}\Lambda}{v\eta}\cos^{2}(\varphi)\right)^{2}}
{\Xi^{4}},
\end{eqnarray}
where $\Xi$ is expressed by Eq.~(\ref{Eq:XiExpression}).

$W_{i}^{(1)}$ can be further written as
\begin{eqnarray}
W_{i}^{(1)}=\frac{\delta\Delta_{i}^{(1)}}{2}\left(\bar{\Psi}_{a}i\gamma_{z}\Psi_{a}\right)\left(\bar{\Psi}_{b}i\gamma_{z}\Psi_{b}\right),
\end{eqnarray}
where
\begin{widetext}
\begin{eqnarray}
\delta\Delta_{C}^{(1)}
&=&2\Delta_{C}\left(\Delta_{C}+\Delta_{M}+\Delta_{AC}+2\Delta_{SO\bot}
+\Delta_{SOz}+\Delta_{PM}+2\Delta_{MN\bot}+\Delta_{MNz}+2\Delta_{AMN\bot}
+\Delta_{AMNz}\right.\nonumber
\\
&&\left.+2\Delta_{CR\bot}+\Delta_{CRz}\right)\frac{\Lambda}{2\pi^{2}v^{2}\sqrt{\eta}}\left(\mathcal{G}_{1}^{\bot}
+\mathcal{G}_{1}^{z}+\mathcal{G}_{1}^{D}+\mathcal{G}_{1}^{m}\right)\ell, \label{Eq:deltaDeltaC1}
\\
\delta\Delta_{M}^{(1)}
&=&2\Delta_{M}\left(\Delta_{C}
+\Delta_{M}-\Delta_{AC}-2\Delta_{SO\bot}-\Delta_{SOz}-\Delta_{PM}+2\Delta_{MN\bot}+\Delta_{MNz}+2\Delta_{AMN\bot}+\Delta_{AMNz}\right.\nonumber
\\
&&\left.-2\Delta_{CR\bot}-2C_{CRz}\right)
\frac{\Lambda}{2\pi^{2}v^{2}\sqrt{\eta}}\left(-\mathcal{G}_{1}^{\bot}
-\mathcal{G}_{1}^{z}-\mathcal{G}_{1}^{D}+\mathcal{G}_{1}^{m}\right)\ell, \label{Eq:deltaDeltaM1}
\\
\delta\Delta_{AC}^{(1)}
&=&2\Delta_{AC}\left(-\Delta_{C}+\Delta_{M}-\Delta_{AC}+2\Delta_{SO\bot}+\Delta_{SOz}+\Delta_{PM}+2\Delta_{MN\bot}+\Delta_{MNz}
-2\Delta_{AMN\bot}-\Delta_{AMNz}\right.\nonumber
\\
&&\left.-2\Delta_{CR\bot}-\Delta_{CRz}\right)
\frac{\Lambda}{2\pi^{2}v^{2}\sqrt{\eta}}\left(-\mathcal{G}_{1}^{\bot}
-\mathcal{G}_{1}^{z}+\mathcal{G}_{1}^{D}+\mathcal{G}_{1}^{m}\right)\ell, \label{Eq:deltaDeltaAC1}
\\
\delta\Delta_{SO\bot}^{(1)}
&=&2\Delta_{SO\bot}\left(-\Delta_{C}
+\Delta_{M}+\Delta_{AC}
+\Delta_{SOz}-\Delta_{PM}-\Delta_{MNz}+\Delta_{AMNz}-\Delta_{CRz}\right)
\frac{\Lambda}{2\pi^{2}v^{2}\sqrt{\eta}}\nonumber
\\
&&\times\left(
-\mathcal{G}_{1}^{z}
-\mathcal{G}_{1}^{D}+\mathcal{G}_{1}^{m}\right)\ell,  \label{Eq:deltaDeltaSOBot1}
\\
\delta\Delta_{SOz}^{(1)}
&=&2\Delta_{SOz}\left(-\Delta_{C}
+\Delta_{M}+\Delta_{AC}+2\Delta_{SO\bot}-\Delta_{SOz}-\Delta_{PM}
-2\Delta_{MN\bot}+\Delta_{MNz}+2\Delta_{AMN\bot}
-\Delta_{AMNz}\right.\nonumber
\\
&&\left.-2\Delta_{CR\bot}+\Delta_{CRz}
\right)\frac{\Lambda}{2\pi^{2}v^{2}\sqrt{\eta}}\left(-\mathcal{G}_{1}^{\bot}
+\mathcal{G}_{1}^{z}-\mathcal{G}_{1}^{D}
+\mathcal{G}_{1}^{m}\right)\ell,                      \label{Eq:deltaDeltaSoz1}
\\
\delta\Delta_{PM}^{(1)}
&=&2\Delta_{PM}\left(-\Delta_{C}
+\Delta_{M}+\Delta_{AC}-2\Delta_{SO\bot}-\Delta_{SOz}-\Delta_{PM}+2\Delta_{MN\bot}+\Delta_{MNz}
-2\Delta_{AMN\bot}-\Delta_{AMNz}\right.\nonumber
\\
&&\left.+2\Delta_{CR\bot}+\Delta_{CRz}\right)\frac{\Lambda}{2\pi^{2}v^{2}\sqrt{\eta}}\left(\mathcal{G}_{1}^{\bot}
+\mathcal{G}_{1}^{z}-\mathcal{G}_{1}^{D}+\mathcal{G}_{1}^{m}\right)\ell,  \label{Eq:deltaDeltaPM1}
\\
\delta\Delta_{MN\bot}^{(1)}
&=&2\Delta_{MN\bot}
\left(\Delta_{C}+\Delta_{M}-\Delta_{AC}+\Delta_{SOz}-\Delta_{PM}-\Delta_{MNz}-\Delta_{AMNz}
+\Delta_{CRz}\right)\frac{\Lambda}{2\pi^{2}v^{2}\sqrt{\eta}}\nonumber
\\
&&\times\left(\mathcal{G}_{1}^{z}-\mathcal{G}_{1}^{D}+\mathcal{G}_{1}^{m}\right)\ell,  \label{Eq:deltaDeltaMNBot1}
\\
\delta\Delta_{MNz}^{(1)}
&=&2\Delta_{MNz}
\left(\Delta_{C}+\Delta_{M}-\Delta_{AC}+2\Delta_{SO\bot}-\Delta_{SOz}-\Delta_{PM}-2\Delta_{MN\bot}+\Delta_{MNz}
-2\Delta_{AMN\bot}+\Delta_{AMNz}\right.\nonumber
\\
&&\left.+2\Delta_{CR\bot}-\Delta_{CRz}\right)\frac{\Lambda}{2\pi^{2}v^{2}\sqrt{\eta}}\left(\mathcal{G}_{1}^{\bot}-\mathcal{G}_{1}^{z}-\mathcal{G}_{1}^{D}+\mathcal{G}_{1}^{m}\right)\ell, \label{Eq:deltaDeltaMNz1}
\\
\delta\Delta_{AMN\bot}^{(1)}
&=&
2\Delta_{AMN\bot}\left(\Delta_{C}+\Delta_{M}+\Delta_{AC}-\Delta_{SOz}+\Delta_{PM}-\Delta_{MNz}-\Delta_{AMNz}
-\Delta_{CRz}\right)\frac{\Lambda}{2\pi^{2}v^{2}\sqrt{\eta}}\nonumber
\\
&&\times\left(
-\mathcal{G}_{1}^{z}+\mathcal{G}_{1}^{D}+\mathcal{G}_{1}^{m}\right)\ell, \label{Eq:deltaDeltaAMNBot1}
\\
\delta\Delta_{AMNz}^{(1)}
&=&2\Delta_{AMNz}\left(\Delta_{C}+\Delta_{M}+\Delta_{AC}-2\Delta_{SO\bot}+\Delta_{SOz}+\Delta_{PM}
-2\Delta_{MN\bot}+\Delta_{MNz}-2\Delta_{AMN\bot}+\Delta_{AMNz}\right.\nonumber
\\
&&\left.-2\Delta_{CR\bot}+\Delta_{CRz}\right)\frac{\Lambda}{2\pi^{2}v^{2}\sqrt{\eta}}
\left(-\mathcal{G}_{1}^{\bot}
+\mathcal{G}_{1}^{z}+\mathcal{G}_{1}^{D}+\mathcal{G}_{1}^{m}\right)\ell,  \label{Eq:deltaDeltaAMNz1}
\\
\delta\Delta_{CR\bot}^{(1)}
&=&2\Delta_{CR\bot}\left(-\Delta_{C}
+\Delta_{M}-\Delta_{AC}-\Delta_{SOz}+\Delta_{PM}-\Delta_{MNz}
+\Delta_{AMNz}
+\Delta_{CRz}\right)\frac{\Lambda}{2\pi^{2}v^{2}\sqrt{\eta}}\nonumber
\\
&&\times\left(\mathcal{G}_{1}^{z}+\mathcal{G}_{1}^{D}+\mathcal{G}_{1}^{m}\right)\ell,   \label{Eq:deltaDeltaCRBot1}
\\
\delta\Delta_{CRz}^{(1)}
&=&2\Delta_{CRz}
\left(-\Delta_{C}+\Delta_{M}-\Delta_{AC}-2\Delta_{SO\bot}+\Delta_{SOz}+\Delta_{PM}
-2\Delta_{MN\bot}+\Delta_{MNz}+2\Delta_{AMN\bot}-\Delta_{AMNz}\right.\nonumber
\\
&&\left.+2\Delta_{CR\bot}
-\Delta_{CRz}\right)\frac{\Lambda}{2\pi^{2}v^{2}\sqrt{\eta}}\left(\mathcal{G}_{1}^{\bot}
-\mathcal{G}_{1}^{z}+\mathcal{G}_{1}^{D}+\mathcal{G}_{1}^{m}\right)\ell.                 \label{Eq:deltaDeltaCRz1}
\end{eqnarray}

It should be noticed that
\begin{eqnarray}
\mathcal{G}_{1}^{\bot}+\mathcal{G}_{1}^{z}+\mathcal{G}_{1}^{D}+\mathcal{G}_{1}^{m}=
\mathcal{G}_{0}^{\bot}+\mathcal{G}_{0}^{z}.
\end{eqnarray}

\subsection{Correction induced by disorder scattering from Feynman diagrams \ref{Fig:VertexCorrection}(b) and \ref{Fig:VertexCorrection}(c)}

Substituting Eq.~(\ref{Eq:FermionPropagator}) into Eq.~(\ref{Eq:VertexCorrectionW23ij}), we obtain
\begin{eqnarray}
W_{i,j}^{(2)+(3)}
&=&-\frac{1}{2}\Delta_{i}\Delta_{j}\left(\bar{\Psi}_{a}\Gamma_{i}
\gamma_{x}\Gamma_{j}\Psi_{a}\right)\left[\bar{\Psi}_{b}\left(\Gamma_{j}\gamma_{x}
\Gamma_{i}-\Gamma_{i}\gamma_{x}\Gamma_{j}\right)\Psi_{b}\right]\frac{\Lambda}{2\pi^{2}v^{2}\sqrt{\eta}}\ell\mathcal{G}_{1}^{\bot}\nonumber
\\
&&-\frac{1}{2}\Delta_{i}\Delta_{j}\left(\bar{\Psi}_{a}\Gamma_{i}
\gamma_{y}\Gamma_{j}\Psi_{a}\right)\left[\bar{\Psi}_{b}\left(\Gamma_{j}\gamma_{y}
\Gamma_{i}-\Gamma_{i}\gamma_{y}\Gamma_{j}\right)\Psi_{b}\right]\frac{\Lambda}{2\pi^{2}v^{2}\sqrt{\eta}}\ell\mathcal{G}_{1}^{\bot}\nonumber
\\
&&-\Delta_{i}\Delta_{j}\left(\bar{\Psi}_{a}\Gamma_{i}\gamma_{z}
\Gamma_{j}\Psi_{a}\right)\left[\bar{\Psi}_{b}\left(\Gamma_{j}\gamma_{z}
\Gamma_{i}-\Gamma_{i}\gamma_{z}\Gamma_{j}\right)\Psi_{b}\right]\frac{\Lambda}{2\pi^{2}v^{2}\sqrt{\eta}}\ell\mathcal{G}_{1}^{z}\nonumber
\\
&&-\Delta_{i}\Delta_{j}\left(\bar{\Psi}_{a}\Gamma_{i}\gamma_{5}\Gamma_{j}\Psi_{a}\right)\left[\bar{\Psi}_{b}\left(\Gamma_{j}\gamma_{5}
\Gamma_{i}+\Gamma_{i}\gamma_{5}\Gamma_{j}\right)\Psi_{b}\right]
\frac{\Lambda}{2\pi^{2}v^{2}\sqrt{\eta}}\ell\mathcal{G}_{1}^{D}\nonumber
\\
&&+\Delta_{i}\Delta_{j}\left(\bar{\Psi}_{a}\Gamma_{i}
\Gamma_{j}\Psi_{a}\right)\left[\bar{\Psi}_{b}\left(\Gamma_{j}
\Gamma_{i}+\Gamma_{i}\Gamma_{j}\right)\Psi_{b}\right]
\frac{\Lambda}{2\pi^{2}v^{2}\sqrt{\eta}}\ell\mathcal{G}_{1}^{m}.
\end{eqnarray}
Concretely, we find that
\begin{eqnarray}
W_{C,C}^{(2)+(3)}
&=&2\Delta_{C}^{2}
\frac{\Lambda}{2\pi^{2}v^{2}\sqrt{\eta}}\mathcal{G}_{1}^{D}\ell\left(\bar{\Psi}_{a}i\gamma_{5}\Psi_{a}\right)\left(\bar{\Psi}_{b}i\gamma_{5}
\Psi_{b}\right)+2\Delta_{C}^{2}
\frac{\Lambda}{2\pi^{2}v^{2}\sqrt{\eta}}\mathcal{G}_{1}^{m}\ell\left(\bar{\Psi}_{a}\Psi_{a}\right)\left(\bar{\Psi}_{b}\Psi_{b}\right),
\\
W_{M,M}^{(2)+(3)}
&=&2\Delta_{M}^{2}
\frac{\Lambda}{2\pi^{2}v^{2}\sqrt{\eta}}\mathcal{G}_{1}^{D}\ell\left(\bar{\Psi}_{a}i\gamma_{5}\Psi_{a}\right)\left(\bar{\Psi}_{b}i\gamma_{5}\Psi_{b}\right)+2\Delta_{M}^{2}
\frac{\Lambda}{2\pi^{2}v^{2}\sqrt{\eta}}\mathcal{G}_{1}^{m}\ell\left(\bar{\Psi}_{a}\Psi_{a}\right)\left(\bar{\Psi}_{b}\Psi_{b}\right),
\\
W_{AC,AC}^{(2)+(3)}
&=&2\Delta_{AC}^{2}
\frac{\Lambda}{2\pi^{2}v^{2}\sqrt{\eta}}\mathcal{G}_{1}^{D}\ell\left(\bar{\Psi}_{a}i\gamma_{5}\Psi_{a}\right)\left(\bar{\Psi}_{b}i\gamma_{5}\Psi_{b}\right)
+2\Delta_{AC}^{2}
\frac{\Lambda}{2\pi^{2}v^{2}\sqrt{\eta}}\mathcal{G}_{1}^{m}\ell\left(\bar{\Psi}_{a}\Psi_{a}\right)\left(\bar{\Psi}_{b}\Psi_{b}\right),
\\
W_{SO\bot,SO\bot}^{(2)+(3)}
&=&4\Delta_{SO\bot}^{2}\frac{\Lambda}{2\pi^{2}v^{2}\sqrt{\eta}}\ell\mathcal{G}_{1}^{z}\left(\bar{\Psi}_{a}\gamma_{0}\gamma_{5}
\Psi_{a}\right)\left(\bar{\Psi}_{b}\gamma_{0}\gamma_{5}\Psi_{b}\right)
+4\Delta_{SO\bot}^{2}
\frac{\Lambda}{2\pi^{2}v^{2}\sqrt{\eta}}\ell\mathcal{G}_{1}^{D}\left(\bar{\Psi}_{a}i\gamma_{5}\Psi_{a}\right)
\left(\bar{\Psi}_{b}i\gamma_{5}\Psi_{b}\right)\nonumber
\\
&&+4\Delta_{SO\bot}^{2}
\frac{\Lambda}{2\pi^{2}v^{2}\sqrt{\eta}}\ell\mathcal{G}_{1}^{m}\left(\bar{\Psi}_{a}\Psi_{a}\right)\left(\bar{\Psi}_{b}\Psi_{b}\right),
\\
W_{SOz,SOz}^{(2)+(3)}
&=&2\Delta_{SOz}^{2}
\frac{\Lambda}{2\pi^{2}v^{2}\sqrt{\eta}}\mathcal{G}_{1}^{D}\ell\left(\bar{\Psi}_{a}i\gamma_{5}\Psi_{a}\right)\left(\bar{\Psi}_{b}i\gamma_{5}
\Psi_{b}\right)
+2\Delta_{SOz}^{2}
\frac{\Lambda}{2\pi^{2}v^{2}\sqrt{\eta}}\mathcal{G}_{1}^{m}\ell\left(\bar{\Psi}_{a}\Psi_{a}\right)\left(\bar{\Psi}_{b}\Psi_{b}\right),
\\
W_{PM,PM}^{(2)+(3)}
&=&2\Delta_{PM}^{2}
\frac{\Lambda}{2\pi^{2}v^{2}\sqrt{\eta}}\mathcal{G}_{1}^{D}\ell
\left(\bar{\Psi}_{a}i\gamma_{5}\Psi_{a}\right)\left(\bar{\Psi}_{b}i\gamma_{5}\Psi_{b}\right)+2\Delta_{PM}^{2}
\frac{\Lambda}{2\pi^{2}v^{2}\sqrt{\eta}}\mathcal{G}_{1}^{m}\ell\left(\bar{\Psi}_{a}\Psi_{a}\right)\left(\bar{\Psi}_{b}\Psi_{b}\right),
\\
W_{MN\bot,MN\bot}^{(2)+(3)}
&=&4\Delta_{MN\bot}^{2}
\frac{\Lambda}{2\pi^{2}v^{2}\sqrt{\eta}}\ell\mathcal{G}_{1}^{z}\left(\bar{\Psi}_{a}\gamma_{0}\gamma_{5}\Psi_{a}\right)\left(\bar{\Psi}_{b}\gamma_{0}\gamma_{5}\Psi_{b}\right)
+4\Delta_{MN\bot}^{2}
\frac{\Lambda}{2\pi^{2}v^{2}\sqrt{\eta}}\ell\mathcal{G}_{1}^{D}\left(\bar{\Psi}_{a}
i\gamma_{5}\Psi_{a}\right)\left(\bar{\Psi}_{b}i\gamma_{5}\Psi_{b}\right)\nonumber
\\
&&+4\Delta_{MN\bot}^{2}
\frac{\Lambda}{2\pi^{2}v^{2}\sqrt{\eta}}\ell\mathcal{G}_{1}^{m}\left(\bar{\Psi}_{a}\Psi_{a}\right)\left(\bar{\Psi}_{b}\Psi_{b}\right),
\\
W_{MNz,MNz}^{(2)+(3)}
&=&2\Delta_{MNz}^{2}
\frac{\Lambda}{2\pi^{2}v^{2}\sqrt{\eta}}\ell\mathcal{G}_{1}^{D}\left(\bar{\Psi}_{a}i\gamma_{5}\Psi_{a}\right)\left(\bar{\Psi}_{b}i\gamma_{5}\Psi_{b}\right)\nonumber
\\
&&+2\Delta_{MNz}^{2}
\frac{\Lambda}{2\pi^{2}v^{2}\sqrt{\eta}}\ell\mathcal{G}_{1}^{m}\left(\bar{\Psi}_{a}
\Psi_{a}\right)\left(\bar{\Psi}_{b}\Psi_{b}\right),
\\
W_{AMN\bot,AMN\bot}^{(2)+(3)}
&=&4\Delta_{AMN\bot}^{2}\frac{\Lambda}{2\pi^{2}v^{2}\sqrt{\eta}}\ell\mathcal{G}_{1}^{z}\left(\bar{\Psi}_{a}\gamma_{0}\gamma_{5}\Psi_{a}\right)\left(\bar{\Psi}_{b}\gamma_{0}\gamma_{5}\Psi_{b}\right)
+4\Delta_{AMN\bot}^{2}
\frac{\Lambda}{2\pi^{2}v^{2}\sqrt{\eta}}\ell\mathcal{G}_{1}^{D}\left(\bar{\Psi}_{a}
i\gamma_{5}\Psi_{a}\right)\left(\bar{\Psi}_{b}i\gamma_{5}\Psi_{b}\right)\nonumber
\\
&&+4\Delta_{AMN\bot}^{2}
\frac{\Lambda}{2\pi^{2}v^{2}\sqrt{\eta}}\ell\mathcal{G}_{1}^{m}\left(\bar{\Psi}_{a}\Psi_{a}\right)\left(\bar{\Psi}_{b}\Psi_{b}\right),
\\
W_{AMNz,AMNz}^{(2)+(3)}
&=&2\Delta_{AMNz}^{2}
\frac{\Lambda}{2\pi^{2}v^{2}\sqrt{\eta}}\ell\mathcal{G}_{1}^{D}\left(\bar{\Psi}_{a}i\gamma_{5}\Psi_{a}\right)\left(\bar{\Psi}_{b}i\gamma_{5}\Psi_{b}\right)
+2\Delta_{AMNz}^{2}
\frac{\Lambda}{2\pi^{2}v^{2}\sqrt{\eta}}\ell\mathcal{G}_{1}^{m}\left(\bar{\Psi}_{a}\Psi_{a}\right)\left(\bar{\Psi}_{b}\Psi_{b}\right),
\\
W_{CR\bot,CR\bot}^{(2)+(3)}
&=&4\Delta_{CR\bot}^{2}
\frac{\Lambda}{2\pi^{2}v^{2}\sqrt{\eta}}\ell\mathcal{G}_{1}^{z}\left(\bar{\Psi}_{a}\gamma_{0}\gamma_{5}\Psi_{a}\right)\left(\bar{\Psi}_{b}\gamma_{0}\gamma_{5}\Psi_{b}\right)
+4\Delta_{CR\bot}^{2}
\frac{\Lambda}{2\pi^{2}v^{2}\sqrt{\eta}}\ell\mathcal{G}_{1}^{D}\left(\bar{\Psi}_{a}
i\gamma_{5}\Psi_{a}\right)\left(\bar{\Psi}_{b}i\gamma_{5}\Psi_{b}\right)\nonumber
\\
&&+4\Delta_{CR\bot}^{2}
\frac{\Lambda}{2\pi^{2}v^{2}\sqrt{\eta}}\ell\mathcal{G}_{1}^{m}\left(\bar{\Psi}_{a}\Psi_{a}\right)\left(\bar{\Psi}_{b}
\Psi_{b}\right),
\\
W_{CRz,CRz}^{(2)+(3)}
&=&2\Delta_{CRz}^{2}
\frac{\Lambda}{2\pi^{2}v^{2}\sqrt{\eta}}\ell\mathcal{G}_{1}^{D}\left(\bar{\Psi}_{a}i\gamma_{5}\Psi_{a}\right)\left(\bar{\Psi}_{b}i\gamma_{5}
\Psi_{b}\right)
+2\Delta_{CRz}^{2}
\frac{\Lambda}{2\pi^{2}v^{2}\sqrt{\eta}}\ell\mathcal{G}_{1}^{m}\left(\bar{\Psi}_{a}\Psi_{a}\right)\left(\bar{\Psi}_{b}\Psi_{b}\right),
\\
W_{C,M}^{(2)+(3)}
&=&\Delta_{C}\Delta_{M}\frac{\Lambda}{2\pi^{2}v^{2}\sqrt{\eta}}\mathcal{G}_{1}^{\bot}\ell\sum_{j=x,y}\left(\bar{\Psi}_{a}\gamma_{0}
\gamma_{j}\Psi_{a}\right)\left(\bar{\Psi}_{b}\gamma_{0}\gamma_{j}
\Psi_{b}\right)\nonumber
\\
&&+2\Delta_{C}\Delta_{M}\frac{\Lambda}{2\pi^{2}v^{2}\sqrt{\eta}}\mathcal{G}_{1}^{z}\ell\left(\bar{\Psi}_{a}\gamma_{0}\gamma_{z}
\Psi_{a}\right)\left(\bar{\Psi}_{b}\gamma_{0}\gamma_{z}\Psi_{b}\right)\nonumber
\\
&&+2\Delta_{C}\Delta_{M}
\frac{\Lambda}{2\pi^{2}v^{2}\sqrt{\eta}}\mathcal{G}_{1}^{m}\ell\left(\bar{\Psi}_{a}\gamma_{0}
\Psi_{a}\right)\left(\bar{\Psi}_{b}\gamma_{0}\Psi_{b}\right),
\\
W_{C,AC}^{(2)+(3)}
&=&0,
\\
W_{C,SO\bot}^{(2)+(3)}
&=&2\Delta_{C}\Delta_{SO\bot}
\frac{\Lambda}{2\pi^{2}v^{2}\sqrt{\eta}}\mathcal{G}_{1}^{\bot}\ell\left(\bar{\Psi}_{a}\Psi_{a}\right)\left(\bar{\Psi}_{b}\Psi_{b}\right)\nonumber
\\
&&+2\Delta_{C}\Delta_{SO\bot}
\frac{\Lambda}{2\pi^{2}v^{2}\sqrt{\eta}}\mathcal{G}_{1}^{D}\ell\sum_{j=x,y}\left(\bar{\Psi}_{a}i\gamma_{5}\gamma_{j}\Psi_{a}\right)\left(\bar{\Psi}_{b}
i\gamma_{5}\gamma_{j}\Psi_{b}\right),
\\
W_{C,SOz}^{(2)+(3)}
&=&2\Delta_{C}\Delta_{SOz}\frac{\Lambda}{2\pi^{2}v^{2}\sqrt{\eta}}\ell\mathcal{G}_{1}^{z}\left(\bar{\Psi}_{a}\Psi_{a}\right)\left(\bar{\Psi}_{b}\Psi_{b}\right)\nonumber
\\
&&+2\Delta_{C}\Delta_{SOz}
\frac{\Lambda}{2\pi^{2}v^{2}\sqrt{\eta}}\ell\mathcal{G}_{1}^{D}\left(\bar{\Psi}_{a}i\gamma_{5}\gamma_{z}\Psi_{a}\right)\left(\bar{\Psi}_{b}i\gamma_{5}\gamma_{z}
\Psi_{b}\right),
\\
W_{C,PM}^{(2)+(3)}
&=&\Delta_{C}\Delta_{PM}\frac{\Lambda}{2\pi^{2}v^{2}\sqrt{\eta}}\mathcal{G}_{1}^{\bot}\ell\left[\left(\bar{\Psi}_{a}i\gamma_{y}\gamma_{z}\Psi_{a}\right)\left(\bar{\Psi}_{b}i\gamma_{y}\gamma_{z}\Psi_{b}\right)
+\left(\bar{\Psi}_{a}i\gamma_{z}\gamma_{x}\Psi_{a}\right)\left(\bar{\Psi}_{b}i\gamma_{z}\gamma_{x}\Psi_{b}\right)\right]\nonumber
\\
&&+2\Delta_{C}\Delta_{PM}\frac{\Lambda}{2\pi^{2}v^{2}\sqrt{\eta}}\mathcal{G}_{1}^{z}\ell\left(\bar{\Psi}_{a}
i\gamma_{x}\gamma_{y}\Psi_{a}\right)\left(\bar{\Psi}_{b}i\gamma_{x}\gamma_{y}\Psi_{b}\right)\nonumber
\\
&&+2\Delta_{C}\Delta_{PM}
\frac{\Lambda}{2\pi^{2}v^{2}\sqrt{\eta}}\mathcal{G}_{1}^{D}\ell\left(\bar{\Psi}_{a}\gamma_{0}\Psi_{a}\right)\left(\bar{\Psi}_{b}
\gamma_{0}\Psi_{b}\right),
\\
W_{C,MN\bot}^{(2)+(3)}
&=&2\Delta_{C}\Delta_{MN\bot}\frac{\Lambda}{2\pi^{2}v^{2}\sqrt{\eta}}\ell\mathcal{G}_{1}^{\bot}\left(\bar{\Psi}_{a}i\gamma_{5}\Psi_{a}\right)\left(\bar{\Psi}_{b}i\gamma_{5}
\Psi_{b}\right)\nonumber
\\
&&+2\Delta_{C}\Delta_{MN\bot}
\frac{\Lambda}{2\pi^{2}v^{2}\sqrt{\eta}}\ell\mathcal{G}_{1}^{m}\sum_{j=x,y}\left(\bar{\Psi}_{a}i\gamma_{5}\gamma_{j}\Psi_{a}\right)\left(\bar{\Psi}_{b}i\gamma_{5}\gamma_{j}\Psi_{b}\right),
\\
W_{C,MNz}^{(2)+(3)}
&=&2\Delta_{C}\Delta_{MNz}\frac{\Lambda}{2\pi^{2}v^{2}\sqrt{\eta}}\ell\mathcal{G}_{1}^{z}\left(\bar{\Psi}_{a}i\gamma_{5}\Psi_{a}\right)\left(\bar{\Psi}_{b}i\gamma_{5}\Psi_{b}\right)\nonumber
\\
&&+2\Delta_{C}\Delta_{MNz}
\frac{\Lambda}{2\pi^{2}v^{2}\sqrt{\eta}}\ell\mathcal{G}_{1}^{m}\left(\bar{\Psi}_{a}i\gamma_{5}\gamma_{z}\Psi_{a}\right)\left(\bar{\Psi}_{b}i\gamma_{5}\gamma_{z}\Psi_{b}\right),
\\
W_{C,AMN\bot}^{(2)+(3)}
&=&2\Delta_{C}\Delta_{AMN\bot}\frac{\Lambda}{2\pi^{2}v^{2}\sqrt{\eta}}\ell\mathcal{G}_{1}^{\bot}\left(\bar{\Psi}_{a}i
\gamma_{z}\Psi_{a}\right)\left(\bar{\Psi}_{b}i\gamma_{z}
\Psi_{b}\right)\nonumber
\\
&&+2\Delta_{C}\Delta_{AMN\bot}\frac{\Lambda}{2\pi^{2}v^{2}\sqrt{\eta}}\ell\mathcal{G}_{1}^{z}\sum_{j=x,y}\left(\bar{\Psi}_{a}i\gamma_{j}\Psi_{a}\right)\left(\bar{\Psi}_{b}i\gamma_{j}\Psi_{b}\right)\nonumber
\\
&&+2\Delta_{C}\Delta_{AMN\bot}
\frac{\Lambda}{2\pi^{2}v^{2}\sqrt{\eta}}\ell\mathcal{G}_{1}^{D}\sum_{j=x,y}\left(\bar{\Psi}_{a}\gamma_{0}\gamma_{j}\Psi_{a}\right)\left(\bar{\Psi}_{b}\gamma_{0}\gamma_{j}
\Psi_{b}\right)\nonumber
\\
&&+2\Delta_{C}\Delta_{AMN\bot}
\frac{\Lambda}{2\pi^{2}v^{2}\sqrt{\eta}}\ell\mathcal{G}_{1}^{m}\left[\left(\bar{\Psi}_{a}i\gamma_{y}\gamma_{z}\Psi_{a}\right)\left(\bar{\Psi}_{b}i\gamma_{y}\gamma_{z}\Psi_{b}\right)
+\left(\bar{\Psi}_{a}i
\gamma_{z}\gamma_{x}\Psi_{a}\right)\left(\bar{\Psi}_{b}i\gamma_{z}\gamma_{x}
\Psi_{b}\right)\right],
\\
W_{C,AMNz}^{(2)+(3)}
&=&\Delta_{C}\Delta_{AMNz}
\frac{\Lambda}{2\pi^{2}v^{2}\sqrt{\eta}}\ell\mathcal{G}_{1}^{\bot}\sum_{j=x,y}\left(\bar{\Psi}_{a}i\gamma_{j}\Psi_{a}\right)\left(\bar{\Psi}_{b}i\gamma_{j}\Psi_{b}\right)\nonumber
\\
&&+2\Delta_{C}\Delta_{AMNz}
\frac{\Lambda}{2\pi^{2}v^{2}\sqrt{\eta}}\ell\mathcal{G}_{1}^{D}\left(\bar{\Psi}_{a}\gamma_{0}\gamma_{z}\Psi_{a}\right)\left(\bar{\Psi}_{b}\gamma_{0}\gamma_{z}
\Psi_{b}\right)\nonumber
\\
&&+2\Delta_{C}\Delta_{AMNz}
\frac{\Lambda}{2\pi^{2}v^{2}\sqrt{\eta}}\ell\mathcal{G}_{1}^{m}\left(\bar{\Psi}_{a}i
\gamma_{x}\gamma_{y}\Psi_{a}\right)\left(\bar{\Psi}_{b}i
\gamma_{x}\gamma_{y}\Psi_{b}\right),
\\
W_{C,CR\bot}^{(2)+(3)}
&=&2\Delta_{C}\Delta_{CR\bot}\frac{\Lambda}{2\pi^{2}v^{2}\sqrt{\eta}}\ell\mathcal{G}_{1}^{\bot}\left(\bar{\Psi}_{a}i\gamma_{5}\gamma_{z}\Psi_{a}\right)\left(\bar{\Psi}_{b}
i\gamma_{5}\gamma_{z}\Psi_{b}\right)\nonumber
\\
&&+2\Delta_{C}\Delta_{CR\bot}\frac{\Lambda}{2\pi^{2}v^{2}\sqrt{\eta}}\ell\mathcal{G}_{1}^{z}\sum_{j=x,y}\left(\bar{\Psi}_{a}i\gamma_{5}\gamma_{j}\Psi_{a}\right)\left(\bar{\Psi}_{b}i\gamma_{5}\gamma_{j}\Psi_{b}\right),
\\
W_{C,CRz}^{(2)+(3)}
&=&\Delta_{C}\Delta_{CRz}\frac{\Lambda}{2\pi^{2}v^{2}\sqrt{\eta}}\ell\mathcal{G}_{1}^{\bot}\sum_{j=x,y}
\left(\bar{\Psi}_{a}i\gamma_{5}\gamma_{j}\Psi_{a}\right)\left(\bar{\Psi}_{b}i\gamma_{5}\gamma_{j}\Psi_{b}\right),
\\
W_{M,AC}^{(2)+(3)}
&=&2\Delta_{M}\Delta_{AC}
\frac{\Lambda}{2\pi^{2}v^{2}\sqrt{\eta}}\ell\mathcal{G}_{1}^{m}\left(\bar{\Psi}_{a}
\gamma_{0}\gamma_{5}\Psi_{a}\right)\left(\bar{\Psi}_{b}\gamma_{0}\gamma_{5}
\Psi_{b}\right),
\\
W_{M,SO\bot}^{(2)+(3)}
&=&2\Delta_{M}\Delta_{SO\bot}\frac{\Lambda}{2\pi^{2}v^{2}\sqrt{\eta}}\ell\mathcal{G}_{1}^{\bot}\left(\bar{\Psi}_{a}
\gamma_{0}\Psi_{a}\right)\left(\bar{\Psi}_{b}\gamma_{0}
\Psi_{b}\right)\nonumber
\\
&&+2\Delta_{M}\Delta_{SO\bot}
\frac{\Lambda}{2\pi^{2}v^{2}\sqrt{\eta}}\ell\mathcal{G}_{1}^{D}\left[\left(\bar{\Psi}_{a}
i\gamma_{y}\gamma_{z}\Psi_{a}\right)\left(\bar{\Psi}_{b}i\gamma_{y}\gamma_{z}\Psi_{b}\right)
+\left(\bar{\Psi}_{a}i\gamma_{z}\gamma_{x}\Psi_{a}\right)\left(\bar{\Psi}_{b}i\gamma_{z}\gamma_{x}\Psi_{b}\right)\right]\nonumber
\\
&&+2\Delta_{M}\Delta_{SO\bot}
\frac{\Lambda}{2\pi^{2}v^{2}\sqrt{\eta}}\ell\mathcal{G}_{1}^{m}\sum_{j=x,y}\left(\bar{\Psi}_{a}
\gamma_{0}\gamma_{j}\Psi_{a}\right)\left(\bar{\Psi}_{b}\gamma_{0}\gamma_{j}
\Psi_{b}\right),
\\
W_{M,SOz}^{(2)+(3)}
&=&2\Delta_{M}\Delta_{SOz}\frac{\Lambda}{2\pi^{2}v^{2}\sqrt{\eta}}\ell\mathcal{G}_{1}^{z}
\left(\bar{\Psi}_{a}
\gamma_{0}\Psi_{a}\right)\left(\bar{\Psi}_{b}
\gamma_{0}\Psi_{b}\right)\nonumber
\\
&&+2\Delta_{M}\Delta_{SOz}
\frac{\Lambda}{2\pi^{2}v^{2}\sqrt{\eta}}\ell\mathcal{G}_{1}^{D}
\left(\bar{\Psi}_{a}i\gamma_{x}\gamma_{y}\Psi_{a}\right)
\left(\bar{\Psi}_{b}i\gamma_{x}\gamma_{y}\Psi_{b}\right)\nonumber
\\
&&+2\Delta_{M}\Delta_{SOz}
\frac{\Lambda}{2\pi^{2}v^{2}\sqrt{\eta}}\ell\mathcal{G}_{1}^{m}
\left(\bar{\Psi}_{a}
\gamma_{0}\gamma_{z}\Psi_{a}\right)\left(\bar{\Psi}_{b}
\gamma_{0}\gamma_{z}\Psi_{b}\right),
\\
W_{M,PM}^{(2)+(3)}
&=&\Delta_{M}\Delta_{PM}\frac{\Lambda}{2\pi^{2}v^{2}\sqrt{\eta}}\ell\mathcal{G}_{1}^{\bot}
\sum_{j=x,y}\left(\bar{\Psi}_{a}
i\gamma_{5}\gamma_{j}\Psi_{a}\right)\left(\bar{\Psi}_{b}
i\gamma_{5}\gamma_{j}\Psi_{b}\right)\nonumber
\\
&&+2\Delta_{M}\Delta_{PM}\frac{\Lambda}{2\pi^{2}v^{2}\sqrt{\eta}}\ell\mathcal{G}_{1}^{z}
\left(\bar{\Psi}_{a}i\gamma_{5}\gamma_{z}
\Psi_{a}\right)\left(\bar{\Psi}_{b}i\gamma_{5}\gamma_{z}\Psi_{b}\right)+2\Delta_{M}\Delta_{PM}
\frac{\Lambda}{2\pi^{2}v^{2}\sqrt{\eta}}\ell\mathcal{G}_{1}^{D}\left(\bar{\Psi}_{a}\Psi_{a}\right)\left(\bar{\Psi}_{b}\Psi_{b}\right)\nonumber
\\
&&+2\Delta_{M}\Delta_{PM}
\frac{\Lambda}{2\pi^{2}v^{2}\sqrt{\eta}}\ell\mathcal{G}_{1}^{m}
\left(\bar{\Psi}_{a}
i\gamma_{5}\Psi_{a}\right)\left(\bar{\Psi}_{b}i\gamma_{5}\Psi_{b}\right),
\\
W_{M,MN\bot}^{(2)+(3)}
&=&2\Delta_{M}\Delta_{MN\bot}\frac{\Lambda}{2\pi^{2}v^{2}\sqrt{\eta}}\ell\mathcal{G}_{1}^{\bot}\left(\bar{\Psi}_{a}i\gamma_{z}\Psi_{a}\right)\left(\bar{\Psi}_{b}i\gamma_{z}\Psi_{b}\right)\nonumber
\\
&&+2\Delta_{M}\Delta_{MN\bot}\frac{\Lambda}{2\pi^{2}v^{2}\sqrt{\eta}}\ell\mathcal{G}_{1}^{z}\sum_{j=x,y}\left(\bar{\Psi}_{a}
i\gamma_{j}\Psi_{a}\right)\left(\bar{\Psi}_{b}i\gamma_{j}\Psi_{b}\right)\nonumber
\\
&&+2\Delta_{M}\Delta_{MN\bot}
\frac{\Lambda}{2\pi^{2}v^{2}\sqrt{\eta}}\ell\mathcal{G}_{1}^{D}\sum_{j=x,y}\left(\bar{\Psi}_{a}
\gamma_{0}\gamma_{j}\Psi_{a}\right)\left(\bar{\Psi}_{b}\gamma_{0}\gamma_{j}\Psi_{b}\right)\nonumber
\\
&&+2\Delta_{M}\Delta_{MN\bot}
\frac{\Lambda}{2\pi^{2}v^{2}\sqrt{\eta}}\ell\mathcal{G}_{1}^{m}\left[\left(\bar{\Psi}_{a}
i\gamma_{y}\gamma_{z}\Psi_{a}\right)\left(\bar{\Psi}_{b}i\gamma_{y}\gamma_{z}\Psi_{b}\right)+\left(\bar{\Psi}_{a}
i\gamma_{z}\gamma_{x}\Psi_{a}\right)\left(\bar{\Psi}_{b}i\gamma_{z}\gamma_{x}\Psi_{b}\right)\right],
\\
W_{M,MNz}^{(2)+(3)}
&=&\Delta_{M}\Delta_{MNz}\frac{\Lambda}{2\pi^{2}v^{2}\sqrt{\eta}}\ell\mathcal{G}_{1}^{\bot}\sum_{j=x,y}\left(\bar{\Psi}_{a}
i\gamma_{j}\Psi_{a}\right)\left(\bar{\Psi}_{b}i\gamma_{j}\Psi_{b}\right)\nonumber
\\
&&+2\Delta_{M}\Delta_{MNz}
\frac{\Lambda}{2\pi^{2}v^{2}\sqrt{\eta}}\ell\mathcal{G}_{1}^{D}\left(\bar{\Psi}_{a}\gamma_{0}\gamma_{z}\Psi_{a}\right)\left(\bar{\Psi}_{b}\gamma_{0}\gamma_{z}\Psi_{b}\right)\nonumber
\\
&&+2\Delta_{M}\Delta_{MNz}
\frac{\Lambda}{2\pi^{2}v^{2}\sqrt{\eta}}\ell\mathcal{G}_{1}^{m}\left(\bar{\Psi}_{a}
i\gamma_{x}\gamma_{y}\Psi_{a}\right)\left(\bar{\Psi}_{b}i\gamma_{x}\gamma_{y}\Psi_{b}\right),
\\
W_{M,AMN\bot}^{(2)+(3)}
&=&2\Delta_{M}\Delta_{AMN\bot}\frac{\Lambda}{2\pi^{2}v^{2}\sqrt{\eta}}\ell\mathcal{G}_{1}^{\bot}\left(\bar{\Psi}_{a}
i\gamma_{5}\Psi_{a}\right)\left(\bar{\Psi}_{b}i\gamma_{5}\Psi_{b}\right)\nonumber
\\
&&+2\Delta_{M}\Delta_{AMN\bot}
\frac{\Lambda}{2\pi^{2}v^{2}\sqrt{\eta}}\ell\mathcal{G}_{1}^{m}\sum_{j=x,y}\left(\bar{\Psi}_{a}
i\gamma_{5}\gamma_{j}\Psi_{a}\right)\left(\bar{\Psi}_{b}i\gamma_{5}\gamma_{j}\Psi_{b}\right),
\\
W_{M,AMNZ}^{(2)+(3)}
&=&2\Delta_{M}\Delta_{AMNz}\frac{\Lambda}{2\pi^{2}v^{2}\sqrt{\eta}}\ell\mathcal{G}_{1}^{z}\left(\bar{\Psi}_{a}
i\gamma_{5}\Psi_{a}\right)\left(\bar{\Psi}_{b}i\gamma_{5}\Psi_{b}\right)\nonumber
\\
&&+2\Delta_{M}\Delta_{AMNz}
\frac{\Lambda}{2\pi^{2}v^{2}\sqrt{\eta}}\ell\mathcal{G}_{1}^{m}\left(\bar{\Psi}_{a}
i\gamma_{5}\gamma_{z}\Psi_{a}\right)\left(\bar{\Psi}_{b}i\gamma_{5}\gamma_{z}\Psi_{b}\right),
\\
W_{M,CR\bot}^{(2)+(3)}
&=&2\Delta_{M}\Delta_{CR\bot}\frac{\Lambda}{2\pi^{2}v^{2}\sqrt{\eta}}\ell\mathcal{G}_{1}^{\bot}\left(\bar{\Psi}_{a}
i\gamma_{x}\gamma_{y}\Psi_{a}\right)\left(\bar{\Psi}_{b}
i\gamma_{x}\gamma_{y}\Psi_{b}\right)\nonumber
\\
&&+2\Delta_{M}\Delta_{CR\bot}\frac{\Lambda}{2\pi^{2}v^{2}\sqrt{\eta}}\ell\mathcal{G}_{1}^{z}\left[\left(\bar{\Psi}_{a}
i\gamma_{y}\gamma_{z}\Psi_{a}\right)\left(\bar{\Psi}_{b}i\gamma_{y}\gamma_{z}\Psi_{b}\right)+\left(\bar{\Psi}_{a}
i\gamma_{z}\gamma_{x}\Psi_{a}\right)\left(\bar{\Psi}_{b}i\gamma_{z}\gamma_{x}\Psi_{b}\right)
\right]\nonumber
\\
&&+2\Delta_{M}\Delta_{CR\bot}
\frac{\Lambda}{2\pi^{2}v^{2}\sqrt{\eta}}\ell\mathcal{G}_{1}^{m}\sum_{j=x,y}\left(\bar{\Psi}_{a}
i\gamma_{j}\Psi_{a}\right)\left(\bar{\Psi}_{b}i\gamma_{j}\Psi_{b}\right),
\\
W_{M,CRz}^{(2)+(3)}
&=&\Delta_{M}\Delta_{CRz}\frac{\Lambda}{2\pi^{2}v^{2}\sqrt{\eta}}\ell\mathcal{G}_{1}^{\bot}
\left[\left(\bar{\Psi}_{a}
i\gamma_{y}\gamma_{z}\Psi_{a}\right)\left(\bar{\Psi}_{b}i\gamma_{y}\gamma_{z}\Psi_{b}\right)
+\left(\bar{\Psi}_{a}
i\gamma_{z}\gamma_{x}\Psi_{a}\right)\left(\bar{\Psi}_{b}i\gamma_{z}\gamma_{x}\Psi_{b}\right)\right]\nonumber
\\
&&+2\Delta_{M}\Delta_{CRz}
\frac{\Lambda}{2\pi^{2}v^{2}\sqrt{\eta}}\ell\mathcal{G}_{1}^{m}\left(\bar{\Psi}_{a}
i\gamma_{z}\Psi_{a}\right)\left(\bar{\Psi}_{b}i\gamma_{z}\Psi_{b}\right),
\\
W_{AC,SO\bot}^{(2)+(3)}
&=&2\Delta_{AC}\Delta_{SO\bot}\frac{\Lambda}{2\pi^{2}v^{2}\sqrt{\eta}}\ell\mathcal{G}_{1}^{\bot}\left(\bar{\Psi}_{a}
\gamma_{0}\gamma_{z}\Psi_{a}\right)
\left(\bar{\Psi}_{b}\gamma_{0}\gamma_{z}\Psi_{b}\right)\nonumber
\\
&&+2\Delta_{AC}\Delta_{SO\bot}
\frac{\Lambda}{2\pi^{2}v^{2}\sqrt{\eta}}\ell\mathcal{G}_{1}^{z}\sum_{j=x,y}\left(\bar{\Psi}_{a}
\gamma_{0}\gamma_{j}\Psi_{a}\right)\left(\bar{\Psi}_{b}\gamma_{0}\gamma_{j}\Psi_{b}\right)\nonumber
\\
&&+2\Delta_{AC}\Delta_{SO\bot}
\frac{\Lambda}{2\pi^{2}v^{2}\sqrt{\eta}}\ell\mathcal{G}_{1}^{D}
\sum_{j=x,y}\left(\bar{\Psi}_{a}i\gamma_{j}\Psi_{a}\right)\left(\bar{\Psi}_{b}i\gamma_{j}\Psi_{b}\right),
\\
W_{AC,SOz}^{(2)+(3)}
&=&\Delta_{AC}\Delta_{SOz}
\frac{\Lambda}{2\pi^{2}v^{2}\sqrt{\eta}}\ell\mathcal{G}_{1}^{\bot}
\sum_{j=x,y}\left(\bar{\Psi}_{a}\gamma_{0}\gamma_{j}\Psi_{a}\right)\left(\bar{\Psi}_{b}\gamma_{0}\gamma_{j}\Psi_{b}\right)\nonumber
\\
&&+2\Delta_{AC}\Delta_{SOz}
\frac{\Lambda}{2\pi^{2}v^{2}\sqrt{\eta}}\ell\mathcal{G}_{1}^{D}
\left(\bar{\Psi}_{a}i\gamma_{z}\Psi_{a}\right)\left(\bar{\Psi}_{b}i\gamma_{z}\Psi_{b}\right),
\\
W_{AC,PM}^{(2)+(3)}
&=&2\Delta_{AC}\Delta_{PM}
\frac{\Lambda}{2\pi^{2}v^{2}\sqrt{\eta}}\ell\mathcal{G}_{1}^{D}
\left(\bar{\Psi}_{a}\gamma_{0}\gamma_{5}\Psi_{a}\right)\left(\bar{\Psi}_{b}i\gamma_{0}\gamma_{5}\Psi_{b}\right),
\\
W_{AC,MN\bot}^{(2)+(3)}
&=&2\Delta_{AC}\Delta_{MN\bot}\frac{\Lambda}{2\pi^{2}v^{2}\sqrt{\eta}}\ell\mathcal{G}_{1}^{\bot}\left(\bar{\Psi}_{a}i
\gamma_{x}\gamma_{y}\Psi_{a}\right)\left(\bar{\Psi}_{b}i
\gamma_{x}\gamma_{y}\Psi_{b}\right)\nonumber
\\
&&+2\Delta_{AC}\Delta_{MN\bot}\frac{\Lambda}{2\pi^{2}v^{2}\sqrt{\eta}}\ell\mathcal{G}_{1}^{z}\left[\left(\bar{\Psi}_{a}i\gamma_{y}\gamma_{z}
\Psi_{a}\right)\left(\bar{\Psi}_{b}i\gamma_{y}\gamma_{z}\Psi_{b}\right)+\left(\bar{\Psi}_{a}i\gamma_{z}\gamma_{x}
\Psi_{a}\right)\left(\bar{\Psi}_{b}i\gamma_{z}
\gamma_{x}\Psi_{b}\right)\right]\nonumber
\\
&&+2\Delta_{AC}\Delta_{MN\bot}
\frac{\Lambda}{2\pi^{2}v^{2}\sqrt{\eta}}\ell\mathcal{G}_{1}^{m}\sum_{j=x,y}\left(\bar{\Psi}_{a}i\gamma_{j}\Psi_{a}\right)\left(\bar{\Psi}_{b}i\gamma_{j}\Psi_{b}\right),
\\
W_{AC,MNz}^{(2)+(3)}
&=&\Delta_{AC}\Delta_{MNz}\frac{\Lambda}{2\pi^{2}v^{2}\sqrt{\eta}}\ell\mathcal{G}_{1}^{\bot}\left[\left(\bar{\Psi}_{a}i\gamma_{y}\gamma_{z}
\Psi_{a}\right)\left(\bar{\Psi}_{b}i\gamma_{y}\gamma_{z}\Psi_{b}\right)+\left(\bar{\Psi}_{a}i\gamma_{z}\gamma_{x}
\Psi_{a}\right)\left(\bar{\Psi}_{b}i\gamma_{z}\gamma_{x}\Psi_{b}\right)\right]\nonumber
\\
&&+2\Delta_{AC}\Delta_{MNz}
\frac{\Lambda}{2\pi^{2}v^{2}\sqrt{\eta}}\ell\mathcal{G}_{1}^{m}\left(\bar{\Psi}_{a}i\gamma_{z}\Psi_{a}\right)\left(\bar{\Psi}_{b}i\gamma_{z}\Psi_{b}\right),
\\
W_{AC,AMN\bot}^{(2)+(3)}
&=&2\Delta_{AC}\Delta_{AMN\bot}
\frac{\Lambda}{2\pi^{2}v^{2}\sqrt{\eta}}\ell\mathcal{G}_{1}^{\bot}\left(\bar{\Psi}_{a}i\gamma_{5}\gamma_{z}\Psi_{a}\right)\left(\bar{\Psi}_{b}i\gamma_{5}\gamma_{z}\Psi_{b}\right)\nonumber
\\
&&+2\Delta_{AC}\Delta_{AMN\bot}\frac{\Lambda}{2\pi^{2}v^{2}\sqrt{\eta}}\ell\mathcal{G}_{1}^{z}\sum_{j=x,y}\left(\bar{\Psi}_{a}i\gamma_{5}\gamma_{j}
\Psi_{a}\right)\left(\bar{\Psi}_{b}i\gamma_{5}\gamma_{j}\Psi_{b}\right),
\\
W_{AC,AMNz}^{(2)+(3)}
&=&\Delta_{AC}\Delta_{AMNz}
\frac{\Lambda}{2\pi^{2}v^{2}\sqrt{\eta}}\ell\mathcal{G}_{1}^{\bot}\sum_{j=x,y}\left(\bar{\Psi}_{a}i\gamma_{5}\gamma_{j}\Psi_{a}\right)\left(\bar{\Psi}_{b}i\gamma_{5}\gamma_{j}\Psi_{b}\right),
\\
W_{AC,CR\bot}^{(2)+(3)}
&=&2\Delta_{AC}\Delta_{CR\bot}\frac{\Lambda}{2\pi^{2}v^{2}\sqrt{\eta}}\ell\mathcal{G}_{1}^{\bot}\left(\bar{\Psi}_{a}i\gamma_{z}\Psi_{a}\right)\left(\bar{\Psi}_{b}i\gamma_{z}\Psi_{b}\right)\nonumber
\\
&&+2\Delta_{AC}\Delta_{CR\bot}\frac{\Lambda}{2\pi^{2}v^{2}\sqrt{\eta}}\ell\mathcal{G}_{1}^{z}\sum_{j=x,y}\left(\bar{\Psi}_{a}i\gamma_{j}
\Psi_{a}\right)\left(\bar{\Psi}_{b}i\gamma_{j}\Psi_{b}\right)\nonumber
\\
&&+2\Delta_{AC}\Delta_{CR\bot}
\frac{\Lambda}{2\pi^{2}v^{2}\sqrt{\eta}}\ell\mathcal{G}_{1}^{D}\sum_{j=x,y}\left(\bar{\Psi}_{a}\gamma_{0}\gamma_{j}\Psi_{a}\right)\left(\bar{\Psi}_{b}\gamma_{0}\gamma_{j}\Psi_{b}\right)\nonumber
\\
&&+2\Delta_{AC}\Delta_{CR\bot}
\frac{\Lambda}{2\pi^{2}v^{2}\sqrt{\eta}}\ell\mathcal{G}_{1}^{m}\left[\left(\bar{\Psi}_{a}i\gamma_{y}\gamma_{z}
\Psi_{a}\right)\left(\bar{\Psi}_{b}i\gamma_{y}\gamma_{z}\Psi_{b}\right)+\left(\bar{\Psi}_{a}i\gamma_{z}\gamma_{x}
\Psi_{a}\right)\left(\bar{\Psi}_{b}i\gamma_{z}\gamma_{x}\Psi_{b}\right)\right],
\\
W_{AC,CRz}^{(2)+(3)}
&=&\Delta_{AC}\Delta_{CRz}
\frac{\Lambda}{2\pi^{2}v^{2}\sqrt{\eta}}\ell\mathcal{G}_{1}^{\bot}\sum_{j=x,y}\left(\bar{\Psi}_{a}i\gamma_{j}\Psi_{a}\right)\left(\bar{\Psi}_{b}i\gamma_{j}\Psi_{b}\right)\nonumber
\\
&&+2\Delta_{AC}\Delta_{CRz}
\frac{\Lambda}{2\pi^{2}v^{2}\sqrt{\eta}}\ell\mathcal{G}_{1}^{D}\left(\bar{\Psi}_{a}\gamma_{0}\gamma_{z}\Psi_{a}\right)\left(\bar{\Psi}_{b}\gamma_{0}\gamma_{z}\Psi_{b}\right)\nonumber
\\
&&+2\Delta_{AC}\Delta_{CRz}
\frac{\Lambda}{2\pi^{2}v^{2}\sqrt{\eta}}\ell\mathcal{G}_{1}^{m}\left(\bar{\Psi}_{a}i\gamma_{x}\gamma_{y}
\Psi_{a}\right)\left(\bar{\Psi}_{b}i\gamma_{x}\gamma_{y}\Psi_{b}\right),
\\
W_{SO\bot,SOz}^{(2)+(3)}
&=&2\Delta_{SO\bot}\Delta_{SOz}
\frac{\Lambda}{2\pi^{2}v^{2}\sqrt{\eta}}\ell\mathcal{G}_{1}^{\bot}\left(\bar{\Psi}_{a}\gamma_{0}\gamma_{5}
\Psi_{a}\right)\left(\bar{\Psi}_{b}\gamma_{0}\gamma_{5}\Psi_{b}\right),
\\
W_{SO\bot,PM}^{(2)+(3)}
&=&2\Delta_{SO\bot}\Delta_{PM}\frac{\Lambda}{2\pi^{2}v^{2}\sqrt{\eta}}\ell\mathcal{G}_{1}^{\bot}\left(\bar{\Psi}_{a}i\gamma_{z}\Psi_{a}\right)
\left(\bar{\Psi}_{b}i\gamma_{z}\Psi_{b}\right)\nonumber
\\
&&+2\Delta_{SO\bot}\Delta_{PM}\frac{\Lambda}{2\pi^{2}v^{2}\sqrt{\eta}}\ell\mathcal{G}_{1}^{z}
\sum_{j=x,y}\left(\bar{\Psi}_{a}i\gamma_{j}\Psi_{a}\right)\left(\bar{\Psi}_{b}i\gamma_{j}
\Psi_{b}\right)\nonumber
\\
&&+2\Delta_{SO\bot}\Delta_{PM}
\frac{\Lambda}{2\pi^{2}v^{2}\sqrt{\eta}}\ell\mathcal{G}_{1}^{D}
\sum_{j=x,y}\left(\bar{\Psi}_{a}\gamma_{0}\gamma_{j}\Psi_{a}\right)\left(\bar{\Psi}_{b}\gamma_{0}\gamma_{j}\Psi_{b}\right)\nonumber
\\
&&+2\Delta_{SO\bot}\Delta_{PM}
\frac{\Lambda}{2\pi^{2}v^{2}\sqrt{\eta}}\ell\mathcal{G}_{1}^{m}\left[\left(\bar{\Psi}_{a}i\gamma_{y}\gamma_{z}\Psi_{a}\right)\left(\bar{\Psi}_{b}i
\gamma_{y}\gamma_{z}\Psi_{b}\right)+\left(\bar{\Psi}_{a}i
\gamma_{z}\gamma_{x}\Psi_{a}\right)\left(\bar{\Psi}_{b}i
\gamma_{z}\gamma_{x}\Psi_{b}\right)\right],
\\
W_{SO\bot,MN\bot}^{(2)+(3)}
&=&4\Delta_{SO\bot}\Delta_{MN\bot}
\frac{\Lambda}{2\pi^{2}v^{2}\sqrt{\eta}}\ell\mathcal{G}_{1}^{\bot}\sum_{j=x,y}\left(\bar{\Psi}_{a}i\gamma_{5}\gamma_{j}\Psi_{a}\right)
\left(\bar{\Psi}_{b}i\gamma_{5}\gamma_{j}\Psi_{b}\right)\nonumber
\\
&&+4\Delta_{SO\bot}\Delta_{MN\bot}
\frac{\Lambda}{2\pi^{2}v^{2}\sqrt{\eta}}\ell\mathcal{G}_{1}^{z}
\left(\bar{\Psi}_{a}i\gamma_{5}\gamma_{z}\Psi_{a}\right)\left(\bar{\Psi}_{b}i
\gamma_{5}\gamma_{z}\Psi_{b}\right)\nonumber
\\
&&+4\Delta_{SO\bot}\Delta_{MN\bot}
\frac{\Lambda}{2\pi^{2}v^{2}\sqrt{\eta}}\ell\mathcal{G}_{1}^{D}
\left(\bar{\Psi}_{a}\Psi_{a}\right)
\left(\bar{\Psi}_{b}\Psi_{b}\right)\nonumber
\\
&&+4\Delta_{SO\bot}\Delta_{MN\bot}
\frac{\Lambda}{2\pi^{2}v^{2}\sqrt{\eta}}\ell\mathcal{G}_{1}^{m}
\left(\bar{\Psi}_{a}i\gamma_{5}\Psi_{a}\right)\left(\bar{\Psi}_{b}i\gamma_{5}\Psi_{b}\right),
\\
W_{SO\bot,MNz}^{(2)+(3)}
&=&2\Delta_{SO\bot}\Delta_{MNz}\frac{\Lambda}{2\pi^{2}v^{2}\sqrt{\eta}}\ell\mathcal{G}_{1}^{\bot}
\left(\bar{\Psi}_{a}i\gamma_{5}\gamma_{z}\Psi_{a}\right)
\left(\bar{\Psi}_{b}i\gamma_{5}\gamma_{z}\Psi_{b}\right)\nonumber
\\
&&+2\Delta_{SO\bot}\Delta_{MNz}
\frac{\Lambda}{2\pi^{2}v^{2}\sqrt{\eta}}\ell\mathcal{G}_{1}^{z}
\sum_{j=x,y}\left(\bar{\Psi}_{a}i\gamma_{5}\gamma_{j}\Psi_{a}\right)\left(\bar{\Psi}_{b}i\gamma_{5}\gamma_{j}\Psi_{b}\right),
\\
W_{SO\bot,AMN\bot}^{(2)+(3)}
&=&4\Delta_{SO\bot}\Delta_{AMN\bot}
\frac{\Lambda}{2\pi^{2}v^{2}\sqrt{\eta}}\ell\mathcal{G}_{1}^{\bot}\left[\left(\bar{\Psi}_{a}i\gamma_{y}\gamma_{z}\Psi_{a}\right)
\left(\bar{\Psi}_{b}i\gamma_{y}\gamma_{z}\Psi_{b}\right)+\left(\bar{\Psi}_{a}i\gamma_{z}\gamma_{x}\Psi_{a}\right)
\left(\bar{\Psi}_{b}i\gamma_{z}\gamma_{x}\Psi_{b}\right)\right]\nonumber
\\
&&+4\Delta_{SO\bot}\Delta_{AMN\bot}\frac{\Lambda}{2\pi^{2}v^{2}\sqrt{\eta}}\ell\mathcal{G}_{1}^{z}
\left(\bar{\Psi}_{a}i
\gamma_{x}\gamma_{y}\Psi_{a}\right)
\left(\bar{\Psi}_{b}i\gamma_{x}\gamma_{y}\Psi_{b}\right)\nonumber
\\
&&+4\Delta_{SO\bot}\Delta_{AMN\bot}
\frac{\Lambda}{2\pi^{2}v^{2}\sqrt{\eta}}\ell\mathcal{G}_{1}^{D}\left(\bar{\Psi}_{a}\gamma_{0}\Psi_{a}\right)\left(\bar{\Psi}_{b}\gamma_{0}\Psi_{b}\right)\nonumber
\\
&&+4\Delta_{SO\bot}\Delta_{AMN\bot}
\frac{\Lambda}{2\pi^{2}v^{2}\sqrt{\eta}}\ell\mathcal{G}_{1}^{m}
\left(\bar{\Psi}_{a}i\gamma_{z}\Psi_{a}\right)\left(\bar{\Psi}_{b}i\gamma_{z}\Psi_{b}\right),
\\
W_{SO\bot,AMNz}^{(2)+(3)}
&=&2\Delta_{SO\bot}\Delta_{AMNz}\frac{\Lambda}{2\pi^{2}v^{2}\sqrt{\eta}}\ell\mathcal{G}_{1}^{\bot}
\left(\bar{\Psi}_{a}i\gamma_{x}\gamma_{y}\Psi_{a}\right)
\left(\bar{\Psi}_{b}i\gamma_{x}\gamma_{y}\Psi_{b}\right)\nonumber
\\
&&+2\Delta_{SO\bot}\Delta_{AMNz}
\frac{\Lambda}{2\pi^{2}v^{2}\sqrt{\eta}}\ell\mathcal{G}_{1}^{z}\left[\left(\bar{\Psi}_{a}i\gamma_{y}\gamma_{z}\Psi_{a}\right)\left(\bar{\Psi}_{b}i\gamma_{y}\gamma_{z}\Psi_{b}\right)
+\left(\bar{\Psi}_{a}i
\gamma_{z}\gamma_{x}\Psi_{a}\right)\left(\bar{\Psi}_{b}i\gamma_{z}\gamma_{x}\Psi_{b}\right)\right]\nonumber
\\
&&+2\Delta_{SO\bot}\Delta_{AMNz}
\frac{\Lambda}{2\pi^{2}v^{2}\sqrt{\eta}}\ell\mathcal{G}_{1}^{m}\sum_{j=x,y}\left(\bar{\Psi}_{a}i\gamma_{j}\Psi_{a}\right)\left(\bar{\Psi}_{b}i\gamma_{j}\Psi_{b}\right),
\\
W_{SO\bot,CR\bot}^{(2)+(3)}
&=&4\Delta_{SO\bot}\Delta_{CR\bot}\frac{\Lambda}{2\pi^{2}v^{2}\sqrt{\eta}}\ell\mathcal{G}_{1}^{z}
\left(\bar{\Psi}_{a}i\gamma_{5}\Psi_{a}\right)\left(\bar{\Psi}_{b}i\gamma_{5}\Psi_{b}\right)\nonumber
\\
&&+4\Delta_{SO\bot}\Delta_{CR\bot}
\frac{\Lambda}{2\pi^{2}v^{2}\sqrt{\eta}}\ell\mathcal{G}_{1}^{D}
\left(\bar{\Psi}_{a}\gamma_{0}
\gamma_{5}\Psi_{a}\right)\left(\bar{\Psi}_{b}\gamma_{0}\gamma_{5}
\Psi_{b}\right)\nonumber
\\
&&+4\Delta_{SO\bot}\Delta_{CR\bot}
\frac{\Lambda}{2\pi^{2}v^{2}\sqrt{\eta}}\ell\mathcal{G}_{1}^{m}
\left(\bar{\Psi}_{a}i\gamma_{5}\gamma_{z}
\Psi_{a}\right)\left(\bar{\Psi}_{b}i
\gamma_{5}\gamma_{z}\Psi_{b}\right),
\\
W_{SO\bot,CRz}^{(2)+(3)}
&=&2\Delta_{SO\bot}\Delta_{CRz}\frac{\Lambda}{2\pi^{2}v^{2}\sqrt{\eta}}\ell\mathcal{G}_{1}^{\bot}\left(\bar{\Psi}_{a}i\gamma_{5}\Psi_{a}\right)\left(\bar{\Psi}_{b}i\gamma_{5}\Psi_{b}\right)\nonumber
\\
&&+2\Delta_{SO\bot}\Delta_{CRz}
\frac{\Lambda}{2\pi^{2}v^{2}\sqrt{\eta}}\ell\mathcal{G}_{1}^{m}\sum_{j=x,y}\left(\bar{\Psi}_{a}i\gamma_{5}\gamma_{j}\Psi_{a}\right)\left(\bar{\Psi}_{b}i\gamma_{5}\gamma_{j}\Psi_{b}\right),
\\
W_{SOz,PM}^{(2)+(3)}
&=&\Delta_{SOz}\Delta_{PM}\frac{\Lambda}{2\pi^{2}v^{2}\sqrt{\eta}}\mathcal{G}_{1}^{\bot}\ell\sum_{j=x,y}\left(\bar{\Psi}_{a}i\gamma_{j}\Psi_{a}\right)\left(\bar{\Psi}_{b}i\gamma_{j}\Psi_{b}\right)\nonumber
\\
&&+2\Delta_{SOz}\Delta_{PM}
\frac{\Lambda}{2\pi^{2}v^{2}\sqrt{\eta}}\mathcal{G}_{1}^{D}\ell\left(\bar{\Psi}_{a}\gamma_{0}\gamma_{z}\Psi_{a}\right)\left(\bar{\Psi}_{b}i\gamma_{0}\gamma_{z}\Psi_{b}\right)\nonumber
\\
&&+2\Delta_{SOz}\Delta_{PM}
\frac{\Lambda}{2\pi^{2}v^{2}\sqrt{\eta}}\mathcal{G}_{1}^{m}\ell
\left(\bar{\Psi}_{a}i
\gamma_{x}\gamma_{y}\Psi_{a}\right)\left(\bar{\Psi}_{b}i\gamma_{x}\gamma_{y}\Psi_{b}\right),
\\
W_{SOz,MN\bot}^{(2)+(3)}
&=&2\Delta_{SOz}\Delta_{MN\bot}\frac{\Lambda}{2\pi^{2}v^{2}\sqrt{\eta}}\ell\mathcal{G}_{1}^{\bot}\left(\bar{\Psi}_{a}i\gamma_{5}\gamma_{z}\Psi_{a}\right)\left(\bar{\Psi}_{b}i\gamma_{5}\gamma_{z}\Psi_{b}\right)\nonumber
\\
&&+2\Delta_{SOz}\Delta_{MN\bot}
\frac{\Lambda}{2\pi^{2}v^{2}\sqrt{\eta}}\ell\mathcal{G}_{1}^{z}\sum_{j=x,y}\left(\bar{\Psi}_{a}i\gamma_{5}\gamma_{j}\Psi_{a}\right)\left(\bar{\Psi}_{b}i\gamma_{5}\gamma_{j}\Psi_{b}\right),
\\
W_{SOz,MNz}^{(2)+(3)}
&=&\Delta_{SOz}\Delta_{MNz}\frac{\Lambda}{2\pi^{2}v^{2}\sqrt{\eta}}\ell\mathcal{G}_{1}^{\bot}
\sum_{j=x,y}\left(\bar{\Psi}_{a}i\gamma_{5}\gamma_{j}
\Psi_{a}\right)\left(\bar{\Psi}_{b}i\gamma_{5}\gamma_{j}\Psi_{b}\right)\nonumber
\\
&&+2\Delta_{SOz}\Delta_{MNz}\frac{\Lambda}{2\pi^{2}v^{2}\sqrt{\eta}}\ell\mathcal{G}_{1}^{z}\left(\bar{\Psi}_{a}i\gamma_{5}
\gamma_{z}\Psi_{a}\right)\left(\bar{\Psi}_{b}i\gamma_{5}\gamma_{z}\Psi_{b}\right)\nonumber
\\
&&+2\Delta_{SOz}\Delta_{MNz}
\frac{\Lambda}{2\pi^{2}v^{2}\sqrt{\eta}}\ell\mathcal{G}_{1}^{D}\left(\bar{\Psi}_{a}\Psi_{a}\right)\left(\bar{\Psi}_{b}\Psi_{b}\right)\nonumber
\\
&&+2\Delta_{SOz}\Delta_{MNz}
\frac{\Lambda}{2\pi^{2}v^{2}\sqrt{\eta}}\ell\mathcal{G}_{1}^{m}\left(\bar{\Psi}_{a}i\gamma_{5}
\Psi_{a}\right)\left(\bar{\Psi}_{b}i\gamma_{5}\Psi_{b}\right),
\\
W_{SOz,AMN\bot}^{(2)+(3)}
&=&2\Delta_{SOz}\Delta_{AMN\bot}\frac{\Lambda}{2\pi^{2}v^{2}\sqrt{\eta}}\ell\mathcal{G}_{1}^{\bot}\left(\bar{\Psi}_{a}i
\gamma_{x}\gamma_{y}\Psi_{a}\right)\left(\bar{\Psi}_{b}i
\gamma_{x}\gamma_{y}\Psi_{b}\right)\nonumber
\\
&&+2\Delta_{SOz}\Delta_{AMN\bot}\frac{\Lambda}{2\pi^{2}v^{2}\sqrt{\eta}}\ell\mathcal{G}_{1}^{z}
\left[\left(\bar{\Psi}_{a}i\gamma_{y}\gamma_{z}\Psi_{a}\right)
\left(\bar{\Psi}_{b}i\gamma_{y}\gamma_{z}\Psi_{b}\right)
+\left(\bar{\Psi}_{a}i\gamma_{z}\gamma_{x}\Psi_{a}\right)
\left(\bar{\Psi}_{b}i\gamma_{z}\gamma_{x}\Psi_{b}\right)\right]\nonumber
\\
&&+2\Delta_{SOz}\Delta_{AMN\bot}
\frac{\Lambda}{2\pi^{2}v^{2}\sqrt{\eta}}\ell\mathcal{G}_{1}^{m}\sum_{j=x,y}\left(\bar{\Psi}_{a}i\gamma_{j}\Psi_{a}\right)\left(\bar{\Psi}_{b}i\gamma_{j}\Psi_{b}\right),
\\
W_{SOz,AMNz}^{(2)+(3)}
&=&\Delta_{SOz}\Delta_{AMNz}\frac{\Lambda}{2\pi^{2}v^{2}\sqrt{\eta}}\ell\mathcal{G}_{1}^{\bot}
\left[\left(\bar{\Psi}_{a}i\gamma_{y}\gamma_{z}\Psi_{a}\right)\left(\bar{\Psi}_{b}i\gamma_{y}\gamma_{z}\Psi_{b}\right)
+\left(\bar{\Psi}_{a}i
\gamma_{z}\gamma_{x}\Psi_{a}\right)\left(\bar{\Psi}_{b}i\gamma_{z}\gamma_{x}\Psi_{b}\right)\right]\nonumber
\\
&&+2\Delta_{SOz}\Delta_{AMNz}\frac{\Lambda}{2\pi^{2}v^{2}\sqrt{\eta}}\ell\mathcal{G}_{1}^{z}
\left(\bar{\Psi}_{a}i
\gamma_{x}\gamma_{y}\Psi_{a}\right)\left(\bar{\Psi}_{b}i\gamma_{x}\gamma_{y}\Psi_{b}\right)\nonumber
\\
&&+2\Delta_{SOz}\Delta_{AMNz}
\frac{\Lambda}{2\pi^{2}v^{2}\sqrt{\eta}}\ell\mathcal{G}_{1}^{D}\left(\bar{\Psi}_{a}\gamma_{0}\Psi_{a}\right)\left(\bar{\Psi}_{b}\gamma_{0}\Psi_{b}\right),
\\
W_{SOz,CR\bot}^{(2)+(3)}
&=&2\Delta_{SOz}\Delta_{CR\bot}\frac{\Lambda}{2\pi^{2}v^{2}\sqrt{\eta}}\ell\mathcal{G}_{1}^{\bot}\left(\bar{\Psi}_{a}i\gamma_{5}\Psi_{a}\right)\left(\bar{\Psi}_{b}i\gamma_{5}\Psi_{b}\right)\nonumber
\\
&&+2\Delta_{SOz}\Delta_{CR\bot}
\frac{\Lambda}{2\pi^{2}v^{2}\sqrt{\eta}}\ell\mathcal{G}_{1}^{m}\sum_{j=x,y}\left(\bar{\Psi}_{a}i\gamma_{5}\gamma_{j}
\Psi_{a}\right)\left(\bar{\Psi}_{b}i\gamma_{5}\gamma_{j}\Psi_{b}\right),
\\
W_{SOz,CRz}^{(2)+(3)}
&=&2\Delta_{SOz}\Delta_{CRz}
\frac{\Lambda}{2\pi^{2}v^{2}\sqrt{\eta}}\ell\mathcal{G}_{1}^{D}\left(\bar{\Psi}_{a}\gamma_{0}\gamma_{5}\Psi_{a}\right)\left(\bar{\Psi}_{b}\gamma_{0}\gamma_{5}\Psi_{b}\right),
\\
W_{PM,MN\bot}^{(2)+(3)}
&=&2\Delta_{PM}\Delta_{MN\bot}\frac{\Lambda}{2\pi^{2}v^{2}\sqrt{\eta}}\ell\mathcal{G}_{1}^{\bot}\left(\bar{\Psi}_{a}
\gamma_{0}\Psi_{a}\right)\left(\bar{\Psi}_{b}\gamma_{0}\Psi_{b}\right)\nonumber
\\
&&+2\Delta_{PM}\Delta_{MN\bot}
\frac{\Lambda}{2\pi^{2}v^{2}\sqrt{\eta}}\ell\mathcal{G}_{1}^{D}\left[\left(\bar{\Psi}_{a}i\gamma_{y}\gamma_{z}\Psi_{a}\right)\left(\bar{\Psi}_{b}i\gamma_{y}\gamma_{z}\Psi_{b}\right)
+\left(\bar{\Psi}_{a}i\gamma_{z}\gamma_{x}\Psi_{a}\right)\left(\bar{\Psi}_{b}i\gamma_{z}\gamma_{x}\Psi_{b}\right)\right]\nonumber
\\
&&+2\Delta_{PM}\Delta_{MN\bot}
\frac{\Lambda}{2\pi^{2}v^{2}\sqrt{\eta}}\ell\mathcal{G}_{1}^{m}\sum_{j=x,y}\left(\bar{\Psi}_{a}
\gamma_{0}\gamma_{j}\Psi_{a}\right)\left(\bar{\Psi}_{b}
\gamma_{0}\gamma_{j}\Psi_{b}\right),
\\
W_{PM,MNz}^{(2)+(3)}
&=&2\Delta_{PM}\Delta_{MNz}\frac{\Lambda}{2\pi^{2}v^{2}\sqrt{\eta}}\ell\mathcal{G}_{1}^{z}\left(\bar{\Psi}_{a}
\gamma_{0}\Psi_{a}\right)\left(\bar{\Psi}_{b}\gamma_{0}\Psi_{b}\right)\nonumber
\\
&&+2\Delta_{PM}\Delta_{MNz}
\frac{\Lambda}{2\pi^{2}v^{2}\sqrt{\eta}}\ell\mathcal{G}_{1}^{D}\left(\bar{\Psi}_{a}i\gamma_{x}\gamma_{y}\Psi_{a}\right)\left(\bar{\Psi}_{b}i\gamma_{x}\gamma_{y}\Psi_{b}\right)\nonumber
\\
&&+2\Delta_{PM}\Delta_{MNz}
\frac{\Lambda}{2\pi^{2}v^{2}\sqrt{\eta}}\ell\mathcal{G}_{1}^{m}\left(\bar{\Psi}_{a}
\gamma_{0}\gamma_{z}\Psi_{a}\right)\left(\bar{\Psi}_{b}\gamma_{0}\gamma_{z}\Psi_{b}\right),
\\
W_{PM,AMN\bot}^{(2)+(3)}
&=&2\Delta_{PM}\Delta_{AMN\bot}\frac{\Lambda}{2\pi^{2}v^{2}\sqrt{\eta}}\ell\mathcal{G}_{1}^{\bot}\left(\bar{\Psi}_{a}\Psi_{a}\right)\left(\bar{\Psi}_{b}\Psi_{b}\right)\nonumber
\\
&&+2\Delta_{PM}\Delta_{AMN\bot}
\frac{\Lambda}{2\pi^{2}v^{2}\sqrt{\eta}}\ell\mathcal{G}_{1}^{D}\sum_{j=x,y}\left(\bar{\Psi}_{a}i\gamma_{5}\gamma_{j}\Psi_{a}\right)\left(\bar{\Psi}_{b}i\gamma_{5}\gamma_{j}\Psi_{b}\right),
\\
W_{PM,AMNz}^{(2)+(3)}
&=&2\Delta_{PM}\Delta_{AMNz}\frac{\Lambda}{2\pi^{2}v^{2}\sqrt{\eta}}\ell\mathcal{G}_{1}^{z}\left(\bar{\Psi}_{a}\Psi_{a}\right)\left(\bar{\Psi}_{b}\Psi_{b}\right)\nonumber
\\
&&+2\Delta_{PM}\Delta_{AMNz}
\frac{\Lambda}{2\pi^{2}v^{2}\sqrt{\eta}}\ell\mathcal{G}_{1}^{D}\left(\bar{\Psi}_{a}i\gamma_{5}\gamma_{z}\Psi_{a}\right)\left(\bar{\Psi}_{b}i\gamma_{5}\gamma_{z}\Psi_{b}\right),
\\
W_{PM,CR\bot}^{(2)+(3)}
&=&2\Delta_{PM}\Delta_{CR\bot}
\frac{\Lambda}{2\pi^{2}v^{2}\sqrt{\eta}}\ell\mathcal{G}_{1}^{\bot}\left(\bar{\Psi}_{a}
\gamma_{0}\gamma_{z}\Psi_{a}\right)\left(\bar{\Psi}_{b}\gamma_{0}\gamma_{z}\Psi_{b}\right)\nonumber
\\
&&+2\Delta_{PM}\Delta_{CR\bot}
\frac{\Lambda}{2\pi^{2}v^{2}\sqrt{\eta}}\ell\mathcal{G}_{1}^{z}\sum_{j=x,y}\left(\bar{\Psi}_{a}
\gamma_{0}\gamma_{j}\Psi_{a}\right)\left(\bar{\Psi}_{b}\gamma_{0}\gamma_{j}\Psi_{b}\right)\nonumber
\\
&&+2\Delta_{PM}\Delta_{CR\bot}\sum_{j=x,y}
\frac{\Lambda}{2\pi^{2}v^{2}\sqrt{\eta}}\ell\mathcal{G}_{1}^{D}\left(\bar{\Psi}_{a}i\gamma_{j}\Psi_{a}\right)\left(\bar{\Psi}_{b}i\gamma_{j}\Psi_{b}\right),
\\
W_{PM,CRz}^{(2)+(3)}
&=&\Delta_{PM}\Delta_{CRz}\frac{\Lambda}{2\pi^{2}v^{2}\sqrt{\eta}}\ell\mathcal{G}_{1}^{\bot}\sum_{j=x,y}\left(\bar{\Psi}_{a}
\gamma_{0}\gamma_{j}\Psi_{a}\right)\left(\bar{\Psi}_{b}\gamma_{0}\gamma_{j}\Psi_{b}\right)\nonumber
\\
&&+2\Delta_{PM}\Delta_{CRz}
\frac{\Lambda}{2\pi^{2}v^{2}\sqrt{\eta}}\ell\mathcal{G}_{1}^{D}\left(\bar{\Psi}_{a}i\gamma_{z}\Psi_{a}\right)\left(\bar{\Psi}_{b}i\gamma_{z}\Psi_{b}\right),
\\
W_{MN\bot,MNz}^{(2)+(3)}
&=&2\Delta_{MN\bot}\Delta_{MNz}\frac{\Lambda}{2\pi^{2}v^{2}\sqrt{\eta}}\ell\mathcal{G}_{1}^{\bot}
\left(\bar{\Psi}_{a}\gamma_{0}\gamma_{5}\Psi_{a}\right)\left(\bar{\Psi}_{b}\gamma_{0}\gamma_{5}\Psi_{b}\right),
\\
W_{MN\bot,AMN\bot}^{(2)+(3)}
&=&4\Delta_{MN\bot}\Delta_{AMN\bot}\frac{\Lambda}{2\pi^{2}v^{2}\sqrt{\eta}}\ell\mathcal{G}_{1}^{\bot}
\sum_{j=x,y}\left(\bar{\Psi}_{a}
\gamma_{0}\gamma_{j}\Psi_{a}\right)\left(\bar{\Psi}_{b}\gamma_{0}\gamma_{j}\Psi_{b}\right)\nonumber
\\
&&+4\Delta_{MN\bot}\Delta_{AMN\bot}\frac{\Lambda}{2\pi^{2}v^{2}\sqrt{\eta}}\ell\mathcal{G}_{1}^{z}
\left(\bar{\Psi}_{a}\gamma_{0}\gamma_{z}\Psi_{a}\right)
\left(\bar{\Psi}_{b}\gamma_{0}\gamma_{z}\Psi_{b}\right)\nonumber
\\
&&+4\Delta_{MN\bot}\Delta_{AMN\bot}
\frac{\Lambda}{2\pi^{2}v^{2}\sqrt{\eta}}\ell\mathcal{G}_{1}^{D}\left(\bar{\Psi}_{a}i\gamma_{z}
\Psi_{a}\right)\left(\bar{\Psi}_{b}i\gamma_{z}\Psi_{b}\right)\nonumber
\\
&&+4\Delta_{MN\bot}\Delta_{AMN\bot}
\frac{\Lambda}{2\pi^{2}v^{2}\sqrt{\eta}}\ell\mathcal{G}_{1}^{m}\left(\bar{\Psi}_{a}\gamma_{0}\Psi_{a}\right)\left(\bar{\Psi}_{b}\gamma_{0}\Psi_{b}\right),
\\
W_{MN\bot,AMNz}^{(2)+(3)}
&=&2\Delta_{MN\bot}\Delta_{AMNz}\frac{\Lambda}{2\pi^{2}v^{2}\sqrt{\eta}}\ell\mathcal{G}_{1}^{\bot}\left(\bar{\Psi}_{a}
\gamma_{0}\gamma_{z}\Psi_{a}\right)\left(\bar{\Psi}_{b}\gamma_{0}\gamma_{z}\Psi_{b}\right)\nonumber
\\
&&+2\Delta_{MN\bot}\Delta_{AMNz}\frac{\Lambda}{2\pi^{2}v^{2}\sqrt{\eta}}\ell\mathcal{G}_{1}^{z}\sum_{j=x,y}\left(\bar{\Psi}_{a}
\gamma_{0}\gamma_{j}\Psi_{a}\right)\left(\bar{\Psi}_{b}\gamma_{0}\gamma_{j}\Psi_{b}\right)\nonumber
\\
&&+2\Delta_{MN\bot}\Delta_{AMNz}
\frac{\Lambda}{2\pi^{2}v^{2}\sqrt{\eta}}\ell\mathcal{G}_{1}^{D}\sum_{j=x,y}\left(\bar{\Psi}_{a}i\gamma_{j}
\Psi_{a}\right)\left(\bar{\Psi}_{b}i\gamma_{j}\Psi_{b}\right),
\\
W_{MN\bot,CR\bot}^{(2)+(3)}
&=&4\Delta_{MN\bot}\Delta_{CR\bot}\left(\bar{\Psi}_{a}\Psi_{a}\right)\left(\bar{\Psi}_{b}\Psi_{b}\right)\frac{\Lambda}{2\pi^{2}v^{2}\sqrt{\eta}}\ell\mathcal{G}_{1}^{z}\nonumber
\\
&&+4\Delta_{MN\bot}\Delta_{CR\bot}\left(\bar{\Psi}_{a}
i\gamma_{5}\gamma_{z}\Psi_{a}\right)\left(\bar{\Psi}_{b}i\gamma_{5}\gamma_{z}\Psi_{b}\right)
\frac{\Lambda}{2\pi^{2}v^{2}\sqrt{\eta}}\ell\mathcal{G}_{1}^{D}\nonumber
\\
&&+4\Delta_{MN\bot}\Delta_{CR\bot}\left(\bar{\Psi}_{a}\gamma_{0}\gamma_{5}
\Psi_{a}\right)\left(\bar{\Psi}_{b}\gamma_{0}\gamma_{5}\Psi_{b}\right)
\frac{\Lambda}{2\pi^{2}v^{2}\sqrt{\eta}}\ell\mathcal{G}_{1}^{m},
\\
W_{MN\bot,CRz}^{(2)+(3)}
&=&2\Delta_{MN\bot}\Delta_{CRz}\frac{\Lambda}{2\pi^{2}v^{2}\sqrt{\eta}}\ell\mathcal{G}_{1}^{\bot}\left(\bar{\Psi}_{a}\Psi_{a}\right)\left(\bar{\Psi}_{b}\Psi_{b}\right)\nonumber
\\
&&+2\Delta_{MN\bot}\Delta_{CRz}
\frac{\Lambda}{2\pi^{2}v^{2}\sqrt{\eta}}\ell\mathcal{G}_{1}^{D}\sum_{j=x,y}\left(\bar{\Psi}_{a}i\gamma_{5}\gamma_{j}
\Psi_{a}\right)\left(\bar{\Psi}_{b}i\gamma_{5}
\gamma_{j}\Psi_{b}\right),
\\
W_{MNz,AMN\bot}^{(2)+(3)}
&=&2\Delta_{MNz}\Delta_{AMN\bot}\frac{\Lambda}{2\pi^{2}v^{2}\sqrt{\eta}}\ell\mathcal{G}_{1}^{\bot}\left(\bar{\Psi}_{a}
\gamma_{0}\gamma_{z}\Psi_{a}\right)\left(\bar{\Psi}_{b}
\gamma_{0}\gamma_{z}\Psi_{b}\right)\nonumber
\\
&&+2\Delta_{MNz}\Delta_{AMN\bot}\frac{\Lambda}{2\pi^{2}v^{2}\sqrt{\eta}}\ell\mathcal{G}_{1}^{z}\sum_{j=x,y}\left(\bar{\Psi}_{a}
\gamma_{0}\gamma_{j}\Psi_{a}\right)\left(\bar{\Psi}_{b}\gamma_{0}\gamma_{j}\Psi_{b}\right)\nonumber
\\
&&+2\Delta_{MNz}\Delta_{AMN\bot}
\frac{\Lambda}{2\pi^{2}v^{2}\sqrt{\eta}}\ell\mathcal{G}_{1}^{D}\sum_{j=x,y}\left(\bar{\Psi}_{a}i\gamma_{j}\Psi_{a}\right)\left(\bar{\Psi}_{b}i\gamma_{j}\Psi_{b}\right),
\\
W_{MNz,AMNz}^{(2)+(3)}
&=&\Delta_{MNz}\Delta_{AMNz}\frac{\Lambda}{2\pi^{2}v^{2}\sqrt{\eta}}\ell\mathcal{G}_{1}^{\bot}\sum_{j=x,y}\left(\bar{\Psi}_{a}
\gamma_{0}\gamma_{j}\Psi_{a}\right)\left(\bar{\Psi}_{b}\gamma_{0}\gamma_{j}\Psi_{b}\right)\nonumber
\\
&&+2\Delta_{MNz}\Delta_{AMNz}\frac{\Lambda}{2\pi^{2}v^{2}\sqrt{\eta}}\ell\mathcal{G}_{1}^{z}\left(\bar{\Psi}_{a}
\gamma_{0}\gamma_{z}\Psi_{a}\right)\left(\bar{\Psi}_{b}\gamma_{0}\gamma_{z}\Psi_{b}\right)\nonumber
\\
&&+2\Delta_{MNz}\Delta_{AMNz}
\frac{\Lambda}{2\pi^{2}v^{2}\sqrt{\eta}}\ell\mathcal{G}_{1}^{m}\left(\bar{\Psi}_{a}\gamma_{0}\Psi_{a}\right)\left(\bar{\Psi}_{b}\gamma_{0}\Psi_{b}\right),
\\
W_{MNz,CR\bot}^{(2)+(3)}
&=&2\Delta_{MNz}\Delta_{CR\bot}\frac{\Lambda}{2\pi^{2}v^{2}\sqrt{\eta}}\ell\mathcal{G}_{1}^{\bot}\left(\bar{\Psi}_{a}\Psi_{a}\right)\left(\bar{\Psi}_{b}\Psi_{b}\right)\nonumber
\\
&&+2\Delta_{MNz}\Delta_{CR\bot}
\frac{\Lambda}{2\pi^{2}v^{2}\sqrt{\eta}}\ell\mathcal{G}_{1}^{D}\sum_{j=x,y}\left(\bar{\Psi}_{a}i\gamma_{5}\gamma_{j}
\Psi_{a}\right)\left(\bar{\Psi}_{b}i\gamma_{5}\gamma_{j}\Psi_{b}\right),
\\
W_{MNz,CRz}^{(2)+(3)}
&=&2\Delta_{MNz}\Delta_{CRz}
\frac{\Lambda}{2\pi^{2}v^{2}\sqrt{\eta}}\ell\mathcal{G}_{1}^{m}\left(\bar{\Psi}_{a}\gamma_{0}\gamma_{5}\Psi_{a}\right)\left(\bar{\Psi}_{b}\gamma_{0}\gamma_{5}\Psi_{b}\right),
\\
W_{AMN\bot,AMNz}^{(2)+(3)}
&=&2\Delta_{AMN\bot}\Delta_{AMNz}
\frac{\Lambda}{2\pi^{2}v^{2}\sqrt{\eta}}\ell\mathcal{G}_{1}^{\bot}\left(\bar{\Psi}_{a}\gamma_{0}\gamma_{5}\Psi_{a}\right)\left(\bar{\Psi}_{b}\gamma_{0}\gamma_{5}\Psi_{b}\right),
\\
W_{AMN\bot,CR\bot}^{(2)+(3)}
&=&4\Delta_{AMN\bot}\Delta_{CR\bot}\frac{\Lambda}{2\pi^{2}v^{2}\sqrt{\eta}}\ell\mathcal{G}_{1}^{z}
\left(\bar{\Psi}_{a}\gamma_{0}\Psi_{a}\right)
\left(\bar{\Psi}_{b}\gamma_{0}\Psi_{b}\right)\nonumber
\\
&&+4\Delta_{AMN\bot}\Delta_{CR\bot}
\frac{\Lambda}{2\pi^{2}v^{2}\sqrt{\eta}}\ell\mathcal{G}_{1}^{D}
\left(\bar{\Psi}_{a}i\gamma_{x}\gamma_{y}
\Psi_{a}\right)\left(\bar{\Psi}_{b}i\gamma_{x}\gamma_{y}\Psi_{b}\right)\nonumber
\\
&&+4\Delta_{AMN\bot}\Delta_{CR\bot}
\frac{\Lambda}{2\pi^{2}v^{2}\sqrt{\eta}}\ell\mathcal{G}_{1}^{m}
\left(\bar{\Psi}_{a}
\gamma_{0}\gamma_{z}\Psi_{a}\right)\left(\bar{\Psi}_{b}\gamma_{0}\gamma_{z}\Psi_{b}\right),
\\
W_{AMN\bot,CRz}^{(2)+(3)}
&=&2\Delta_{AMN\bot}\Delta_{CRz}
\frac{\Lambda}{2\pi^{2}v^{2}\sqrt{\eta}}\ell\mathcal{G}_{1}^{\bot}\left(\bar{\Psi}_{a}\gamma_{0}\Psi_{a}\right)\left(\bar{\Psi}_{b}\gamma_{0}\Psi_{b}\right)\nonumber
\\
&&+2\Delta_{AMN\bot}\Delta_{CRz}
\frac{\Lambda}{2\pi^{2}v^{2}\sqrt{\eta}}\ell\mathcal{G}_{1}^{D}
\left[\left(\bar{\Psi}_{a}i\gamma_{y}
\gamma_{z}\Psi_{a}\right)\left(\bar{\Psi}_{b}i\gamma_{y}\gamma_{z}\Psi_{b}\right)+\left(\bar{\Psi}_{a}i\gamma_{z}\gamma_{x}
\Psi_{a}\right)\left(\bar{\Psi}_{b}i\gamma_{z}\gamma_{x}\Psi_{b}\right)\right]\nonumber
\\
&&+2\Delta_{AMN\bot}\Delta_{CRz}
\frac{\Lambda}{2\pi^{2}v^{2}\sqrt{\eta}}\ell\mathcal{G}_{1}^{m}
\sum_{j=x,y}\left(\bar{\Psi}_{a}
\gamma_{0}\gamma_{j}\Psi_{a}\right)\left(\bar{\Psi}_{b}\gamma_{0}\gamma_{j}\Psi_{b}\right),
\\
W_{AMNz,CR\bot}^{(2)+(3)}
&=&2\Delta_{AMNz}\Delta_{CR\bot}\frac{\Lambda}{2\pi^{2}v^{2}\sqrt{\eta}}\ell\mathcal{G}_{1}^{\bot}\left(\bar{\Psi}_{a}\gamma_{0}\Psi_{a}\right)\left(\bar{\Psi}_{b}\gamma_{0}\Psi_{b}\right)\nonumber
\\
&&+2\Delta_{AMNz}\Delta_{CR\bot}
\frac{\Lambda}{2\pi^{2}v^{2}\sqrt{\eta}}\ell\mathcal{G}_{1}^{D}\left[\left(\bar{\Psi}_{a}i\gamma_{y}\gamma_{z}
\Psi_{a}\right)\left(\bar{\Psi}_{b}i\gamma_{y}\gamma_{z}\Psi_{b}\right)+\left(\bar{\Psi}_{a}i\gamma_{z}
\gamma_{x}\Psi_{a}\right)\left(\bar{\Psi}_{b}i\gamma_{z}\gamma_{x}\Psi_{b}\right)
\right]\nonumber
\\
&&+2\Delta_{AMNz}\Delta_{CR\bot}
\frac{\Lambda}{2\pi^{2}v^{2}\sqrt{\eta}}\ell\mathcal{G}_{1}^{m}\sum_{j=x,y}\left(\bar{\Psi}_{a}
\gamma_{0}\gamma_{j}\Psi_{a}\right)\left(\bar{\Psi}_{b}\gamma_{0}\gamma_{j}\Psi_{b}\right),
\\
W_{AMNz,CRz}^{(2)+(3)}
&=&0
\\
W_{CR\bot,CRz}^{(2)+(3)}
&=&2\Delta_{CR\bot}\Delta_{CRz}\frac{\Lambda}{2\pi^{2}v^{2}\sqrt{\eta}}\ell\mathcal{G}_{1}^{\bot}
\left(\bar{\Psi}_{a}
\gamma_{0}\gamma_{5}\Psi_{a}\right)\left(\bar{\Psi}_{b}\gamma_{0}\gamma_{5}\Psi_{b}\right).
\end{eqnarray}

$W^{(2)+(3)}$ can be written as
\begin{eqnarray}
W^{(2)+(3)}&=&\sum_{i}\sum_{i\le j}W_{i,j}^{(2)+(3)}\nonumber
\\
&=&\frac{\delta\Delta_{i}^{(2)+(3)}}{2}\left(\bar{\Psi}_{a}\Gamma_{i}\Psi_{a}\right)
\left(\bar{\Psi}_{b}\Gamma_{i}\Psi_{b}\right),
\end{eqnarray}
where
\begin{eqnarray}
\delta\Delta_{C}^{(2)+(3)}&=&4\left[\left(\Delta_{M}\Delta_{SO\bot}+\Delta_{PM}\Delta_{MN\bot}+\Delta_{AMN\bot}\Delta_{CRz}+\Delta_{AMNz}\Delta_{CR\bot}\right)\mathcal{G}_{1}^{\bot}
+\left(\Delta_{M}\Delta_{SOz}+\Delta_{PM}\Delta_{MNz}\right.\right.\nonumber
\\
&&\left.+2\Delta_{AMN\bot}\Delta_{CR\bot}\right)\mathcal{G}_{1}^{z}+\left(\Delta_{C}\Delta_{PM}+2\Delta_{SO\bot}\Delta_{AMN\bot}+\Delta_{SOz}\Delta_{AMNz}\right)\mathcal{G}_{1}^{D}\nonumber
\\
&&\left.+\left(\Delta_{C}\Delta_{M}+2\Delta_{MN\bot}\Delta_{AMN\bot}+\Delta_{MNz}\Delta_{AMNz}\right)\mathcal{G}_{1}^{m}\right]\frac{\Lambda}{2\pi^{2}v^{2}\sqrt{\eta}}\ell, \label{Eq:deltaDeltaC2and3}
\\
\delta\Delta_{M}^{(2)+(3)}&=&4\left[\left(\Delta_{C}\Delta_{SO\bot}+\Delta_{PM}\Delta_{AMN\bot}+\Delta_{MN\bot}\Delta_{CRz}+\Delta_{MNz}\Delta_{CR\bot}\right)\mathcal{G}_{1}^{\bot}
+\left(\Delta_{C}\Delta_{SOz}+\Delta_{PM}\Delta_{AMNz}\right.\right.\nonumber
\\
&&\left.+2\Delta_{MN\bot}\Delta_{CR\bot}\right)\mathcal{G}_{1}^{z}
+\left(\Delta_{M}\Delta_{PM}+2\Delta_{SO\bot}\Delta_{MN\bot}+\Delta_{SOz}\Delta_{MNz}\right)\mathcal{G}_{1}^{D}\nonumber
\\
&&+\left(\Delta_{C}^{2}+\Delta_{M}^{2}+\Delta_{AC}^{2}+2\Delta_{SO\bot}^{2}+\Delta_{SOz}^{2}+\Delta_{PM}^{2}
+2\Delta_{MN\bot}^{2}+\Delta_{MNz}^{2}+2\Delta_{AMN\bot}^{2}+\Delta_{AMNz}^{2}\right.\nonumber
\\
&&\left.\left.+2\Delta_{CR\bot}^{2}+\Delta_{CRz}^{2}\right)\mathcal{G}_{1}^{m}\right]
\frac{\Lambda}{2\pi^{2}v^{2}\sqrt{\eta}}\ell, \label{Eq:deltaDeltaM2and3}
\\
\delta\Delta_{AC}^{(2)+(3)}&=&4\left[\left(\Delta_{SO\bot}\Delta_{SOz}+\Delta_{MN\bot}\Delta_{MNz}+\Delta_{AMN\bot}\Delta_{AMNz}+\Delta_{CR\bot}\Delta_{CRz}\right)\mathcal{G}_{1}^{\bot}\right.\nonumber
\\
&&+2\left(\Delta_{SO\bot}^{2}+\Delta_{MN\bot}^{2}+\Delta_{AMN\bot}^{2}+\Delta_{CR\bot}^{2}\right)\mathcal{G}_{1}^{z}
+\left(\Delta_{AC}\Delta_{PM}+2\Delta_{SO\bot}\Delta_{CR\bot}+\Delta_{SOz}\Delta_{CRz}\right)\mathcal{G}_{1}^{D}\nonumber
\\
&&\left.+\left(\Delta_{M}\Delta_{AC}+\Delta_{MNz}\Delta_{CRz}+2\Delta_{MN\bot}\Delta_{CR\bot}\right)\mathcal{G}_{1}^{m}\right]\frac{\Lambda}{2\pi^{2}v^{2}\sqrt{\eta}}\ell, \label{Eq:deltaDeltaAC2and3}
\\
\delta\Delta_{SO\bot}^{(2)+(3)}&=&2\left[\left(\Delta_{C}\Delta_{M}+\Delta_{AC}\Delta_{SOz}+\Delta_{PM}\Delta_{CRz}+4\Delta_{MN\bot}\Delta_{AMN\bot}+\Delta_{MNz}\Delta_{AMNz}\right)\mathcal{G}_{1}^{\bot}
\right.\nonumber
\\
&&+2\left(\Delta_{AC}\Delta_{SO\bot}+\Delta_{PM}\Delta_{CR\bot}+\Delta_{MN\bot}\Delta_{AMNz}+\Delta_{MNz}\Delta_{AMN\bot}\right)\mathcal{G}_{1}^{z}\nonumber
\\
&&+2\left(\Delta_{C}\Delta_{AMN\bot}+\Delta_{M}\Delta_{MN\bot}+\Delta_{AC}\Delta_{CR\bot}+\Delta_{SO\bot}\Delta_{PM}\right)\mathcal{G}_{1}^{D}\nonumber
\\
&&\left.+2\left(\Delta_{M}\Delta_{SO\bot}+\Delta_{PM}\Delta_{MN\bot}+\Delta_{AMN\bot}\Delta_{CRz}+\Delta_{AMNz}\Delta_{CR\bot}\right)\mathcal{G}_{1}^{m}\right]
\frac{\Lambda}{2\pi^{2}v^{2}\sqrt{\eta}}\ell,  \label{Eq:deltaDeltaSOBot2and3}
\\
\delta\Delta_{SOz}^{(2)+(3)}&=&4\left[\left(\Delta_{AC}\Delta_{SO\bot}+\Delta_{PM}\Delta_{CR\bot}+\Delta_{MN\bot}\Delta_{AMNz}+\Delta_{MNz}\Delta_{AMN\bot}\right)\mathcal{G}_{1}^{\bot}\right.\nonumber
\\
&&+\left(\Delta_{C}\Delta_{M}+2\Delta_{MN\bot}\Delta_{AMN\bot}+\Delta_{MNz}\Delta_{AMNz}\right)\mathcal{G}_{1}^{z}\nonumber
\\
&&+\left(\Delta_{C}\Delta_{AMNz}+\Delta_{M}\Delta_{MNz}+\Delta_{AC}\Delta_{CRz}+\Delta_{SOz}\Delta_{PM}\right)\mathcal{G}_{1}^{D}\nonumber
\\
&&\left.+\left(\Delta_{M}\Delta_{SOz}+\Delta_{PM}\Delta_{MNz}+2\Delta_{AMN\bot}\Delta_{CR\bot}\right)\mathcal{G}_{1}^{m}\right]
\frac{\Lambda}{2\pi^{2}v^{2}\sqrt{\eta}}\ell, \label{Eq:deltaDeltaSOz2and3}
\\
\delta\Delta_{PM}^{(2)+(3)}&=&4\left[\left(\Delta_{C}\Delta_{MN\bot}+\Delta_{M}\Delta_{AMN\bot}+\Delta_{SO\bot}\Delta_{CRz}+\Delta_{SOz}\Delta_{CR\bot}\right)\mathcal{G}_{1}^{\bot}\right.\nonumber
\\
&&+\left(\Delta_{C}\Delta_{MNz}+\Delta_{M}\Delta_{AMNz}+2\Delta_{SO\bot}\Delta_{CR\bot}\right)\mathcal{G}_{1}^{z}+\left(\Delta_{C}^{2}+\Delta_{M}^{2}+\Delta_{AC}^{2}+2\Delta_{SO\bot}^{2}+\Delta_{SOz}^{2}\right.\nonumber
\\
&&\left.+\Delta_{PM}^{2}+2\Delta_{MN\bot}^{2}+\Delta_{MNZ}^{2}+2\Delta_{AMN\bot}^{2}+\Delta_{AMNz}^{2}+2\Delta_{CR\bot}^{2}+\Delta_{CRz}^{2}\right)\mathcal{G}_{1}^{D}\nonumber
\\
&&\left.+\left(\Delta_{M}\Delta_{PM}+2\Delta_{SO\bot}\Delta_{MN\bot}+\Delta_{SOz}\Delta_{MNz}\right)\mathcal{G}_{1}^{m}\right]
\frac{\Lambda}{2\pi^{2}v^{2}\sqrt{\eta}}\ell, \label{Eq:deltaDeltaPM2and3}
\\
\delta\Delta_{MN\bot}^{(2)+(3)}&=&2\left[\left(\Delta_{C}\Delta_{PM}+\Delta_{M}\Delta_{CRz}+\Delta_{AC}\Delta_{MNz}+4\Delta_{SO\bot}\Delta_{AMN\bot}+\Delta_{SOz}\Delta_{AMNz}\right)\mathcal{G}_{1}^{\bot}\right.\nonumber
\\
&&+2\left(\Delta_{M}\Delta_{CR\bot}+\Delta_{AC}\Delta_{MN\bot}+\Delta_{SO\bot}\Delta_{AMNz}+\Delta_{SOz}\Delta_{AMN\bot}\right)\mathcal{G}_{1}^{z}\nonumber
\\
&&+2\left(\Delta_{M}\Delta_{SO\bot}+\Delta_{PM}\Delta_{MN\bot}+\Delta_{AMN\bot}\Delta_{CRz}+\Delta_{AMNz}\Delta_{CR\bot}\right)\mathcal{G}_{1}^{D}\nonumber
\\
&&\left.+2\left(\Delta_{C}\Delta_{AMN\bot}+\Delta_{M}\Delta_{MN\bot}+\Delta_{AC}\Delta_{CR\bot}+\Delta_{SO\bot}\Delta_{PM}\right)\mathcal{G}_{1}^{m}\right]
\frac{\Lambda}{2\pi^{2}v^{2}\sqrt{\eta}}\ell,  \label{Eq:deltaDeltaMNBot2and3}
\\
\delta\Delta_{MNz}^{(2)+(3)}&=&4\left[\left(\Delta_{M}\Delta_{CR\bot}+\Delta_{AC}\Delta_{MN\bot}+\Delta_{SO\bot}\Delta_{AMNz}+\Delta_{SOz}\Delta_{AMN\bot}\right)\mathcal{G}_{1}^{\bot}\right.\nonumber
\\
&&+\left(\Delta_{C}\Delta_{PM}+2\Delta_{SO\bot}\Delta_{AMN\bot}+\Delta_{SOz}\Delta_{AMNz}\right)\mathcal{G}_{1}^{z}\nonumber
\\
&&+\left(\Delta_{M}\Delta_{SOz}+\Delta_{PM}\Delta_{MNz}+2\Delta_{AMN\bot}\Delta_{CR\bot}\right)\mathcal{G}_{1}^{D}\nonumber
\\
&&\left.+\left(\Delta_{C}\Delta_{AMNz}+\Delta_{M}\Delta_{MNz}+\Delta_{AC}\Delta_{CRz}+\Delta_{SOz}\Delta_{PM}\right)\mathcal{G}_{1}^{m}\right]
\frac{\Lambda}{2\pi^{2}v^{2}\sqrt{\eta}}\ell, \label{Eq:deltaDeltaMNz2and3}
\\
\delta\Delta_{AMN\bot}^{(2)+(3)}&=&2\left[\left(\Delta_{C}\Delta_{CRz}+\Delta_{M}\Delta_{PM}+\Delta_{AC}\Delta_{AMNz}+4\Delta_{SO\bot}\Delta_{MN\bot}+\Delta_{SOz}\Delta_{MNz}\right)\mathcal{G}_{1}^{\bot}\right.\nonumber
\\
&&+2\left(\Delta_{C}\Delta_{CR\bot}+\Delta_{AC}\Delta_{AMN\bot}+\Delta_{SO\bot}\Delta_{MNz}+\Delta_{SOz}\Delta_{MN\bot}\right)\mathcal{G}_{1}^{z}\nonumber
\\
&&+2\left(\Delta_{C}\Delta_{SO\bot}+\Delta_{PM}\Delta_{AMN\bot}+\Delta_{MN\bot}\Delta_{CRz}+\Delta_{MNz}\Delta_{CR\bot}\right)\mathcal{G}_{1}^{D}\nonumber
\\
&&\left.+2\left(\Delta_{C}\Delta_{MN\bot}+\Delta_{M}\Delta_{AMN\bot}+\Delta_{SO\bot}\Delta_{CRz}+\Delta_{SOz}\Delta_{CR\bot}\right)\mathcal{G}_{1}^{m}\right]
\frac{\Lambda}{2\pi^{2}v^{2}\sqrt{\eta}}\ell, \label{Eq:deltaDeltaAMNBot2and3}
\\
\delta\Delta_{AMNz}^{(2)+(3)}&=&4\left[\left(\Delta_{C}\Delta_{CR\bot}+\Delta_{AC}\Delta_{AMN\bot}+\Delta_{SO\bot}\Delta_{MNz}+\Delta_{SOz}\Delta_{MN\bot}\right)\mathcal{G}_{1}^{\bot}\right.\nonumber
\\
&&+\left(\Delta_{M}\Delta_{PM}+2\Delta_{SO\bot}\Delta_{MN\bot}+\Delta_{SOz}\Delta_{MNz}\right)\mathcal{G}_{1}^{z}\nonumber
\\
&&+\left(\Delta_{C}\Delta_{SOz}+\Delta_{PM}\Delta_{AMNz}+2\Delta_{MN\bot}\Delta_{CR\bot}\right)\mathcal{G}_{1}^{D}\nonumber
\\
&&\left.+\left(\Delta_{C}\Delta_{MNz}+\Delta_{M}\Delta_{AMNz}+2\Delta_{SO\bot}\Delta_{CR\bot}\right)\mathcal{G}_{1}^{m}\right]
\frac{\Lambda}{2\pi^{2}v^{2}\sqrt{\eta}}\ell,  \label{Eq:deltaDeltaAMNz2and3}
\\
\delta\Delta_{CR\bot}^{(2)+(3)}&=&2\left[\left(\Delta_{C}\Delta_{AMNz}+\Delta_{M}\Delta_{MNz}+\Delta_{AC}\Delta_{CRz}+\Delta_{SOz}\Delta_{PM}\right)\mathcal{G}_{1}^{\bot}\right.\nonumber
\\
&&+2\left(\Delta_{C}\Delta_{AMN\bot}+\Delta_{M}\Delta_{MN\bot}+\Delta_{AC}\Delta_{CR\bot}+\Delta_{SO\bot}\Delta_{PM}\right)\mathcal{G}_{1}^{z}\nonumber
\\
&&+2\left(\Delta_{AC}\Delta_{SO\bot}+\Delta_{PM}\Delta_{CR\bot}+\Delta_{MN\bot}\Delta_{AMNz}+\Delta_{MNz}\Delta_{AMN\bot}\right)\mathcal{G}_{1}^{D}\nonumber
\\
&&\left.+2\left(\Delta_{M}\Delta_{CR\bot}+\Delta_{AC}\Delta_{MN\bot}+\Delta_{SO\bot}\Delta_{AMNz}+\Delta_{SOz}\Delta_{AMN\bot}\right)\mathcal{G}_{1}^{m}\right]
\frac{\Lambda}{2\pi^{2}v^{2}\sqrt{\eta}}\ell, \label{Eq:deltaDeltaCRBot2and3}
\\
\delta\Delta_{CRz}^{(2)+(3)}&=&4\left[\left(\Delta_{C}\Delta_{AMN\bot}+\Delta_{M}\Delta_{MN\bot}+\Delta_{AC}\Delta_{CR\bot}+\Delta_{SO\bot}\Delta_{PM}\right)\mathcal{G}_{1}^{\bot}\right.\nonumber
\\
&&+\left(\Delta_{AC}\Delta_{SOz}+\Delta_{PM}\Delta_{CRz}+2\Delta_{MN\bot}\Delta_{AMN\bot}\right)\mathcal{G}_{1}^{D}\nonumber
\\
&&\left.+\left(\Delta_{M}\Delta_{CRz}+\Delta_{AC}\Delta_{MNz}+2\Delta_{SO\bot}\Delta_{AMN\bot}\right)\mathcal{G}_{1}^{m}\right]
\frac{\Lambda}{2\pi^{2}v^{2}\sqrt{\eta}}\ell. \label{Eq:deltaDeltaCRz2and3}
\end{eqnarray}

\end{widetext}

\subsection{Correction induced by long-range Coulomb interaction from Feynman diagram \ref{Fig:VertexCorrection}~(d)}

Substituting the expressions of fermion and boson propagators into Eq.~(\ref{Eq:VertexCorrectionDisW4i}), we  get
\begin{eqnarray}
W_{i}^{(4)}
&=&\Delta_{i}\left(\bar{\Psi}_{a}\Gamma_{i}\Psi_{a}\right)\left(\bar{\Psi}_{b}\Gamma_{i}\Psi_{b}\right)\alpha\frac{1}{2}\left(\mathcal{F}_{0}^{\bot}+\mathcal{F}_{0}^{z}\right)\ell\nonumber
\\
&&+\Delta_{i}\left(\bar{\Psi}_{a}\Gamma_{i}\Psi_{a}\right)\left(\bar{\Psi}_{b}\gamma_{0}\gamma_{x}\Gamma_{i}\gamma_{x}\gamma_{0}\Psi_{b}\right)
\alpha\mathcal{F}_{3}^{\bot}\ell\nonumber
\\
&&+\Delta_{i}\left(\bar{\Psi}_{a}\Gamma_{i}\Psi_{a}\right)\left(\bar{\Psi}_{b}\gamma_{0}\gamma_{y}
\Gamma_{i}\gamma_{y}\gamma_{0}\Psi_{b}\right)\alpha\mathcal{F}_{3}^{\bot}\ell\nonumber
\\
&&+\Delta_{i}\left(\bar{\Psi}_{a}\Gamma_{i}\Psi_{a}\right)\left(\bar{\Psi}_{b}\gamma_{0}\gamma_{z}
\Gamma_{i}\gamma_{z}\gamma_{0}\Psi_{b}\right)\alpha\mathcal{F}_{3}^{z}\ell
\nonumber
\\
&&+\Delta_{i}\left(\bar{\Psi}_{a}\Gamma_{i}\Psi_{a}\right)\left(\bar{\Psi}_{b}\gamma_{0}\gamma_{5}\Gamma_{i}\gamma_{5}\gamma_{0}\Psi_{b}\right)\alpha
\mathcal{F}_{3}^{D}\ell\nonumber
\\
&&-\Delta_{i}\left(\bar{\Psi}_{a}\Gamma_{i}\Psi_{a}\right)\left(\bar{\Psi}_{b}\gamma_{0}
\Gamma_{i}\gamma_{0}\Psi_{b}\right)\alpha\mathcal{F}_{3}^{m}\ell,
\end{eqnarray}
where
\begin{eqnarray}
\mathcal{F}_{3}^{\bot}&=&\frac{1}{16\pi}\int_{0}^{\pi}\sin(\varphi)d\varphi\int_{0}^{2\pi}d\theta\frac{\sin^{2}(\varphi)}{\Xi^{3}},
\\
\mathcal{F}_{3}^{z}&=&\frac{1}{8\pi}\int_{0}^{\pi}\sin(\varphi)d\varphi\int_{0}^{2\pi}d\theta\nonumber
\\
&&\times\frac{\left(\frac{v_{z}}{v\sqrt{\eta}}\right)^{2}\cos^{2}(\varphi)}{\Xi^{3}},
\\
\mathcal{F}_{3}^{D}&=&\frac{1}{8\pi}\int_{0}^{\pi}\sin(\varphi)d\varphi\int_{0}^{2\pi}d\theta\nonumber
\\
&&\times\frac{\frac{D^{2}\Lambda^{2}}{v^{2}}\sin^{2}(\varphi)\cos^{2}(2\theta)}
{\Xi^{3}},
\\
\mathcal{F}_{3}^{m}&=&\frac{1}{8\pi}\int_{0}^{\pi}\sin(\varphi)d\varphi\int_{0}^{2\pi}d\theta\nonumber
\\
&&\times\frac{\left(\frac{m}{v\Lambda}-\frac{B_{\bot}\Lambda}{v}\sin^{2}(\varphi)-\frac{B_{z}\Lambda}{v\eta}\cos^{2}(\varphi)\right)^{2}}{\Xi^{3}}.
\end{eqnarray}
$\Xi$ is given by Eq.~(\ref{Eq:XiExpression}).

$W_{i}^{(4)}$ can be also further written as
\begin{eqnarray}
W_{i}^{(4)}&=&\frac{\delta\Delta_{i}^{4}}{2}\left(\bar{\Psi}_{a}\Gamma_{i}\Psi_{a}\right)\left(\bar{\Psi}_{b}\Gamma_{i}\Psi_{b}\right),
\end{eqnarray}
where
\begin{eqnarray}
\delta\Delta_{C}^{(4)}
&=&2\Delta_{C}\alpha
\Big[\frac{1}{2}\left(\mathcal{F}_{0}^{\bot}+\mathcal{F}_{0}^{z}\right)-2\mathcal{F}_{3}^{\bot}-\mathcal{F}_{3}^{z}\nonumber
\\
&&-\mathcal{F}_{3}^{D}
-\mathcal{F}_{3}^{m}\Big]\ell,  \label{Eq:deltaDeltaC4}
\\
\delta\Delta_{M}^{(4)}
&=&2\Delta_{M}\alpha
\Big[\frac{1}{2}\left(\mathcal{F}_{0}^{\bot}+\mathcal{F}_{0}^{z}\right)+2\mathcal{F}_{3}^{\bot}+\mathcal{F}_{3}^{z}\nonumber
\\
&&+\mathcal{F}_{3}^{D}
-\mathcal{F}_{3}^{m}\Big]\ell,  \label{Eq:deltaDeltaM4}
\\
\delta\Delta_{AC}^{(4)}
&=&2\Delta_{AC}
\alpha\Big[\frac{1}{2}\left(\mathcal{F}_{0}^{\bot}+\mathcal{F}_{0}^{z}\right)-2\mathcal{F}_{3}^{\bot}-\mathcal{F}_{3}^{z}\nonumber
\\
&&+\mathcal{F}_{3}^{D}
+\mathcal{F}_{3}^{m}\Big]\ell,  \label{Eq:deltaDeltaAC4}
\\
\delta\Delta_{SO\bot}^{(4)}
&=&2\Delta_{SO\bot}
\alpha\Big[\frac{1}{2}\left(\mathcal{F}_{0}^{\bot}+\mathcal{F}_{0}^{z}\right)-\mathcal{F}_{3}^{z}-\mathcal{F}_{3}^{D}\nonumber
\\
&&+\mathcal{F}_{3}^{m}\Big]\ell, \label{Eq:deltaDeltaSOBot4}
\\
\delta\Delta_{SOz}^{(4)}
&=&2\Delta_{SOz}
\alpha\Big[\frac{1}{2}\left(\mathcal{F}_{0}^{\bot}+\mathcal{F}_{0}^{z}\right)-2\mathcal{F}_{3}^{\bot}+\mathcal{F}_{3}^{z}\nonumber
\\
&&-\mathcal{F}_{3}^{D}
+\mathcal{F}_{3}^{m}\Big]\ell,  \label{Eq:deltaDeltaSOz4}
\\
\delta\Delta_{PM}^{(4)}
&=&2\Delta_{PM}
\alpha\Big[\frac{1}{2}\left(\mathcal{F}_{0}^{\bot}+\mathcal{F}_{0}^{z}\right)+2\mathcal{F}_{3}^{\bot}
+\mathcal{F}_{3}^{z}\nonumber
\\
&&-\mathcal{F}_{3}^{D}+\mathcal{F}_{3}^{m}\Big]\ell, \label{Eq:deltaDeltaPM4}
\\
\delta\Delta_{MN\bot}^{(4)}
&=&2\Delta_{MN\bot}
\alpha\Big[\frac{1}{2}\left(\mathcal{F}_{0}^{\bot}+\mathcal{F}_{0}^{z}\right)-\mathcal{F}_{3}^{z}
+\mathcal{F}_{3}^{D}\nonumber
\\
&&-\mathcal{F}_{3}^{m}\Big]\ell,  \label{Eq:deltaDeltaMNBot4}
\\
\delta\Delta_{MNz}^{(4)}
&=&2\Delta_{MNz}\alpha
\Big[\frac{1}{2}\left(\mathcal{F}_{0}^{\bot}+\mathcal{F}_{0}^{z}\right)-2\mathcal{F}_{3}^{\bot}+\mathcal{F}_{3}^{z}\nonumber
\\
&&+\mathcal{F}_{3}^{D}-\mathcal{F}_{3}^{m}\Big]\ell, \label{Eq:deltaDeltaMNz4}
\\
\delta\Delta_{AMN\bot}^{(4)}
&=&2\Delta_{AMN\bot}\alpha
\Big[\frac{1}{2}\left(\mathcal{F}_{0}^{\bot}+\mathcal{F}_{0}^{z}\right)+\mathcal{F}_{3}^{z}-\mathcal{F}_{3}^{D}\nonumber
\\
&&-\mathcal{F}_{3}^{m}\Big]\ell,  \label{Eq:deltaDeltaAMNBot4}
\\
\delta\Delta_{AMNz}^{(4)}
&=&2\Delta_{AMNz}\alpha
\Big[\frac{1}{2}\left(\mathcal{F}_{0}^{\bot}+\mathcal{F}_{0}^{z}\right)+2\mathcal{F}_{3}^{\bot}-\mathcal{F}_{3}^{z}\nonumber
\\
&&-\mathcal{F}_{3}^{D}
-\mathcal{F}_{3}^{m}\Big]\ell,  \label{Eq:deltaDeltaAMNz4}
\\
\delta\Delta_{CR\bot}^{(4)}
&=&2\Delta_{CR\bot}\alpha
\Big[\frac{1}{2}\left(\mathcal{F}_{0}^{\bot}+\mathcal{F}_{0}^{z}\right)+\mathcal{F}_{3}^{z}+\mathcal{F}_{3}^{D}\nonumber
\\
&&+\mathcal{F}_{3}^{m}\Big]\ell, \label{Eq:deltaDeltaCRBot4}
\\
\delta\Delta_{CRz}^{(4)}
&=&2\Delta_{CRz}\alpha
\Big[\frac{1}{2}\left(\mathcal{F}_{0}^{\bot}+\mathcal{F}_{0}^{z}\right)+2\mathcal{F}_{3}^{\bot}-\mathcal{F}_{3}^{z}\nonumber
\\
&&+\mathcal{F}_{3}^{D}+\mathcal{F}_{3}^{m}\Big]\ell, \label{Eq:deltaDeltaCRz4}
\end{eqnarray}
It should be notice that
\begin{eqnarray}
2\mathcal{F}_{3}^{\bot}+\mathcal{F}_{3}^{z}+\mathcal{F}_{3}^{D}+\mathcal{F}_{3}^{m}&=&\frac{1}{2}\left(\mathcal{F}_{0}^{\bot}+\mathcal{F}_{0}^{z}\right).
\end{eqnarray}
Therefore,
\begin{eqnarray}
\delta\Delta_{C}^{(4)}=0.
\end{eqnarray}

\subsection{Correction induced by long-range Coulomb interaction from Feynman diagram \ref{Fig:VertexCorrection}~(e)}

Substituting Eqs.~(\ref{Eq:FermionPropagator}) and (\ref{Eq:BosonPropagator}) into Eq.~(\ref{Eq:VertexCorrectionDisW5i}), one can obtain
\begin{eqnarray}
W_{C}^{(5)}
&=&\Delta_{C}\left(\bar{\Psi}_{a}\gamma_{0}\Psi_{a}\right)\Pi(\mathbf{k})D_{0}(\mathbf{k})
\left(\bar{\Psi}_{b}\gamma_{0}\Psi_{b}\right)\nonumber
\\
&=&\Delta_{C}\left(-C_{\bot}k_{\bot}^{2}\ell-C_{z}\eta k_{z}^{2}\ell\right)
\frac{1}{k_{\bot}^{2}+\eta k_{z}^{2}}\nonumber
\\
&&\times\left(\bar{\Psi}_{a}\gamma_{0}\Psi_{a}\right)\left(\bar{\Psi}_{b}\gamma_{0}\Psi_{b}\right).
\end{eqnarray}
It can be approximated as
\begin{eqnarray}
W_{C}^{(5)}&\approx&-\Delta_{C}\frac{C_{\bot}+C_{z}}{2}\ell\left(\bar{\Psi}_{a}\gamma_{0}\Psi_{a}\right)
\left(\bar{\Psi}_{b}\gamma_{0}\Psi_{b}\right)\nonumber
\\
&=&-\Delta_{C}\alpha\frac{\mathcal{F}_{2}^{\bot}+\mathcal{F}_{2}^{z}}{2}\ell\nonumber
\\
&&\times\left(\bar{\Psi}_{a}\gamma_{0}\Psi_{a}\right)
\left(\bar{\Psi}_{b}\gamma_{0}\Psi_{b}\right).
\end{eqnarray}
It is easy to verify that
\begin{eqnarray}
W_{M}^{(5)}&=&0,
\\
W_{AC}^{(5)}&=&0,
\\
W_{SO\bot}^{(5)}&=&0,
\\
W_{SOz}^{(5)}&=&0,
\\
W_{PM}^{(5)}&=&0,
\\
W_{MN\bot}^{(5)}&=&0,
\\
W_{MNz}^{(5)}&=&0,
\\
W_{AMN\bot}^{(5)}&=&0,
\\
W_{AMNz}^{(5)}&=&0,
\\
W_{CR\bot}^{(5)}&=&0,
\\
W_{CRz}^{(5)}&=&0.
\end{eqnarray}

$W_{i}^{(5)}$ can be written as
\begin{eqnarray}
W_{i}^{(5)}&=&\frac{\delta\Delta_{i}^{(5)}}{2}\left(\bar{\Psi}_{a}\Gamma_{i}\Psi_{a}\right)
\left(\bar{\Psi}_{b}\Gamma_{i}\Psi_{b}\right),
\end{eqnarray}
where
\begin{eqnarray}
\delta\Delta_{C}^{(5)}&=&-\Delta_{C}\alpha\left(\mathcal{F}_{2}^{\bot}+\mathcal{F}_{2}^{z}\right)\ell, \label{Eq:deltaDeltaC5}
\\
\delta\Delta_{M}^{(5)}&=&0, \label{Eq:deltaDeltaM5}
\\
\delta\Delta_{AC}^{(5)}&=&0, \label{Eq:deltaDeltaAC5}
\\
\delta\Delta_{SO\bot}^{(5)}&=&0, \label{Eq:deltaDeltaSOBot5}
\\
\delta\Delta_{SOz}^{(5)}&=&0, \label{Eq:deltaDeltaSOz5}
\\
\delta\Delta_{PM}^{(5)}&=&0, \label{Eq:deltaDeltaPM5}
\\
\delta\Delta_{MN\bot}^{(5)}&=&0, \label{Eq:deltaDeltaMNBot5}
\\
\delta\Delta_{MNz}^{(5)}&=&0, \label{Eq:deltaDeltaMNz5}
\\
\delta\Delta_{AMN\bot}^{(5)}&=&0, \label{Eq:deltaDeltaAMNBot5}
\\
\delta\Delta_{AMNz}^{(5)}&=&0, \label{Eq:deltaDeltaAMNz5}
\\
\delta\Delta_{CR\bot}^{(5)}&=&0, \label{Eq:deltaDeltaCRBot5}
\\
\delta\Delta_{CRz}^{(5)}&=&0. \label{Eq:deltaDeltaCRz5}
\end{eqnarray}

\subsection{Summary of the corrections}

The total contribution is given by
\begin{eqnarray}
\delta\Delta_{i}&=&\delta\Delta_{i}^{(1)}
+\delta\Delta_{i}^{(2)+(3)}+\delta\Delta_{i}^{(4)}
+\delta\Delta_{i}^{(5)}. \label{Eq:deltaDeltaTotal}
\end{eqnarray}
The concrete expressions of $\Delta_{i}$ can be obtained through Eqs.~(\ref{Eq:deltaDeltaC1})-(\ref{Eq:deltaDeltaCRz1}),
Eqs.~(\ref{Eq:deltaDeltaC2and3})-(\ref{Eq:deltaDeltaCRz2and3}), (\ref{Eq:deltaDeltaC4})-(\ref{Eq:deltaDeltaCRz4}),
(\ref{Eq:deltaDeltaC5})-(\ref{Eq:deltaDeltaCRz5}) into Eq.~(\ref{Eq:deltaDeltaTotal}).

\section{Derivation of RG equations \label{Appendix:DerivationRGEs}}

The free action of fermions is
\begin{eqnarray}
S_{\Psi}&=&\int\frac{dk_{0}}{2\pi}\frac{d^{3}\mathbf{k}}{(2\pi)^{3}}\bar{\Psi}(k_{0},\mathbf{k})
\Big\{i\left[k_{0}\gamma_{0}+v\left(k_{x}\gamma_{x}+k_{y}\gamma_{y}\right)\right.\nonumber
\\
&&\left.+v_{z}k_{z}\gamma_{z}+D\left(k_{x}^{2}-k_{y}^{2}\right)\gamma_{5}\right]+m-B_{\bot}k_{\bot}^{2}\nonumber
\\
&&-B_{z}k_{z}^{2}
\Big\}\Psi(k_{0},\mathbf{k}).
\end{eqnarray}
Considering the corrections of fermion self-energies induced by Coulomb interaction and disorder, the action of fermions becomes
\begin{eqnarray}
S_{\Psi}&=&\int\frac{dk_{0}}{2\pi}\frac{d^{3}\mathbf{k}}{(2\pi)^{3}}\bar{\Psi}(k_{0},\mathbf{k})
\Big\{i\left[k_{0}\gamma_{0}+v\left(k_{x}\gamma_{x}+k_{y}\gamma_{y}\right)\right.\nonumber
\\
&&\left.+v_{z}k_{z}\gamma_{z}+D\left(k_{x}^{2}-k_{y}^{2}\right)\gamma_{5}\right]+m-B_{\bot}k_{\bot}^{2}\nonumber
\nonumber
\\
&&-B_{z}k_{z}^{2}-\Sigma(k_{0},\mathbf{k})-\Sigma_{dis}(k_{0})
\Big\}\Psi(k_{0},\mathbf{k})\nonumber
\\
&\approx&\int\frac{dk_{0}}{2\pi}\frac{d^{3}\mathbf{k}}{(2\pi)^{3}}\bar{\Psi}(k_{0},\mathbf{k})
\Big\{i\left[k_{0}\gamma_{0}e^{C_{0}^{dis}\ell}\right.\nonumber\nonumber
\\
&&+v\left(k_{x}\gamma_{x}+k_{y}\gamma_{y}\right)e^{C_{v}\ell}+
v_{z}k_{z}\gamma_{z}e^{C_{v_{z}}\ell}\nonumber
\\
&&\left.+D\left(k_{x}^{2}-k_{y}^{2}\right)\gamma_{5}e^{C_{D}\ell}\right]+me^{\left(C_{m}+C_{m}^{dis}\right)\ell}
\nonumber
\\
&&-B_{\bot}k_{\bot}^{2}e^{-C_{B_{\bot}}\ell}-B_{z}k_{z}^{2}e^{-C_{B_{z}}\ell}
\Big\}\Psi(k_{0},\mathbf{k}).
\end{eqnarray}
Employing the transformations
\begin{eqnarray}
k_{0}&=&k_{0}'e^{-\ell},\label{Eq:k0Scaling}
\\
k_{x}&=&k_{x}'e^{-\ell}, \label{Eq:kxScaling}
\\
k_{y}&=&k_{y}'e^{-\ell}, \label{Eq:kyScaling}
\\
k_{z}&=&k_{z}'e^{-\ell}, \label{Eq:kzScaling}
\\
\Psi&=&\Psi' e^{\frac{1}{2}\left(5-C_{0}^{dis}\right)\ell}, \label{Eq:PsiScaling}
\\
v&=&v'e^{-\left(C_{v}-C_{0}^{dis}\right)\ell}, \label{Eq:vScaling}
\\
v_{z}&=&v_{z}'e^{-\left(C_{v_{z}}-C_{0}^{dis}\right)\ell}, \label{Eq:vzScaling}
\\
m&=&m'e^{-\left(1+C_{m}+C_{m}^{dis}-C_{0}^{dis}\right)\ell}, \label{Eq:mScaling}
\\
B_{\bot}&=&B_{\bot}'e^{\left(1+C_{B_{\bot}}+C_{0}^{dis}\right)\ell}, \label{Eq:BbotScaling}
\\
B_{z}&=&B_{z}'e^{\left(1+C_{B_{z}}+C_{0}^{dis}\right)\ell}, \label{Eq:BzScaling}
\\
D&=&D'e^{\left(1-C_{D}+C_{0}^{dis}\right)\ell}, \label{Eq:DScaling}
\end{eqnarray}
the action of fermions can be further written as
\begin{eqnarray}
S_{\Psi'}&=&\int\frac{dk_{0}'}{2\pi}\frac{d^{3}\mathbf{k}'}{(2\pi)^{3}}\bar{\Psi}'(k_{0}',\mathbf{k}')
\Big\{i\left[k_{0}'\gamma_{0}+v'\left(k_{x}'\gamma_{x}\right.\right.\nonumber\nonumber
\\
&&\left.\left.+k_{y}'\gamma_{y}\right)+
v_{z}'k_{z}'\gamma_{z}e^{C_{v_{z}}\ell}+D'\left(k_{x}'^{2}-k_{y}'^{2}\right)\gamma_{5}\right]
\nonumber
\\
&&+m'-B_{\bot}'k_{\bot}'^{2}-B_{z}'k_{z}'^{2}
\Big\}\Psi'(k_{0}',\mathbf{k}'),
\end{eqnarray}
which recovers the original form of fermion action.

The free action of boson $\phi$ is
\begin{eqnarray}
S_{\phi}=\int\frac{dk_{0}}{2\pi}\frac{d^3\mathbf{k}}{(2\pi)^{3}}\phi(k_{0},\mathbf{k})
\left(k_{\bot}^{2}+\eta k_{z}^{2}
\right)\phi(k_{0},\mathbf{k}).
\end{eqnarray}
Including the correction to self-energy of boson, the action of $\phi$ becomes
\begin{eqnarray}
S_{\phi}&=&\int\frac{dk_{0}}{2\pi}\frac{d^3\mathbf{k}}{(2\pi)^{3}}\phi(k_{0},\mathbf{k})
\left(k_{\bot}^{2}+\eta k_{z}^{2}-\Pi(0,\mathbf{k})
\right)\nonumber
\\
&&\times\phi(k_{0},\mathbf{k})\nonumber
\\
&\approx&\int\frac{dk_{0}}{2\pi}\frac{d^3\mathbf{k}}{(2\pi)^{3}}\phi(k_{0},\mathbf{k})
\left(k_{\bot}^{2}e^{C_{\bot}\ell}+\eta k_{z}^{2}e^{C_{z}\ell}
\right)\nonumber
\\
&&\times\phi(k_{0},\mathbf{k}).
\end{eqnarray}
Using the transformations as shown in Eqs.~(\ref{Eq:k0Scaling})-(\ref{Eq:kzScaling}), and
\begin{eqnarray}
\phi&=&\phi'e^{\left(3-\frac{C_{\bot}}{2}\right)\ell}, \label{Eq:phiScaling}
\\
\eta &=&\eta'e^{\left(C_{\bot}-C_{z}\right)\ell}, \label{Eq:etaScaling}
\end{eqnarray}
the action of boson can be written as
\begin{eqnarray}
S_{\phi'}&=&\int\frac{dk_{0}'}{2\pi}\frac{d^3\mathbf{k}'}{(2\pi)^{3}}\phi'(k_{0}',\mathbf{k}')
\left(k_{\bot}'^{2}+\eta' k_{z}'^{2}
\right)\nonumber
\\
&&\times\phi(k_{0}',\mathbf{k}'),
\end{eqnarray}
which has the same form as the original boson action.

The action of fermion-boson coupling is
\begin{eqnarray}
S_{\Psi\phi}&=&ig\int\frac{dk_{0,1}}{2\pi}\frac{d^3\mathbf{k}_{1}}{(2\pi)^{3}}\frac{dk_{0,2}}{2\pi}
\frac{d^3\mathbf{k}_{2}}{(2\pi)^{3}}\bar{\Psi}(k_{0,1},\mathbf{k}_{1})\nonumber
\\
&&\times\gamma_{0}\Psi(k_{0,2},\mathbf{k}_{2})\phi(k_{0,1}-k_{0,2},\mathbf{k}_{1}
-\mathbf{k}_{2}).
\end{eqnarray}
Considering the vertex correction, the action has the from
\begin{eqnarray}
S_{\Psi\phi}&=&i\left(g+\delta g\right)\int\frac{dk_{0,1}}{2\pi}\frac{d^3\mathbf{k}_{1}}{(2\pi)^{3}}\frac{dk_{0,2}}{2\pi}
\frac{d^3\mathbf{k}_{2}}{(2\pi)^{3}}\bar{\Psi}(k_{0,1},\mathbf{k}_{1})\nonumber
\\
&&\times\gamma_{0}\Psi(k_{0,2},\mathbf{k}_{2})\phi(k_{0,1}-k_{0,2},\mathbf{k}_{1}
-\mathbf{k}_{2})\nonumber
\\
&\approx&ige^{C_{0}^{dis}\ell}\int\frac{dk_{0,1}}{2\pi}\frac{d^3\mathbf{k}_{1}}{(2\pi)^{3}}\frac{dk_{0,2}}{2\pi}
\frac{d^3\mathbf{k}_{2}}{(2\pi)^{3}}\bar{\Psi}(k_{0,1},\mathbf{k}_{1})\nonumber
\\
&&\times\gamma_{0}\Psi(k_{0,2},\mathbf{k}_{2})\phi(k_{0,1}-k_{0,2},\mathbf{k}_{1}
-\mathbf{k}_{2}).
\end{eqnarray}
 Applying the transformations as shown in Eqs.~(\ref{Eq:k0Scaling})-(\ref{Eq:PsiScaling}), (\ref{Eq:phiScaling}), and
\begin{eqnarray}
g=g'e^{\frac{C_{\bot}}{2}\ell}, \label{Eq:gScaling}
\end{eqnarray}
the action can be written as
\begin{eqnarray}
S_{\Psi'\phi'}
&=&ig'\int\frac{dk_{0,1}'}{2\pi}\frac{d^3\mathbf{k}_{1}'}{(2\pi)^{3}}\frac{dk_{0,2}'}{2\pi}
\frac{d^3\mathbf{k}_{2}'}{(2\pi)^{3}}
\bar{\Psi}'(k_{0,1}',\mathbf{k}_{1}')\nonumber
\\
&&\times\gamma_{0}\Psi'(k_{0,2}',\mathbf{k}_{2}')\phi(k_{0,1}'-k_{0,2}',\mathbf{k}_{1}'
-\mathbf{k}_{2}'),
\end{eqnarray}
which recovers the original form of the action of fermion-boson coupling.

The action of fermion-disorder coupling is given by
\begin{eqnarray}
S_{dis}&=&-\sum_{j=0}^{3}\frac{\Delta_{j}}{2} \int\frac{dk_{0,1}
dk_{0,2}d^{3}\mathbf{k}_{1} d^{3}\mathbf{k}_{2}
d^{3}\mathbf{k}_{3}}{(2\pi)^{11}}\nonumber
\\
&&\times\bar{\Psi}_{a}(k_{0,1},\mathbf{k}_{1})\Gamma_{j}
\Psi_{a}(k_{0,1},\mathbf{k}_{2})
\bar{\Psi}_{b}(k_{0,2},\mathbf{k}_{3})\Gamma_{j}\nonumber
\\
&&\times\Psi_{b}(k_{0,2},\mathbf{k}_{1}+\mathbf{k}_{2}+\mathbf{k}_{3}).
\end{eqnarray}
Including the corrections to one-loop order, the action of fermion-disorder coupling becomes
\begin{eqnarray}
S_{dis}&=&-\sum_{j=0}^{3}\frac{\left(\Delta_{j}+\delta\Delta_{j}\right)}{2} \int\frac{dk_{0,1}
dk_{0,2}d^{3}\mathbf{k}_{1} d^{3}\mathbf{k}_{2}
d^{3}\mathbf{k}_{3}}{(2\pi)^{11}}\nonumber
\\
&&\times\bar{\Psi}_{a}(k_{0,1},\mathbf{k}_{1})\Gamma_{j}
\Psi_{a}(k_{0,1},\mathbf{k}_{2})
\bar{\Psi}_{b}(k_{0,2},\mathbf{k}_{3})\Gamma_{j}\nonumber
\\
&&\times\Psi_{b}(k_{0,2},\mathbf{k}_{1}+\mathbf{k}_{2}+\mathbf{k}_{3}).
\end{eqnarray}
Using the transformations as shown in Eqs.~(\ref{Eq:k0Scaling})-(\ref{Eq:PsiScaling}), we obtain
\begin{eqnarray}
S_{dis}
&=&-\sum_{j=0}^{3}\frac{\left(\Delta_{j}+\delta\Delta_{j}\right)}{2}e^{\left(-1-2C_{0}^{dis}\right)\ell}\nonumber
\\
&&\times\int\frac{dk_{0,1}'
dk_{0,2}'d^{3}\mathbf{k}_{1}' d^{3}\mathbf{k}_{2}'
d^{3}\mathbf{k}_{3}'}{(2\pi)^{11}}
\bar{\Psi}_{a}'(k_{0,1}',\mathbf{k}_{1}')\Gamma_{j}\nonumber
\\
&&\times\Psi_{a}'(k_{0,1}',\mathbf{k}_{2}')
\bar{\Psi}_{b}'(k_{0,2}',\mathbf{k}_{3}')\Gamma_{j}\nonumber
\\
&&\times\Psi_{b}'(k_{0,2}',\mathbf{k}_{1}'+\mathbf{k}_{2}'+\mathbf{k}_{3}')\nonumber
\\
&\approx&-\sum_{j=0}^{3}\frac{\left[\Delta_{j}+\Delta_{j}\left(-1-2C_{0}^{dis}\right)\ell+\delta\Delta_{j}\right]}{2}\nonumber
\\
&&\times\int\frac{dk_{0,1}'
dk_{0,2}'d^{3}\mathbf{k}_{1}' d^{3}\mathbf{k}_{2}'
d^{3}\mathbf{k}_{3}'}{(2\pi)^{11}}
\bar{\Psi}_{a}'(k_{0,1}',\mathbf{k}_{1}')\Gamma_{j}\nonumber
\\
&&\times\Psi_{a}'(k_{0,1}',\mathbf{k}_{2}')
\bar{\Psi}_{b}'(k_{0,2}',\mathbf{k}_{3}')\Gamma_{j}\nonumber
\\
&&\times\Psi_{b}'(k_{0,2}',\mathbf{k}_{1}'+\mathbf{k}_{2}'+\mathbf{k}_{3}').
\end{eqnarray}
Adopting the transformation
\begin{eqnarray}
\Delta_{j}'&=&\Delta_{j}+\Delta_{j}\left(-1-2C_{0}^{dis}\right)\ell+\delta\Delta_{j}, \label{Eq:DeltajScaling}
\end{eqnarray}
the action for fermion-disorder coupling can be written as
\begin{eqnarray}
S_{dis}
&&-\sum_{j=0}^{3}\frac{\Delta_{j}'}{2}\int\frac{dk_{0,1}'
dk_{0,2}'d^{3}\mathbf{k}_{1}' d^{3}\mathbf{k}_{2}'
d^{3}\mathbf{k}_{3}'}{(2\pi)^{11}}
\bar{\Psi}_{a}'(k_{0,1}',\mathbf{k}_{1}')\nonumber
\\
&&\times\Gamma_{j}
\Psi_{a}'(k_{0,1}',\mathbf{k}_{2}')
\bar{\Psi}_{b}'(k_{0,2}',\mathbf{k}_{3}')\Gamma_{j}\nonumber
\\
&&\times\Psi_{b}'(k_{0,2}',\mathbf{k}_{1}'+\mathbf{k}_{2}'+\mathbf{k}_{3}'),
\end{eqnarray}
which recovers the original from of the action of fermion-disorder coupling.

From Eqs.~(\ref{Eq:vScaling})-(\ref{Eq:DScaling}), (\ref{Eq:etaScaling}), (\ref{Eq:gScaling}), and (\ref{Eq:DeltajScaling}), we obtain the RG equations
\begin{eqnarray}
\frac{dv}{d\ell}&=&\left(C_{v}-C_{0}^{dis}\right)v,
\\
\frac{dv_{z}}{d\ell}&=&\left(C_{v_{z}}-C_{0}^{dis}\right)v_{z},
\\
\frac{dm}{d\ell}&=&\left(1+C_{m}+C_{m}^{dis}-C_{0}^{dis}\right)m,
\\
\frac{dB_{\bot}}{d\ell}&=&\left(-1-C_{B_{\bot}}-C_{0}^{dis}\right)B_{\bot},
\\
\frac{dB_{z}}{d\ell}&=&\left(-1-C_{B_{z}}-C_{0}^{dis}\right)B_{z},
\\
\frac{dD}{d\ell}&=&\left(-1+C_{D}-C_{0}^{dis}\right)D,
\\
\frac{d\eta}{d\ell}&=&\left(C_{z}-C_{\bot}\right)\eta,
\\
\frac{dg}{d\ell}&=&-\frac{C_{\bot}}{2}g,
\\
\frac{d\alpha}{d\ell}
&=&-\alpha\left[C_{v}-C_{0}^{dis}
+\frac{1}{2}\left(C_{\bot}+C_{z}\right)\right],
\\
\frac{d\zeta}{d\ell}
&=&\left[C_{v_{z}}-C_{v}
-\frac{1}{2}\left(C_{z}-C_{\bot}\right)\right]\zeta,
\\
\frac{d\Delta_{j}}{d\ell}&=&\Delta_{j}\left(-1-2C_{0}^{dis}\right)+\frac{d\delta\Delta_{j}}{d\ell},
\end{eqnarray}
where $\zeta$ is defined as
\begin{eqnarray}
\zeta&=&\frac{v_{z}}{v\sqrt{\eta}}.
\end{eqnarray}

Adopting the transformations
\begin{eqnarray}
\frac{m}{v\Lambda}\rightarrow m,
\\
\frac{B_{\bot}\Lambda}{v}\rightarrow B_{\bot},
\\
\frac{B_{z}\Lambda}{v\eta}\rightarrow B_{z},
\\
\frac{D\Lambda}{v}\rightarrow D,
\\
\Delta_{j}\frac{\Lambda}{2\pi^{2} v^{2}\sqrt{\eta}}\rightarrow\Delta_{j},
\end{eqnarray}
the RG equations can be further compactly written as
\begin{eqnarray}
\frac{dv}{d\ell}&=&\left(\alpha\mathcal{R}_{v}-C_{0}^{dis}\right)v,
\\
\frac{dv_{z}}{d\ell}&=&\left(\alpha\mathcal{R}_{v_{z}}-C_{0}^{dis}\right)v_{z},
\\
\frac{dm}{d\ell}
&=&m+\alpha\mathcal{R}_{m}+C_{dis}^{m}m,
\\
\frac{dB_{\bot}}{d\ell}
&=&-B_{\bot}+\alpha\mathcal{R}_{B_{\bot}}
\\
\frac{dB_{z}}{d\ell}
&=&-B_{z}+\alpha\mathcal{R}_{B_{z}}
\\
\frac{dD}{d\ell}
&=&-D+\alpha \mathcal{R}_{D}D,
\\
\frac{d\eta}{d\ell}
&=&\alpha\mathcal{R}_{\eta}\eta,
\\
\frac{dg}{d\ell}&=&-\alpha \mathcal{R}_{g}g,
\\
\frac{d\alpha}{d\ell}
&=&-\alpha^{2}\mathcal{R}_{\alpha}+\alpha C_{0}^{dis},
\\
\frac{d\zeta}{d\ell}
&=&\alpha\mathcal{R}_{\gamma}\zeta,
\end{eqnarray}
\begin{widetext}
\begin{eqnarray}
\frac{d\Delta_{C}}{d\ell}
&=&-\Delta_{C}+2\Delta_{C}\left(\Delta_{C}+\Delta_{M}+\Delta_{AC}+2\Delta_{SO\bot}
+\Delta_{SOz}+\Delta_{PM}+2\Delta_{MN\bot}+\Delta_{MNz}+2\Delta_{AMN\bot}\right.\nonumber
\\
&&\left.
+\Delta_{AMNz}+2\Delta_{CR\bot}+\Delta_{CRz}\right)\left(\mathcal{G}_{1}^{\bot}
+\mathcal{G}_{1}^{z}+\mathcal{G}_{1}^{D}+\mathcal{G}_{1}^{m}\right)\nonumber
\\
&&+4\left[\left(\Delta_{M}\Delta_{SO\bot}+\Delta_{PM}\Delta_{MN\bot}+\Delta_{AMN\bot}\Delta_{CRz}+\Delta_{AMNz}\Delta_{CR\bot}\right)\mathcal{G}_{1}^{\bot}
+\left(\Delta_{M}\Delta_{SOz}+\Delta_{PM}\Delta_{MNz}\right.\right.\nonumber
\\
&&\left.+2\Delta_{AMN\bot}\Delta_{CR\bot}\right)\mathcal{G}_{1}^{z}+\left(\Delta_{C}\Delta_{PM}+2\Delta_{SO\bot}\Delta_{AMN\bot}+\Delta_{SOz}\Delta_{AMNz}\right)\mathcal{G}_{1}^{D}\nonumber
\\
&&\left.+\left(\Delta_{C}\Delta_{M}+2\Delta_{MN\bot}\Delta_{AMN\bot}+\Delta_{MNz}\Delta_{AMNz}\right)\mathcal{G}_{1}^{m}\right]\nonumber
\\
&&-\Delta_{C}\alpha\left(2\mathcal{F}_{0}^{\bot}+\frac{1}{2}\mathcal{F}_{2}^{\bot}+\frac{3}{2}\mathcal{F}_{2}^{z}\right),
\\
\frac{d\Delta_{M}}{d\ell}
&=&-\Delta_{M}+2\Delta_{M}\left(\Delta_{C}
+\Delta_{M}-\Delta_{AC}-2\Delta_{SO\bot}-\Delta_{SOz}-\Delta_{PM}+2\Delta_{MN\bot}+\Delta_{MNz}+2\Delta_{AMN\bot}\right.\nonumber
\\
&&\left.+\Delta_{AMNz}-2\Delta_{CR\bot}-2C_{CRz}\right)
\left(-\mathcal{G}_{1}^{\bot}
-\mathcal{G}_{1}^{z}-\mathcal{G}_{1}^{D}+\mathcal{G}_{1}^{m}\right)\nonumber
\\
&&+4\left[\left(\Delta_{C}\Delta_{SO\bot}+\Delta_{PM}\Delta_{AMN\bot}+\Delta_{MN\bot}\Delta_{CRz}+\Delta_{MNz}\Delta_{CR\bot}\right)\mathcal{G}_{1}^{\bot}
+\left(\Delta_{C}\Delta_{SOz}+\Delta_{PM}\Delta_{AMNz}\right.\right.\nonumber
\\
&&\left.+2\Delta_{MN\bot}\Delta_{CR\bot}\right)\mathcal{G}_{1}^{z}
+\left(\Delta_{M}\Delta_{PM}+2\Delta_{SO\bot}\Delta_{MN\bot}+\Delta_{SOz}\Delta_{MNz}\right)\mathcal{G}_{1}^{D}\nonumber
\\
&&+\left(\Delta_{C}^{2}+\Delta_{M}^{2}+\Delta_{AC}^{2}+2\Delta_{SO\bot}^{2}+\Delta_{SOz}^{2}+\Delta_{PM}^{2}
+2\Delta_{MN\bot}^{2}+\Delta_{MNz}^{2}+2\Delta_{AMN\bot}^{2}+\Delta_{AMNz}^{2}\right.\nonumber
\\
&&\left.\left.+2\Delta_{CR\bot}^{2}+\Delta_{CRz}^{2}\right)\mathcal{G}_{1}^{m}\right]
\nonumber
\\
&&+2\Delta_{M}\alpha
\left[\frac{1}{2}\left(-\mathcal{F}_{0}^{\bot}+\mathcal{F}_{0}^{z}\right)+\frac{1}{4}\mathcal{F}_{2}^{\bot}
-\frac{1}{4}\mathcal{F}_{2}^{z}+2\mathcal{F}_{3}^{\bot}+\mathcal{F}_{3}^{z}+\mathcal{F}_{3}^{D}
-\mathcal{F}_{3}^{m}\right],
\\
\frac{d\Delta_{AC}}{d\ell}&=&-\Delta_{AC}+2\Delta_{AC}\left(-\Delta_{C}+\Delta_{M}-\Delta_{AC}+2\Delta_{SO\bot}+\Delta_{SOz}+\Delta_{PM}+2\Delta_{MN\bot}+\Delta_{MNz}
\right.\nonumber
\\
&&\left.-2\Delta_{AMN\bot}-\Delta_{AMNz}-2\Delta_{CR\bot}-\Delta_{CRz}\right)
\left(-\mathcal{G}_{1}^{\bot}
-\mathcal{G}_{1}^{z}+\mathcal{G}_{1}^{D}+\mathcal{G}_{1}^{m}\right)\nonumber
\\
&&+4\left[\left(\Delta_{SO\bot}\Delta_{SOz}+\Delta_{MN\bot}\Delta_{MNz}+\Delta_{AMN\bot}\Delta_{AMNz}+\Delta_{CR\bot}\Delta_{CRz}\right)\mathcal{G}_{1}^{\bot}\right.\nonumber
\\
&&+2\left(\Delta_{SO\bot}^{2}+\Delta_{MN\bot}^{2}+\Delta_{AMN\bot}^{2}+\Delta_{CR\bot}^{2}\right)\mathcal{G}_{1}^{z}
+\left(\Delta_{AC}\Delta_{PM}+2\Delta_{SO\bot}\Delta_{CR\bot}+\Delta_{SOz}\Delta_{CRz}\right)\mathcal{G}_{1}^{D}\nonumber
\\
&&\left.+\left(\Delta_{M}\Delta_{AC}+\Delta_{MNz}\Delta_{CRz}+2\Delta_{MN\bot}\Delta_{CR\bot}\right)\mathcal{G}_{1}^{m}\right]\nonumber
\\
&&+2\Delta_{AC}
\alpha\left[\frac{1}{2}\left(-\mathcal{F}_{0}^{\bot}+\mathcal{F}_{0}^{z}\right)+\frac{1}{4}\mathcal{F}_{2}^{\bot}
-\frac{1}{4}\mathcal{F}_{2}^{z}-2\mathcal{F}_{3}^{\bot}-\mathcal{F}_{3}^{z}+\mathcal{F}_{3}^{D}
+\mathcal{F}_{3}^{m}\right],
\\
\frac{d\Delta_{SO\bot}}{d\ell}&=&-\Delta_{SO\bot}+2\Delta_{SO\bot}\left(-\Delta_{C}
+\Delta_{M}+\Delta_{AC}
+\Delta_{SOz}-\Delta_{PM}-\Delta_{MNz}+\Delta_{AMNz}-\Delta_{CRz}\right)
\nonumber
\\
&&\times\left(
-\mathcal{G}_{1}^{z}
-\mathcal{G}_{1}^{D}+\mathcal{G}_{1}^{m}\right)\nonumber
\\
&&+2\left[\left(\Delta_{C}\Delta_{M}+\Delta_{AC}\Delta_{SOz}+\Delta_{PM}\Delta_{CRz}+4\Delta_{MN\bot}\Delta_{AMN\bot}+\Delta_{MNz}\Delta_{AMNz}\right)\mathcal{G}_{1}^{\bot}
\right.\nonumber
\\
&&+2\left(\Delta_{AC}\Delta_{SO\bot}+\Delta_{PM}\Delta_{CR\bot}+\Delta_{MN\bot}\Delta_{AMNz}+\Delta_{MNz}\Delta_{AMN\bot}\right)\mathcal{G}_{1}^{z}\nonumber
\\
&&+2\left(\Delta_{C}\Delta_{AMN\bot}+\Delta_{M}\Delta_{MN\bot}+\Delta_{AC}\Delta_{CR\bot}+\Delta_{SO\bot}\Delta_{PM}\right)\mathcal{G}_{1}^{D}\nonumber
\\
&&\left.+2\left(\Delta_{M}\Delta_{SO\bot}+\Delta_{PM}\Delta_{MN\bot}+\Delta_{AMN\bot}\Delta_{CRz}+\Delta_{AMNz}\Delta_{CR\bot}\right)\mathcal{G}_{1}^{m}\right]
\nonumber
\\
&&+2\Delta_{SO\bot}
\alpha\left[\frac{1}{2}\left(-\mathcal{F}_{0}^{\bot}+\mathcal{F}_{0}^{z}\right)+\frac{1}{4}\mathcal{F}_{2}^{\bot}
-\frac{1}{4}\mathcal{F}_{2}^{z}-\mathcal{F}_{3}^{z}-\mathcal{F}_{3}^{D}+\mathcal{F}_{3}^{m}\right],
\\
\frac{d\Delta_{SOz}}{d\ell}&=&-\Delta_{SOz}+2\Delta_{SOz}\left(-\Delta_{C}
+\Delta_{M}+\Delta_{AC}+2\Delta_{SO\bot}-\Delta_{SOz}-\Delta_{PM}
-2\Delta_{MN\bot}+\Delta_{MNz}\right.\nonumber
\\
&&\left.+2\Delta_{AMN\bot}
-\Delta_{AMNz}-2\Delta_{CR\bot}+\Delta_{CRz}
\right)\left(-\mathcal{G}_{1}^{\bot}
+\mathcal{G}_{1}^{z}-\mathcal{G}_{1}^{D}
+\mathcal{G}_{1}^{m}\right)\nonumber
\\
&&+4\left[\left(\Delta_{AC}\Delta_{SO\bot}+\Delta_{PM}\Delta_{CR\bot}+\Delta_{MN\bot}\Delta_{AMNz}+\Delta_{MNz}\Delta_{AMN\bot}\right)\mathcal{G}_{1}^{\bot}\right.\nonumber
\\
&&+\left(\Delta_{C}\Delta_{M}+2\Delta_{MN\bot}\Delta_{AMN\bot}+\Delta_{MNz}\Delta_{AMNz}\right)\mathcal{G}_{1}^{z}\nonumber
\\
&&+\left(\Delta_{C}\Delta_{AMNz}+\Delta_{M}\Delta_{MNz}+\Delta_{AC}\Delta_{CRz}+\Delta_{SOz}\Delta_{PM}\right)\mathcal{G}_{1}^{D}\nonumber
\\
&&\left.+\left(\Delta_{M}\Delta_{SOz}+\Delta_{PM}\Delta_{MNz}+2\Delta_{AMN\bot}\Delta_{CR\bot}\right)\mathcal{G}_{1}^{m}\right]
\nonumber
\\
&&+2\Delta_{SOz}
\alpha\left[\frac{1}{2}\left(-\mathcal{F}_{0}^{\bot}+\mathcal{F}_{0}^{z}\right)+\frac{1}{4}\mathcal{F}_{2}^{\bot}
-\frac{1}{4}\mathcal{F}_{2}^{z}-2\mathcal{F}_{3}^{\bot}+\mathcal{F}_{3}^{z}-\mathcal{F}_{3}^{D}
+\mathcal{F}_{3}^{m}\right],
\\
\frac{d\Delta_{PM}}{d\ell}&=&-\Delta_{PM}+2\Delta_{PM}\left(-\Delta_{C}
+\Delta_{M}+\Delta_{AC}-2\Delta_{SO\bot}-\Delta_{SOz}-\Delta_{PM}+2\Delta_{MN\bot}+\Delta_{MNz}
\right.\nonumber
\\
&&\left.-2\Delta_{AMN\bot}-\Delta_{AMNz}+2\Delta_{CR\bot}+\Delta_{CRz}\right)\left(\mathcal{G}_{1}^{\bot}
+\mathcal{G}_{1}^{z}-\mathcal{G}_{1}^{D}+\mathcal{G}_{1}^{m}\right)\nonumber
\\
&&+4\left[\left(\Delta_{C}\Delta_{MN\bot}+\Delta_{M}\Delta_{AMN\bot}+\Delta_{SO\bot}\Delta_{CRz}+\Delta_{SOz}\Delta_{CR\bot}\right)\mathcal{G}_{1}^{\bot}\right.\nonumber
\\
&&+\left(\Delta_{C}\Delta_{MNz}+\Delta_{M}\Delta_{AMNz}+2\Delta_{SO\bot}\Delta_{CR\bot}\right)\mathcal{G}_{1}^{z}+\left(\Delta_{C}^{2}+\Delta_{M}^{2}+\Delta_{AC}^{2}+2\Delta_{SO\bot}^{2}+\Delta_{SOz}^{2}\right.\nonumber
\\
&&\left.+\Delta_{PM}^{2}+2\Delta_{MN\bot}^{2}+\Delta_{MNZ}^{2}+2\Delta_{AMN\bot}^{2}+\Delta_{AMNz}^{2}+2\Delta_{CR\bot}^{2}+\Delta_{CRz}^{2}\right)\mathcal{G}_{1}^{D}\nonumber
\\
&&\left.+\left(\Delta_{M}\Delta_{PM}+2\Delta_{SO\bot}\Delta_{MN\bot}+\Delta_{SOz}\Delta_{MNz}\right)\mathcal{G}_{1}^{m}\right]
\nonumber
\\
&&+2\Delta_{PM}
\alpha\left[\frac{1}{2}\left(-\mathcal{F}_{0}^{\bot}+\mathcal{F}_{0}^{z}\right)+\frac{1}{4}\mathcal{F}_{2}^{\bot}
-\frac{1}{4}\mathcal{F}_{2}^{z}+2\mathcal{F}_{3}^{\bot}
+\mathcal{F}_{3}^{z}-\mathcal{F}_{3}^{D}+\mathcal{F}_{3}^{m}\right],
\\
\frac{d\Delta_{MN\bot}}{d\ell}&=&-\Delta_{MN\bot}
+2\Delta_{MN\bot}
\left(\Delta_{C}+\Delta_{M}-\Delta_{AC}+\Delta_{SOz}-\Delta_{PM}-\Delta_{MNz}-\Delta_{AMNz}
+\Delta_{CRz}\right)\nonumber
\\
&&\times\left(\mathcal{G}_{1}^{z}-\mathcal{G}_{1}^{D}+\mathcal{G}_{1}^{m}\right)\nonumber
\\
&&+2\left[\left(\Delta_{C}\Delta_{PM}+\Delta_{M}\Delta_{CRz}+\Delta_{AC}\Delta_{MNz}+4\Delta_{SO\bot}\Delta_{AMN\bot}+\Delta_{SOz}\Delta_{AMNz}\right)\mathcal{G}_{1}^{\bot}\right.\nonumber
\\
&&+2\left(\Delta_{M}\Delta_{CR\bot}+\Delta_{AC}\Delta_{MN\bot}+\Delta_{SO\bot}\Delta_{AMNz}+\Delta_{SOz}\Delta_{AMN\bot}\right)\mathcal{G}_{1}^{z}\nonumber
\\
&&+2\left(\Delta_{M}\Delta_{SO\bot}+\Delta_{PM}\Delta_{MN\bot}+\Delta_{AMN\bot}\Delta_{CRz}+\Delta_{AMNz}\Delta_{CR\bot}\right)\mathcal{G}_{1}^{D}\nonumber
\\
&&\left.+2\left(\Delta_{C}\Delta_{AMN\bot}+\Delta_{M}\Delta_{MN\bot}+\Delta_{AC}\Delta_{CR\bot}+\Delta_{SO\bot}\Delta_{PM}\right)\mathcal{G}_{1}^{m}\right]
\nonumber
\\
&&+2\Delta_{MN\bot}
\alpha\left[\frac{1}{2}\left(-\mathcal{F}_{0}^{\bot}+\mathcal{F}_{0}^{z}\right)+\frac{1}{4}\mathcal{F}_{2}^{\bot}
-\frac{1}{4}\mathcal{F}_{2}^{z}-\mathcal{F}_{3}^{z}
+\mathcal{F}_{3}^{D}-\mathcal{F}_{3}^{m}\right],
\\
\frac{d\Delta_{MNz}}{d\ell}&=&-\Delta_{MNz}+2\Delta_{MNz}
\left(\Delta_{C}+\Delta_{M}-\Delta_{AC}+2\Delta_{SO\bot}-\Delta_{SOz}-\Delta_{PM}-2\Delta_{MN\bot}+\Delta_{MNz}
\right.\nonumber
\\
&&\left.-2\Delta_{AMN\bot}+\Delta_{AMNz}+2\Delta_{CR\bot}-\Delta_{CRz}\right)\left(\mathcal{G}_{1}^{\bot}-\mathcal{G}_{1}^{z}-\mathcal{G}_{1}^{D}+\mathcal{G}_{1}^{m}\right)\nonumber
\\
&&+4\left[\left(\Delta_{M}\Delta_{CR\bot}+\Delta_{AC}\Delta_{MN\bot}+\Delta_{SO\bot}\Delta_{AMNz}+\Delta_{SOz}\Delta_{AMN\bot}\right)\mathcal{G}_{1}^{\bot}\right.\nonumber
\\
&&+\left(\Delta_{C}\Delta_{PM}+2\Delta_{SO\bot}\Delta_{AMN\bot}+\Delta_{SOz}\Delta_{AMNz}\right)\mathcal{G}_{1}^{z}\nonumber
\\
&&+\left(\Delta_{M}\Delta_{SOz}+\Delta_{PM}\Delta_{MNz}+2\Delta_{AMN\bot}\Delta_{CR\bot}\right)\mathcal{G}_{1}^{D}\nonumber
\\
&&\left.+\left(\Delta_{C}\Delta_{AMNz}+\Delta_{M}\Delta_{MNz}+\Delta_{AC}\Delta_{CRz}+\Delta_{SOz}\Delta_{PM}\right)\mathcal{G}_{1}^{m}\right]
\nonumber
\\
&&+2\Delta_{MNz}\alpha
\left[\frac{1}{2}\left(-\mathcal{F}_{0}^{\bot}+\mathcal{F}_{0}^{z}\right)+\frac{1}{4}\mathcal{F}_{2}^{\bot}
-\frac{1}{4}\mathcal{F}_{2}^{z}-2\mathcal{F}_{3}^{\bot}+\mathcal{F}_{3}^{z}
+\mathcal{F}_{3}^{D}-\mathcal{F}_{3}^{m}\right],
\\
\frac{d\Delta_{AMN\bot}}{d\ell}&=&-\Delta_{AMN\bot}+2\Delta_{AMN\bot}\left(\Delta_{C}+\Delta_{M}+\Delta_{AC}-\Delta_{SOz}+\Delta_{PM}-\Delta_{MNz}-\Delta_{AMNz}
-\Delta_{CRz}\right)\nonumber
\\
&&\times\left(
-\mathcal{G}_{1}^{z}+\mathcal{G}_{1}^{D}+\mathcal{G}_{1}^{m}\right)\nonumber
\\
&&+2\left[\left(\Delta_{C}\Delta_{CRz}+\Delta_{M}\Delta_{PM}+\Delta_{AC}\Delta_{AMNz}+4\Delta_{SO\bot}\Delta_{MN\bot}+\Delta_{SOz}\Delta_{MNz}\right)\mathcal{G}_{1}^{\bot}\right.\nonumber
\\
&&+2\left(\Delta_{C}\Delta_{CR\bot}+\Delta_{AC}\Delta_{AMN\bot}+\Delta_{SO\bot}\Delta_{MNz}+\Delta_{SOz}\Delta_{MN\bot}\right)\mathcal{G}_{1}^{z}\nonumber
\\
&&+2\left(\Delta_{C}\Delta_{SO\bot}+\Delta_{PM}\Delta_{AMN\bot}+\Delta_{MN\bot}\Delta_{CRz}+\Delta_{MNz}\Delta_{CR\bot}\right)\mathcal{G}_{1}^{D}\nonumber
\\
&&\left.+2\left(\Delta_{C}\Delta_{MN\bot}+\Delta_{M}\Delta_{AMN\bot}+\Delta_{SO\bot}\Delta_{CRz}+\Delta_{SOz}\Delta_{CR\bot}\right)\mathcal{G}_{1}^{m}\right]
\nonumber
\\
&&+2\Delta_{AMN\bot}\alpha
\left[\frac{1}{2}\left(-\mathcal{F}_{0}^{\bot}+\mathcal{F}_{0}^{z}\right)+\frac{1}{4}\mathcal{F}_{2}^{\bot}
-\frac{1}{4}\mathcal{F}_{2}^{z}+\mathcal{F}_{3}^{z}-\mathcal{F}_{3}^{D}
-\mathcal{F}_{3}^{m}\right],
\\
\frac{d\Delta_{AMNz}}{d\ell}&=&-\Delta_{AMNz}+2\Delta_{AMNz}\left(\Delta_{C}+\Delta_{M}+\Delta_{AC}-2\Delta_{SO\bot}+\Delta_{SOz}+\Delta_{PM}
-2\Delta_{MN\bot}+\Delta_{MNz}\right.\nonumber
\\
&&\left.-2\Delta_{AMN\bot}+\Delta_{AMNz}-2\Delta_{CR\bot}+\Delta_{CRz}\right)
\left(-\mathcal{G}_{1}^{\bot}
+\mathcal{G}_{1}^{z}+\mathcal{G}_{1}^{D}+\mathcal{G}_{1}^{m}\right)\nonumber
\\
&&+4\left[\left(\Delta_{C}\Delta_{CR\bot}+\Delta_{AC}\Delta_{AMN\bot}+\Delta_{SO\bot}\Delta_{MNz}+\Delta_{SOz}\Delta_{MN\bot}\right)\mathcal{G}_{1}^{\bot}\right.\nonumber
\\
&&+\left(\Delta_{M}\Delta_{PM}+2\Delta_{SO\bot}\Delta_{MN\bot}+\Delta_{SOz}\Delta_{MNz}\right)\mathcal{G}_{1}^{z}\nonumber
\\
&&+\left(\Delta_{C}\Delta_{SOz}+\Delta_{PM}\Delta_{AMNz}+2\Delta_{MN\bot}\Delta_{CR\bot}\right)\mathcal{G}_{1}^{D}\nonumber
\\
&&\left.+\left(\Delta_{C}\Delta_{MNz}+\Delta_{M}\Delta_{AMNz}+2\Delta_{SO\bot}\Delta_{CR\bot}\right)\mathcal{G}_{1}^{m}\right]
\nonumber
\\
&&+2\Delta_{AMNz}\alpha
\left[\frac{1}{2}\left(-\mathcal{F}_{0}^{\bot}+\mathcal{F}_{0}^{z}\right)+\frac{1}{4}\mathcal{F}_{2}^{\bot}
-\frac{1}{4}\mathcal{F}_{2}^{z}+2\mathcal{F}_{3}^{\bot}-\mathcal{F}_{3}^{z}-\mathcal{F}_{3}^{D}
-\mathcal{F}_{3}^{m}\right],
\\
\frac{d\Delta_{CR\bot}}{d\ell}&=&-\Delta_{CR\bot}+2\Delta_{CR\bot}\left(-\Delta_{C}
+\Delta_{M}-\Delta_{AC}-\Delta_{SOz}+\Delta_{PM}-\Delta_{MNz}
+\Delta_{AMNz}
+\Delta_{CRz}\right)\nonumber
\\
&&\times\left(\mathcal{G}_{1}^{z}+\mathcal{G}_{1}^{D}+\mathcal{G}_{1}^{m}\right)
\nonumber
\\
&&+2\left[\left(\Delta_{C}\Delta_{AMNz}+\Delta_{M}\Delta_{MNz}+\Delta_{AC}\Delta_{CRz}+\Delta_{SOz}\Delta_{PM}\right)\mathcal{G}_{1}^{\bot}\right.\nonumber
\\
&&+2\left(\Delta_{C}\Delta_{AMN\bot}+\Delta_{M}\Delta_{MN\bot}+\Delta_{AC}\Delta_{CR\bot}+\Delta_{SO\bot}\Delta_{PM}\right)\mathcal{G}_{1}^{z}\nonumber
\\
&&+2\left(\Delta_{AC}\Delta_{SO\bot}+\Delta_{PM}\Delta_{CR\bot}+\Delta_{MN\bot}\Delta_{AMNz}+\Delta_{MNz}\Delta_{AMN\bot}\right)\mathcal{G}_{1}^{D}\nonumber
\\
&&\left.+2\left(\Delta_{M}\Delta_{CR\bot}+\Delta_{AC}\Delta_{MN\bot}+\Delta_{SO\bot}\Delta_{AMNz}+\Delta_{SOz}\Delta_{AMN\bot}\right)\mathcal{G}_{1}^{m}\right]
\nonumber
\\
&&+2\Delta_{CR\bot}\alpha
\left[\frac{1}{2}\left(-\mathcal{F}_{0}^{\bot}+\mathcal{F}_{0}^{z}\right)+\frac{1}{4}\mathcal{F}_{2}^{\bot}
-\frac{1}{4}\mathcal{F}_{2}^{z}+\mathcal{F}_{3}^{z}+\mathcal{F}_{3}^{D}+\mathcal{F}_{3}^{m}\right],
\\
\frac{d\Delta_{CRz}}{d\ell}&=&-\Delta_{CRz}+2\Delta_{CRz}
\left(-\Delta_{C}+\Delta_{M}-\Delta_{AC}-2\Delta_{SO\bot}+\Delta_{SOz}+\Delta_{PM}
-2\Delta_{MN\bot}+\Delta_{MNz}\right.\nonumber
\\
&&\left.+2\Delta_{AMN\bot}-\Delta_{AMNz}+2\Delta_{CR\bot}
-\Delta_{CRz}\right)\left(\mathcal{G}_{1}^{\bot}
-\mathcal{G}_{1}^{z}+\mathcal{G}_{1}^{D}+\mathcal{G}_{1}^{m}\right)\nonumber
\\
&&+4\left[\left(\Delta_{C}\Delta_{AMN\bot}+\Delta_{M}\Delta_{MN\bot}+\Delta_{AC}\Delta_{CR\bot}+\Delta_{SO\bot}\Delta_{PM}\right)\mathcal{G}_{1}^{\bot}\right.\nonumber
\\
&&+\left(\Delta_{AC}\Delta_{SOz}+\Delta_{PM}\Delta_{CRz}+2\Delta_{MN\bot}\Delta_{AMN\bot}\right)\mathcal{G}_{1}^{D}\nonumber
\\
&&\left.+\left(\Delta_{M}\Delta_{CRz}+\Delta_{AC}\Delta_{MNz}+2\Delta_{SO\bot}\Delta_{AMN\bot}\right)\mathcal{G}_{1}^{m}\right]
\nonumber
\\
&&+2\Delta_{CRz}\alpha
\left[\frac{1}{2}\left(-\mathcal{F}_{0}^{\bot}+\mathcal{F}_{0}^{z}\right)+\frac{1}{4}\mathcal{F}_{2}^{\bot}
-\frac{1}{4}\mathcal{F}_{2}^{z}+2\mathcal{F}_{3}^{\bot}-\mathcal{F}_{3}^{z}
+\mathcal{F}_{3}^{D}+\mathcal{F}_{3}^{m}\right],
\end{eqnarray}
\end{widetext}
where
\begin{eqnarray}
\mathcal{R}_{v}&=&\mathcal{F}_{0}^{\bot},
\\
\mathcal{R}_{v_{z}}&=&2\mathcal{F}_{0}^{z},
\\
\mathcal{R}_{m}&=&m\mathcal{F}_{0}^{z}
-B_{\bot}\mathcal{F}_{0}^{\bot}-B_{z} \mathcal{F}_{0}^{z},
\\
\mathcal{R}_{B_{\bot}}&=&-m\left(\mathcal{F}_{1}^{\bot}+\mathcal{F}_{1}^{z}\right)+B_{\bot}\left(\mathcal{F}_{1}^{\bot}-\mathcal{F}_{0}^{\bot}\right)\nonumber
\\
&&+B_{z}\mathcal{F}_{1}^{z},
\\
\mathcal{R}_{B_{z}}&=&-m\left[\mathcal{F}_{0}^{\bot}+\mathcal{F}_{0}^{z}
-2\left(\mathcal{F}_{1}^{\bot}+\mathcal{F}_{1}^{z}\right)\right]\nonumber
\\
&&+B_{\bot}\left(\mathcal{F}_{0}^{\bot}-2\mathcal{F}_{1}^{\bot}\right)\nonumber
\\
&&+B_{z}\left(-\mathcal{F}_{0}^{\bot}+\mathcal{F}_{0}^{z}-2\mathcal{F}_{1}^{z}
-\mathcal{F}_{2}^{z}+\mathcal{F}_{2}^{\bot}\right),
\\
\mathcal{R}_{D}&=&\left(\mathcal{F}_{1}^{D}-\mathcal{F}_{0}^{\bot}\right),
\\
\mathcal{R}_{\eta}&=&\left( \mathcal{F}_{2}^{z}-\mathcal{F}_{2}^{\bot}\right),
\\
\mathcal{R}_{g}&=&\frac{1}{2}\mathcal{F}_{2}^{\bot},
\\
\mathcal{R}_{\alpha}&=&\mathcal{F}_{0}^{\bot}
+\frac{1}{2}\left(\mathcal{F}_{2}^{\bot}+\mathcal{F}_{2}^{z}\right),
\\
\mathcal{R}_{\zeta}&=&2\mathcal{F}_{0}^{z}-\mathcal{F}_{0}^{\bot}
-\frac{1}{2}\left(\mathcal{F}_{2}^{z}- \mathcal{F}_{2}^{\bot}\right),
\\
C_{0}^{dis}&=&\left(\Delta_{C}+\Delta_{M}+\Delta_{AC}+2\Delta_{SO\bot}+\Delta_{SOz}\right.\nonumber
\\
&&+\Delta_{PM}+2\Delta_{MN\bot}+\Delta_{MNz}+2\Delta_{AMN\bot}\nonumber
\\
&&\left.+\Delta_{AMNz}+2\Delta_{CR\bot}
+\Delta_{CRz}\right)\nonumber
\\
&&\times\left(\mathcal{G}_{0}^{\bot}+\mathcal{G}_{0}^{z}\right), \label{Eq:C0DisFinal}
\\
C_{m}^{dis}&=&-\left(\Delta_{C}+\Delta_{M}-\Delta_{AC}-2\Delta_{SO\bot}-\Delta_{SOz}\right.\nonumber
\\
&&-\Delta_{PM}
+2\Delta_{MN\bot}+\Delta_{MNz}+2\Delta_{AMN\bot}\nonumber
\\
&&\left.+\Delta_{AMNz}-2\Delta_{CR\bot}-\Delta_{CRz}\right)\nonumber
\\
&&\times\frac{1}{m}\left[m\left(\mathcal{G}_{0}^{\bot}+\mathcal{G}_{0}^{z}\right)
-B_{\bot}\mathcal{G}_{0}^{\bot}
-B_{z}\mathcal{G}_{0}^{z}\right], \label{Eq:CmDisFinal}
\end{eqnarray}
with
\begin{eqnarray}
\mathcal{F}_{0}^{\bot}&=&\frac{1}{4\pi}\int_{0}^{\pi}\sin(\varphi)d\varphi\int_{0}^{2\pi}d\theta\frac{\sin^{2}(\varphi)}
{\Xi},
\\
\mathcal{F}_{0}^{z}&=&\frac{1}{4\pi}\int_{0}^{\pi}\sin(\varphi)d\varphi\int_{0}^{2\pi}d\theta\frac{\cos^{2}(\varphi)}
{\Xi},
\\
\mathcal{F}_{1}^{\bot}&=&-\frac{1}{4\pi}\int_{0}^{\pi}\sin(\varphi)d\varphi\int_{0}^{2\pi}d\theta\nonumber
\\
&&\times\frac{\sin^{2}(\varphi)\cos(2\varphi)}
{\Xi},
\\
\mathcal{F}_{1}^{z}&=&-\frac{1}{4\pi}\int_{0}^{\pi}\sin(\varphi)d\varphi\int_{0}^{2\pi}d\theta\nonumber
\\
&&\times\frac{\cos^{2}(\varphi)\cos(2\varphi)}
{\Xi},
\\
\mathcal{F}_{1}^{D}&=&\frac{1}{2\pi}\int_{0}^{\pi}\sin(\varphi)d\varphi\int_{0}^{2\pi}d\theta\nonumber
\\
&&\times\frac{\sin^{4}(\varphi)\cos^{2}(2\theta)}
{\Xi},
\end{eqnarray}
\begin{widetext}
\begin{eqnarray}
\mathcal{F}_{2}^{x}&=&\frac{1}{4\pi}\int_{0}^{\pi}\sin(\varphi)d\varphi
\int_{0}^{2\pi}d\theta
\left\{\frac{1+4\left(D^{2}+B_{\bot}^{2}\right)
\sin^{2}(\varphi)\cos^{2}(\theta)}{\Xi^{3}}
\right.\nonumber
\\
&&\left.-\frac{\left[1+2D^{2}\sin^{2}(\varphi)\cos(2\theta)-2mB_{\bot}
+2B_{\bot}^{2}\sin^{2}(\varphi)+2B_{\bot}B_{z}\cos^{2}(\varphi)\right]^{2}\sin^{2}(\varphi)
\cos^{2}(\theta)}
{\Xi^{5}}
\right\},
\\
\mathcal{F}_{2}^{y}&=&\frac{1}{4\pi}\int_{0}^{\pi}\sin(\varphi)d\varphi
\int_{0}^{2\pi}d\theta
\left\{\frac{1+4\left(D^{2}+B_{\bot}^{2}\right)\sin^{2}(\varphi)\sin^{2}(\theta)}
{\Xi^{3}}\right.\nonumber
\\
&&\left.-\frac{\left[1-2D^{2}\sin^{2}(\varphi)\cos(2\theta)-2m
B_{\bot}+2B_{\bot}^{2}\sin^{2}(\varphi)
+2B_{\bot}B_{z}\cos^{2}(\varphi)\right]^{2}\sin^{2}(\varphi)\sin^{2}(\theta)}
{\Xi^{5}}\right\},
\\
\mathcal{F}_{2}^{z}&=&\frac{1}{4\pi}\int_{0}^{\pi}\sin(\varphi)d\varphi
\int_{0}^{2\pi}d\theta\left\{\frac{\zeta^{2}+4B_{z}^{2}\cos^{2}(\varphi)}
{\Xi^{3}}-\frac{\left(\zeta^{2}-2mB_{z}
+2B_{z}^{2}\cos^{2}(\varphi)+2B_{z}B_{\bot}
\sin^{2}(\varphi)\right)^{2}\cos^{2}(\varphi)}{\Xi^{5}}\right\},\nonumber
\\
\end{eqnarray}
\end{widetext}
\begin{eqnarray}
\mathcal{F}_{3}^{\bot}&=&\frac{1}{16\pi}\int_{0}^{\pi}\sin(\varphi)d\varphi\int_{0}^{2\pi}d\theta\frac{\sin^{2}(\varphi)}{\Xi^{3}},
\\
\mathcal{F}_{3}^{z}&=&\frac{1}{8\pi}\int_{0}^{\pi}\sin(\varphi)d\varphi\int_{0}^{2\pi}d\theta\frac{\zeta^{2}\cos^{2}(\varphi)}{\Xi^{3}},
\\
\mathcal{F}_{3}^{D}&=&\frac{1}{8\pi}\int_{0}^{\pi}\sin(\varphi)d\varphi\int_{0}^{2\pi}d\theta\nonumber
\\
&&\times\frac{D^{2}\sin^{4}(\varphi)\cos^{2}(2\theta)}
{\Xi^{3}},
\\
\mathcal{F}_{3}^{m}&=&\frac{1}{8\pi}\int_{0}^{\pi}\sin(\varphi)d\varphi\int_{0}^{2\pi}d\theta\nonumber
\\
&&\times\frac{\left(m-B_{\bot}\sin^{2}(\varphi)-B_{z}\cos^{2}(\varphi)\right)^{2}}{\Xi^{3}},
\\
\mathcal{G}_{0}^{\bot}&=&\frac{1}{4\pi}\int_{0}^{\pi}d\varphi\sin(\varphi)\int_{0}^{2\pi}d\theta \frac{\sin^{2}(\varphi)}{\Xi^{2}},
\\
\mathcal{G}_{0}^{z}&=&\frac{1}{4\pi}\int_{0}^{\pi}d\varphi\sin(\varphi)\int_{0}^{2\pi}d\theta\frac{\cos^{2}(\varphi)}{\Xi^{2}},
\\
\mathcal{G}_{1}^{\bot}&=&\frac{1}{4\pi}\int_{0}^{\pi}d\varphi\sin(\varphi)\int_{0}^{2\pi}d\theta\frac{\sin^{2}(\varphi)}
{\Xi^{4}},
\\
\mathcal{G}_{1}^{z}&=&\frac{1}{4\pi}\int_{0}^{\pi}d\varphi\sin(\varphi)\int_{0}^{2\pi}d\theta \frac{\zeta^{2}\cos^{2}(\varphi)}{\Xi^{4}},
\\
\mathcal{G}_{1}^{D}&=&\frac{1}{4\pi}\int_{0}^{\pi}d\varphi\sin(\varphi)\int_{0}^{2\pi}d\theta \nonumber
\\
&&\times\frac{D^{2}\sin^{4}(\varphi)\cos^{2}(2\theta)}
{\Xi^{4}},
\\
\mathcal{G}_{1}^{m}&=&\frac{1}{4\pi}\int_{0}^{\pi}d\varphi\sin(\varphi)\int_{0}^{2\pi}d\theta\nonumber
\\
&&\times\frac{\left(m-B_{\bot}\sin^{2}(\varphi)
-B_{z}\cos^{2}(\varphi)\right)^{2}}
{\Xi^{4}},
\end{eqnarray}
where
\begin{eqnarray}
\Xi&=&\Big[\sin^{2}(\varphi)+\zeta^{2}\cos^{2}(\varphi)+D^{2}\sin^{4}(\varphi)\cos^{2}(2\theta)\nonumber
\\
&&+\left(m-B_{\bot}\sin^{2}(\varphi)-B_{z}\cos^{2}(\varphi)\right)^{2}\Big]^{\frac{1}{2}}.
\end{eqnarray}

Employing the relation
\begin{eqnarray}
\mathcal{F}_{1}^{\bot}+\mathcal{F}_{1}^{z}&=&\mathcal{F}_{0}^{\bot}-\mathcal{F}_{0}^{z},
\end{eqnarray}
we can find that $\mathcal{R}_{B_{z}}$ can be also written as
\begin{eqnarray}
\mathcal{R}_{B_{z}}
&=&-m\left(-\mathcal{F}_{0}^{\bot}+3\mathcal{F}_{0}^{z}\right)
+B_{\bot}\left(\mathcal{F}_{0}^{\bot}-2\mathcal{F}_{1}^{\bot}\right)\nonumber
\\
&&+B_{z}\left(-\mathcal{F}_{1}^{\bot}-3\mathcal{F}_{1}^{z}
-\mathcal{F}_{2}^{z}+\mathcal{F}_{2}^{\bot}\right).
\end{eqnarray}

\end{document}